\documentclass[epj,nopacs,onecolumn]{svjour}
\usepackage[sort&compress,numbers]{natbib}
\usepackage{amssymb}
\usepackage{graphicx}
\usepackage{geometry}
\usepackage[dvipsnames,table]{xcolor}
\usepackage{slashed}
\usepackage{hyperref} 
\usepackage{xspace}
\usepackage[tight]{subfigure}
\usepackage{amsmath}
\usepackage{amsfonts}
\usepackage{mathrsfs}
\usepackage{comment}
\usepackage{afterpage}
\usepackage{verbatim}
\usepackage{booktabs}
\usepackage{stackengine}
\usepackage{mathtools}
\usepackage{pgffor}
\usepackage{array}
\usepackage[title,titletoc]{appendix}
\usepackage{tabularx}

\newcolumntype{C}[1]{>{\centering\arraybackslash}p{#1}}

\onecolumn

\def\reffi#1{\mbox{Figure~\ref{#1}}}
\def\reffis#1{\mbox{Figures~\ref{#1}}}

\def\refse#1{\mbox{Section~\ref{#1}}}
\def\refses#1{\mbox{Sections~\ref{#1}}}

\def\citere#1{\mbox{Ref.~\cite{#1}}}
\def\citeres#1{\mbox{Refs.~\cite{#1}}}

\newcommand{\newc}{\newcommand}
\newc{\beq}{\begin{equation}}
\newc{\eeq}{\end{equation}}
\newc{\beqn}{\begin{eqnarray}}
\newc{\eeqn}{\end{eqnarray}}
\newc{\bit}{\begin{itemize}}
\newc{\eit}{\end{itemize}}
\newc{\ben}{\begin{enumerate}}
\newc{\een}{\end{enumerate}}
\newc{\bce}{\begin{center}}
\newc{\ece}{\end{center}}
\newc{\bfi}{\begin{figure}}
\newc{\efi}{\end{figure}}

\newcommand{\rd}{\mathrm d}

\newcommand{\rT}{{\mathrm{T}}}
\newcommand{\rR}{{\mathrm{R}}}
\newcommand{\rL}{{\mathrm{L}}}
\newcommand{\rF}{{\mathrm{F}}}

\newcommand{\ie}{\emph{i.e.}\ }
\newcommand{\eg}{\emph{e.g.}\ }

\newcommand{\GeV}{\ensuremath{\,\text{GeV}}\xspace}
\newcommand{\TeV}{\ensuremath{\,\text{TeV}}\xspace}
\newcommand{\fb}{{\ensuremath\unskip\,\text{fb}}\xspace}

\newcommand{\PH}{\ensuremath{\text{H}}\xspace}

\newcommand{\Pp}{\ensuremath{\text{p}}}
\newcommand{\Pe}{\ensuremath{\text{e}}\xspace}
\newcommand{\Pb}{\ensuremath{\text{b}}\xspace}

\newcommand{\Pg}{\ensuremath{\text{g}}}

\newcommand{\PW}{\ensuremath{\text{W}}\xspace}
\newcommand{\PZ}{\ensuremath{\text{Z}}\xspace}

\newcommand{\MWOS}{\ensuremath{M_\PW^\text{OS}}\xspace}
\newcommand{\MW}{\ensuremath{M_\PW}\xspace}
\newcommand{\MZOS}{\ensuremath{M_\PZ^\text{OS}}\xspace}
\newcommand{\MZ}{\ensuremath{M_\PZ}\xspace}

\newcommand{\GZOS}{\ensuremath{\Gamma_\PZ^\text{OS}}\xspace}

\newcommand{\GWOS}{\ensuremath{\Gamma_\PW^\text{OS}}\xspace}

\newcommand{\MVOS}{\ensuremath{M_{V}^\text{OS}}\xspace}%
\newcommand{\GVOS}{\ensuremath{\Gamma_{V}^\text{OS}}\xspace}%

\newcommand{\recola}{{\sc Recola}\xspace}
\newcommand{\Sherpa}{{\sc Sherpa}\xspace}

\newcommand{\mocanlo}{{\sc MoCaNLO}\xspace}
\newcommand{\powheg}{{\sc Powheg-Box-Res}\xspace}
\newcommand{\collier}{{\sc Collier}\xspace}

\newcolumntype{.}{D{.}{.}{-1}}
\newcolumntype{d}[1]{D{.}{.}{#1}}
\colorlet{tableoverheadcolor}{gray!37.5}
\colorlet{tableheadcolor}{gray!25}
\colorlet{tablerowcolor}{gray!12.5}

\def\draftdate{\relax}
\def\mda{\relax}
\def\mua{\relax}
\def\mla{\relax}
\def\draft{
\def\thtystars{******************************}
\def\sixtystars{\thtystars\thtystars}
\typeout{}
\typeout{\sixtystars**}
\typeout{* Draft mode!
         For final version remove \protect\draft\space in source file *}
\typeout{\sixtystars**}
\typeout{}
\def\draftdate{\today}
\def\mua{\marginpar[\boldmath\hfil$\uparrow$]%
                   {\boldmath$\uparrow$\hfil}\color{black}%
                    \typeout{marginpar: $\uparrow$}\ignorespaces}
\def\mda{\color{red}\marginpar[\boldmath\hfil$\downarrow$]%
                   {\boldmath$\downarrow$\hfil}%
                    \typeout{marginpar: $\downarrow$}\ignorespaces}
\def\mla{\marginpar[\boldmath\hfil$\rightarrow$]%
                   {\boldmath$\leftarrow $\hfil}%
                    \typeout{marginpar: $\leftrightarrow$}\ignorespaces}
\def\Mua{\marginpar[\boldmath\hfil$\Uparrow$]%
                   {\boldmath$\Uparrow$\hfil}\color{black}%
                    \typeout{marginpar: $\uparrow$}\ignorespaces}
\def\Mda{\color{red}\marginpar[\boldmath\hfil$\Downarrow$]%
                   {\boldmath$\Downarrow$\hfil}%
                    \typeout{marginpar: $\downarrow$}\ignorespaces}
\def\Mla{\marginpar[\boldmath\hfil\textcolor{red}{$\Rightarrow$}]%
                   {\boldmath\textcolor{red}{$\Leftarrow $}\hfil}%
                    \typeout{marginpar: $\leftrightarrow$}\ignorespaces}
\overfullrule 5pt
\oddsidemargin 15mm
\marginparwidth 29mm
}

\newcommand{\mc}{\mathcal}
\newcommand{\as}{\alpha_{\textrm{s}}}
\newcommand{\pt}[1]{p_{\rT,{#1}}}
\newcommand{\nnb}{\nonumber}

\newcommand{\tl}{\theta^*_{\Pe^+}}
\newcommand{\rU}{{\rm U}}

\newcommand{\brabar}{\scalebox{.3}{(}\raisebox{-1.7pt}{--}\scalebox{.3}{)}} 

\newcommand{\noun}[1]{{\scshape #1}}
\newcommand{\POWHEG}{\noun{Powheg}}

\newcommand{\POWHEGBOXRES}{\noun{Powheg-Box-Res}}

\begin{document}
\title{
  Polarised-boson pairs at the LHC with NLOPS accuracy\\[-3cm]
  \hspace*{14.8cm}\mbox{\small {MPP-2023-257}}\\[3cm]
}

\author{Giovanni Pelliccioli$^{(a)}$ \and Giulia Zanderighi$^{(a,b)}$}
    \institute{
  (a) Max-Planck-Institut f\"ur Physik, Boltzmannstrasse 8, 85748 Garching, Germany\\
   (b) Technische Universit\"at M\"unchen, James-Franck-Strasse 1, 85748 Garching, Germany   
    }

\abstract{
  We present a calculation of inclusive diboson ($\rm WZ, ZZ, WW$) processes at the LHC in the presence of intermediate polarised weak bosons decaying leptonically,
  matching next-to-leading-order accuracy in QCD with parton-shower effects.
  Starting from recent developments in polarised-boson simulation based on the helicity selection at the amplitude level, we have carried out the implementation in the
  \POWHEGBOXRES{} framework, and validated it against independent fixed-order calculations. A phenomenological analysis in realistic LHC setups,
  as well as a comparison with recent ATLAS measurements, are presented.
}

\authorrunning{\emph{
  G.~Pelliccioli \and G.~Zanderighi
}}
\titlerunning{\emph{
  Polarised-boson pairs at the LHC with NLOPS accuracy
}}

\maketitle
\tableofcontents
\section{Introduction}\label{sec:intro}

Isolating the longitudinal mode of massive electroweak (EW) gauge
bosons is a pivotal means of investigating the mechanism responsible
for breaking the EW symmetry.  In the context of the
Standard-Model (SM), the $\PW$ and $\PZ$ bosons obtain a mass and a
longitudinal polarisation mode as a result of their interaction with
the Higgs field.
This implies that quantifying the production rate of longitudinal
bosons in high-energy scattering processes serves as a remarkably
sensitive indicator for the existence of new physics effects that
could disrupt the delicate interplay between the SM's gauge and
scalar sectors.

Due to the inherently unstable nature of EW bosons, determining their
polarisation state is challenging. In principle, this determination is
restricted to interpreting the distribution shapes of their decay
products, which are typically unpolarised. Nevertheless, a deeper
understanding of the polarisation structure in high-energy processes
can be achieved through the definition of polarised signals also for
intermediate bosons, and their full simulation with Monte Carlo (MC)
generators.

The experimental investigation of the polarisation structure of
multi-boson processes with Run-2 data has already yielded several
results in inclusive diboson production and in vector-boson
scattering~\cite{Aaboud:2019gxl,Sirunyan:2020gvn,CMS:2021icx,ATLAS:2022oge,ATLAS:2023zrv}.
These results are consistent with the predictions of the Standard
Model.
Compared to the Run-1 analyses,
the latest approach for
polarisation measurements involves a fitting procedure using polarised
templates generated with Monte Carlo tools. This approach allows for a
more refined understanding of the spin structure of a process,
including off-shell and interference effects. Achieving this level of
detail requires a precise and accurate theoretical description of the
production and decay of polarised electroweak bosons.

In order to meet the experimental needs, a natural definition of
polarised-boson signals has been recently proposed, based on the
separation of helicity states in resonant amplitudes (either SM or
beyond-the-SM) and ensuring gauge invariance by means of a pole
\cite{Stuart:1991cc,Stuart:1991xk,Aeppli:1993cb,Aeppli:1993rs,Denner:2000bj,Denner:2005fg,Denner:2019vbn}
or narrow-width
\cite{Richardson:2001df,Uhlemann:2008pm,Artoisenet:2012st}
approximation. Initially proposed for vector-boson scattering
\cite{Ballestrero:2017bxn,BuarqueFranzosi:2019boy} at leading order
(LO), the simulation of intermediate polarised bosons in LHC processes
has been extended to next-to-leading (NLO) accuracy
\cite{Denner:2020bcz,Denner:2020eck} and next-to-next-to-leading order
(NNLO) accuracy \cite{Poncelet:2021jmj} in QCD for inclusive diboson
with leptonic decays.  The calculation of NLO EW corrections, which is
more involved owing to photons that can be radiated off both the
production and the decay partons, has been achieved for $\PZ\PZ$
\cite{Denner:2021csi} and $\PW\PZ$
\cite{Le:2022lrp,Le:2022ppa,Dao:2023pkl} inclusive production with
leptonic decay. A similar structure of the radiative corrections
concerns the NLO QCD corrections to diboson in the semi-leptonic decay
channel, which has been studied recently \cite{Denner:2022riz}.

Although the SM predictions for off-shell diboson production (with
leptonic decays) matched to parton shower (PS) have reached NNLO QCD
\cite{Alioli:2016xab,Re:2018vac,Alioli:2021egp,Alioli:2021wpn,Buonocore:2021fnj,Lombardi:2021rvg,Lindert:2022qdd}
and NLO EW accuracy
\cite{Brauer:2020kfv,Chiesa:2020ttl,Lindert:2022qdd}, the event
generators that could be used by ATLAS and CMS in Run-2 analyses
\cite{Aaboud:2019gxl,Sirunyan:2020gvn,CMS:2021icx,ATLAS:2022oge,ATLAS:2023zrv}
to simulate polarised signals with PS matching only feature LO
accuracy
\cite{Ballestrero:2007xq,Alwall:2014hca,Ballestrero:2017bxn,BuarqueFranzosi:2019boy}.
This situation has induced the experimental collaborations to employ
reweighting techniques to account for higher-order effects.  Beside
being involved, a reweighting procedure introduces an additional
uncertainty which is difficult to estimate.

An effort is therefore needed from the theory side to narrow the gap
between SM predictions and experimental needs, towards an accurate and
realistic modelling of polarised bosons at the LHC. A first step
toward this goal has been recently provided by the automation of
intermediate polarised-boson simulation at approximate NLO (nLO) QCD
accuracy matched to PS in the {\sc Sherpa} framework \cite{SherpaPOL}.

In this paper we perform a further step in this direction by reaching
exact NLO QCD accuracy matched to PS (NLOPS) for polarised diboson
processes.

We present an implementation in the \POWHEGBOXRES{} framework
\cite{Nason:2004rx,Frixione:2007vw,Alioli:2010xd,Jezo:2015aia}, based on the public code
of \citere{Chiesa:2020ttl}, of inclusive diboson production at NLOPS accuracy in the fully leptonic
decay channel, with intermediate polarised $\PW$ and $\PZ$
bosons. This represents the first calculation at this accuracy for
polarised-boson processes, and paves the way towards the inclusion of
higher orders in the QCD (NNLO) and EW (NLO) coupling, as well as
towards the extension to more complicated multi-boson signatures at
the LHC.

This paper is structured as follows. In \refse{sec:details} we
describe the details of the implementation and present a validation
against fixed-order results. In \refses{sec:WZ} we perform a
phenomenological study of doubly polarised WZ production. In
\refses{sec:ZZ} and \ref{sec:WW} we show results for the ZZ and
$\PW^+\PW^-$ channels, respectively. The conclusions and outlook are
presented in \refse{sec:con}.


%
\section{Calculation details}\label{sec:details}
In this section we first recall how polarised signals for diboson
production and decay at the LHC are defined. We then describe the
technical changes required to implement polarised processes in the
\POWHEGBOXRES{} framework and present a validation of our results against
fixed-order NLO predictions.

\subsection{Pole approximation and polarisation selection}\label{subsec:polepol}

In order to have a theoretically sound, \ie gauge invariant,
definition of polarised signals, the target bosons must be on their
mass shell. Therefore, the modeling of the production and decay of
polarised bosons requires some additional care, compared to full
off-shell calculations.  Practically speaking, enabling a MC generator
to simulate intermediate EW bosons with fixed polarisation state
requires two main ingredients:
\begin{enumerate}
\item the selection of the resonant contribution to the amplitude in a
  gauge-invariant manner, through an on-shell projection;
\item the selection of individual polarisation states in the resonant
  amplitude.
\end{enumerate}

\begin{figure}[t]
  \centering
  \includegraphics[width=0.5\textwidth]{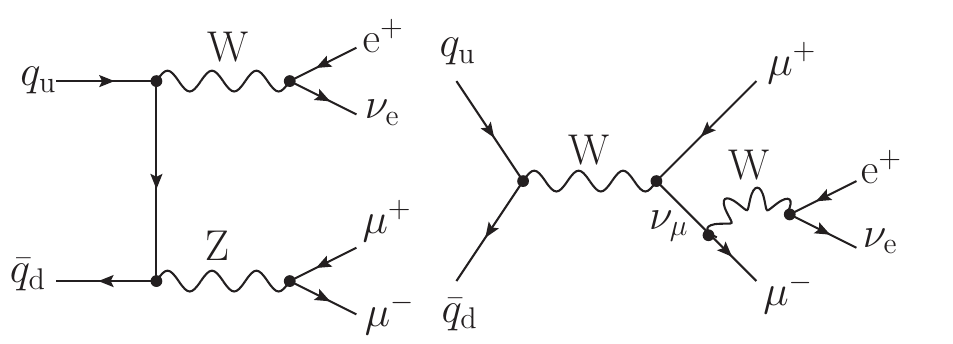}
  \caption{
    Sample tree-level resonant (left) and non-resonant (right) diagrams contributing to WZ production and decay at the LHC.
    Only resonant diagrams must be retained in the DPA procedure.
  }\label{fig:resdiag}
\end{figure}

As depicted in \reffi{fig:resdiag}, out of the complete,
gauge-invariant set of diagrams contributing to a full off-shell
process at a given perturbative order, only resonant contributions
must be retained, dropping non-resonant ones.  For diboson production,
this amounts to selecting diagrams that are factorisable into a
production part, two $s$-channel EW-boson propagators, and one decay
part for each boson. Therefore, the notation \emph{resonant} is
understood as \emph{doubly resonant} for diboson processes. In the
't~Hooft-Feynman gauge this reads,
\beqn
\mc A_{\rm full}(x_1,x_2;\,k_{1\ldots 4}) &=& \mc A_{\rm non-res}(x_1,x_2;\,k_{1\ldots 4}) + \mc A_{\rm res}(x_1,x_2;\,k_{1\ldots 4})\,\longrightarrow
\mc A_{\rm res}(x_1,x_2;\,k_{1\ldots 4})\,,
\eeqn
where,
\beqn
\mc A_{\rm res}(x_1,x_2;\,k_{1\ldots 4}) &=&\mc P_{\mu\nu}(x_1,x_2;\,k_{12},k_{34})\,\nnb\\
&\times&
\frac{-{\rm i}\,g^{\mu\alpha}}{k_{12}^2-M_{1}^2+{\rm i} \Gamma_{1} M_{1}}
\frac{-{\rm i}\,g^{\nu\beta}}{k_{34}^2-M_{2}^2+{\rm i} \Gamma_{2} M_{2}}
\mc D_{\alpha}(k_1,k_2)\,\mc D_{\beta}(k_3,k_4)\,.
\eeqn
We use the notation $k_{ij}=k_i+k_j$,  $M_{1},M_2$ and $\Gamma_1,\Gamma_2$ are the pole masses and decay widths of the two bosons, respectively.
The tensor $\mc P_{\mu\nu}$ describes the sub-amplitude for the production of two bosons, while $\mc D_\alpha, \mc D_\beta$ describe the
decay sub-amplitude for the first and second boson, respectively. The variables $x_1,x_2$ represent the incoming-parton energy fractions.

In order to ensure the EW gauge invariance of the calculation (the
resonant-diagram sub-set is not \emph{per s\'e} gauge invariant), we
employ the double pole approximation (DPA) as used in
\citeres{Denner:2020bcz,Denner:2020eck,Denner:2021csi,Denner:2022riz}.
Starting from off-shell kinematics, the amplitude numerator is
projected on shell (according to the pole masses of the two bosons),
while the propagator denominators are kept with the original off-shell
kinematics.
In formulae,
\beqn\label{eq:dpaappl}
\mc A_{\rm res}(x_1,x_2;\,k_{1\ldots 4}) &\longrightarrow&
\mc A_{\rm res}(x_1,x_2;\,\tilde{k}_{1\ldots 4}) \,=\,\mc P_{\mu\nu}(x_1,x_2;\,\tilde{k}_{12},\tilde{k}_{34})\,\nnb\\
&\times&
\frac{-{\rm i}\,g^{\mu\alpha}}{k_{12}^2-M_{1}^2+{\rm i} \Gamma_{1} M_{1}}
\frac{-{\rm i}\,g^{\nu\beta}}{k_{34}^2-M_{2}^2+{\rm i} \Gamma_{2} M_{2}}
\mc D_{\alpha}(\tilde{k}_1,\tilde{k}_2)\,\mc D_{\beta}(\tilde{k}_3,\tilde{k}_4)\,,
\eeqn
where the momenta ${\tilde{k}_{1\ldots4}}$ fulfil $\tilde{k}^2_{12}=(\tilde{k}_1+\tilde{k}_2)^2=M_1^2$ and $\tilde{k}^2_{34}=(\tilde{k}_3+\tilde{k}_4)^2=M_2^2$, where $M_1$ and $M_2$ are the masses of the two gauge bosons considered. 
The DPA on-shell mapping
\beq
{\rm \Phi}_{4}=\{x_1,x_2;{k}_{1\ldots4}\}\,\overset{\rm DPA}{\longrightarrow}\, \tilde{\rm \Phi}_{4}=\{x_1,x_2;\tilde{k}_{1\ldots4}\}\,,
\eeq
is not unique, but the details of a specific choice
do not affect the numerical results substantially~\cite{Ballestrero:2017bxn}.
A common feature to all such mappings is the presence of a natural
threshold for the invariant mass of the four decay products,
\beq\label{eq:threshold}
(k_1+k_2+k_3+k_4)^2 > (M_{1}+M_{2})^2\,,
\eeq
in order to kinematically allow the production of two on-shell bosons.
For our purposes, we choose the on-shell mapping used in \citere{Denner:2021csi}, which
preserves
\begin{itemize}
\item the total momentum of the diboson system, $k_1+k_2+k_3+k_4$;
\item the spatial direction of $k_{12}$ (and  of $k_{34}$) single boson in the diboson CM frame;
\item the spatial direction of $k_1$ (and of $k_2$) in the rest frame of the system $k_1+k_2$;
\item the spatial direction of $k_3$ (and of $k_4$) in the rest frame of the system $k_3+k_4$;
\item initial-state energy fractions $x_1,x_2$;
\item the momentum associated to the additional radiation at NLO.
\end{itemize}
This turns out to simplify the DPA implementation,
especially for what concerns subtracted-real corrections \cite{Denner:2020bcz}. Notice in fact that the on-shell projection does not
modify the initial-state energy fractions.
Additionally, owing to the conservation of decay angles (direction of decay products in the decayed-boson rest frame),
it represents an optimal choice
for polarisation purposes.

Once the resonant contributions are evaluated in the DPA according to Eq.~\eqref{eq:dpaappl} and squared, the unpolarised-signal
squared matrix element is obtained. In order to select a specific polarisation mode
for the bosons, the DPA amplitudes undergo the replacement of the Lorentz-invariant
tensor structure of the propagator numerators with individual polarisation-vector
contributions. For a single boson with momentum $k$, in the 't~Hooft-Feynman gauge, this reads,
\beq\label{eq:ampnum}
-g_{\mu\nu}\,=\, \sum_{\lambda'} \varepsilon_\mu^{(\lambda')}(k)\,\varepsilon^{(\lambda')\,*}_\nu(k)\longrightarrow\,\varepsilon_\mu^{(\lambda)}(k)\,\varepsilon^{(\lambda)\,*}_\nu(k)\,,\qquad \lambda=\rL,+,-\,,
\eeq 
where the labels $\{\rL,+,-\}$ stand for the longitudinal, right-handed and left-handed physical polarisation states.
Notice that in left side of Eq.~\eqref{eq:ampnum} $\lambda'$ runs over four polarisation states, namely the three physical ones and
a fourth one. This fourth polarisation term is unphysical and always cancels against Goldstone-boson contributions, at any order in perturbation theory.
The squared of the obtained amplitude (for a given $\lambda$ physical state)
multiplied by a flux factor, PDF and phase-space weights, gives the wanted polarised cross section (differential in any observable).

It is a common choice to use a unique transverse state ($\rT$) defined as the coherent sum of the left and right-handed ones,
\ie including their interference.

We also stress that the polarisation vectors in Eq.~\eqref{eq:ampnum} depend on the Lorentz frame where the helicity states
of the bosons are defined. This implies that polarised signals are computed in a specific Lorentz frame.
The most natural choice~\cite{Denner:2020eck,Denner:2021csi,Le:2022lrp,Le:2022ppa} for inclusive diboson production
is the boson-pair centre-of-mass (CM) frame, which is also the choice made in this work.

As a last comment of this section, it is worth noting that in order to define a polarised signal beyond LO,
the DPA application and the polarisation selection have to be applied to all contributions to the cross section,
\ie at NLO they have to be applied to Born, virtual, real and subtraction-counterterm contributions.
As for the polarisation selection, since we choose to define polarisation vectors in the boson-pair CM frame,
care has to be taken in real contributions. In fact, while the diboson CM frame
equals the partonic CM frame (where momenta are evaluated, in the \POWHEG{} code) in Born-like contributions,
they differ by a boost in real contributions. Therefore, it is essential to evaluate real matrix elements
with momenta boosted into the diboson CM frame.

\subsection{Implementation}

We have implemented the whole DPA/helicity-selection machinery in a branch of the \sloppy\powheg code presented
in \citere{Chiesa:2020ttl}.
We have interfaced it with the \recola1 library \cite{Actis:2012qn,Actis:2016mpe}, which
provides tree-level and one-loop SM amplitudes through the \collier tensor-integral reduction and
evaluation \cite{Denner:2016kdg}. \recola1 has recently been modified to enable the selection
of specific helicity states for intermediate resonances, facilitating a number of fixed-order calculations
\cite{Denner:2020bcz,Denner:2020eck,Poncelet:2021jmj,Denner:2021csi,Denner:2022riz}
\footnote{The usage of version 1.4.4 is recommended. It can be downloaded at this link \url{https://recola.hepforge.org/downloads/recola-1.4.4.tar.gz}
and the documentation regarding internal helicity selection can be found here: \url{https://recola.gitlab.io/recola2/api/polsel.html}.}.

In \powheg, the QCD soft and collinear singularities are subtracted in the
FKS scheme~\cite{Frixione:1995ms,Frixione:2007vw,Jezo:2015aia}. It is then crucial that the DPA
procedure preserves the local cancellation between real contributions and subtraction counterterms
in the radiative phase space.
This is achieved applying first the FKS-subtraction mapping, and second the DPA on-shell mapping,
to evaluate the subtracted-real kinematics,
\beq
   \!\! {\rm\Phi_{4\ell}}=\{x_1,x_2;k_{1\ldots4}\} \,\overset{\rm FKS}{\longrightarrow}\,({\bar{\rm\Phi}_{4\ell},\rm\Phi_{\rm rad}})=\{\bar x_1,\bar x_2;{\bar{k}}_{1\ldots4},k_{\rm rad}\}  \,\overset{\rm DPA}{\longrightarrow}\,
   ({\tilde{\bar{\rm\Phi}}_{4\ell},\rm\Phi_{\rm rad}})=\{\bar x_1,\bar x_2;\tilde{\bar{k}}_{1\ldots4},k_{\rm rad}\}\,.
\eeq
It is easy to see that in the case of QCD initial-state radiation (ISR), the only one present in the considered processes, the
cancellation of real-phase-space singularities is guaranteed by the specific choice we make for the DPA mapping (see discussion after Eq.~\eqref{eq:threshold}),
which preserves the colour-singlet total momentum, and therefore commutes with the FKS mapping (a combination of boosts applied democratically to all four decay leptons).
This implies that the subtraction of infrared singularities proceeds in a similar manner as in the full off-shell calculation.
In order to extend this method to hadronic decays at NLO QCD, as well as to NLO EW corrections to leptonic decays,
a separate treatment of ISR and final-state radiation (FSR) is needed to
avoid disrupting the infrared-singularity subtraction. Although not trivial, this has been achieved in
the dipole scheme \cite{Denner:2021csi,Le:2022lrp,Le:2022ppa,Dao:2023pkl,Denner:2022riz} and can be generalised to other subtraction schemes.
We leave this further development for future work.

In the diboson calculation presented here, the \POWHEG{} matching procedure \cite{Nason:2004rx,Frixione:2007vw,Jezo:2015aia}
works as in the off-shell case, up to small technical subtleties.
The usual matching formula, tailored to four-lepton processes in the DPA, reads (the label $\ell$ includes here both charged leptons and neutrinos),
\beq\label{eq:pwgformula}
\langle \mc O \rangle\,=\,\int\!{\rm d\Phi}_{4\ell}\,\tilde{\rm B}({\rm \tilde\Phi}_{4\ell})\,\left[
  {\mc O}({\rm \tilde{\rm \Phi}}_{4\ell}){\rm \Delta}({t_0})
  +\int_{t>t_0}\!{\rm d\Phi}_{\rm rad}\mc O({\tilde{\bar{\rm \Phi}}}_{4\ell},{\rm \Phi_{\rm rad}})\,
  \frac{\rm R({\rm \tilde{\bar{\rm \Phi}}}_{4\ell},\Phi_{\rm rad})}{\rm B(\tilde\Phi_{4\ell})}\,{\rm \Delta}(t)
  \right]\,,
\eeq
where the Sudakov form factor ${\rm \Delta}(t)$ is given by,
\beq
{\rm \Delta}(t) = \exp{\left[-\int_{t'>t}\!{\rm d\Phi}'_{\rm rad}\frac{\rm R({\rm \tilde{\bar{\rm \Phi}}}_{4\ell},\Phi'_{\rm rad})}{\rm B(\tilde\Phi_{4\ell})}\right]}\,,
\eeq
and the ordering variable $t$ is the radiation transverse momentum. 
The $\tilde{\rm B}$ factor includes Born, subtracted-virtual and subtracted-real (including collinear remnants) contributions, 
\beq
\tilde{\rm B}({\rm \tilde \Phi_{4\ell}}) = {\rm B}({\rm \tilde\Phi_{4\ell}}) + {\rm V}_{\rm reg}({\rm \tilde\Phi_{4\ell}}) +
\int\!\rd{\rm \Phi}_{\rm rad}\left[{\rm R(\tilde{\rm \bar\Phi}_{4\ell},\Phi_{\rm rad})- {\rm CT}(\tilde{\rm \bar\Phi}_{4\ell},{\Phi_{\rm rad}})}\right]\,,
\eeq
where $\rm V_{\rm reg}$ stands for the sum of finite virtual contributions and the integrated version of the FKS-subtraction counterterms (CT).
In Eq.~\eqref{eq:pwgformula}, it is understood that while the DPA-mapped kinematics, \ie $\tilde{\rm \Phi}_{4\ell}$ for Born-like contributions and $(\tilde{\bar{\rm \Phi}}_{4\ell},{\rm \Phi}_{\rm rad})$ for real ones, is used for the matrix-element numerators, the original off-shell kinematics, \ie ${\rm \Phi}_{4\ell}$ for Born-like contributions and $(\bar{\rm \Phi}_{4\ell},{\rm \Phi}_{\rm rad})$ for real ones,
is used in matrix-element denominators as well as in phase-space weights. Analogously, possible selection cuts are always applied to the off-shell kinematics.

Compared to the off-shell calculation, the only subtlety concerns the additional kinematic constraint due to the DPA application described by Eq.~\eqref{eq:threshold},
namely $M_{4\ell} > M_{1}+M_{2}\,$.
Practically speaking, 
since in the \POWHEG{} approach the real kinematics is
obtained starting from the  Born-like one,
the Born-like phase-space points that do not satisfy Eq.~\eqref{eq:threshold}
are discarded by setting the corresponding Monte Carlo jacobian to zero, avoiding numerical instabilities in the evaluation of
the ratio $\rm R/B$ in Eq.~\eqref{eq:pwgformula}.

\subsection{Default numerical input}\label{subsec:setup}
The following diboson processes are considered at NLO QCD accuracy matched to PS,
\begin{eqnarray}\label{eq:resprocess}
(\PW\PW)\,:\qquad \Pp\Pp &   \rightarrow          \PW^+_{\lambda}\,(\ell^+\nu_\ell)     &\PW^-_{\lambda'}\,(\ell'^-\bar{\nu}_{\ell'})+X\nnb\,,\\
\,(\PW\PZ)\,\,:\qquad \Pp\Pp &  \rightarrow       \PW^\pm_{\lambda}\,(\ell^\pm\overset{\brabar}{\nu}_\ell)&\,\,\PZ_{\lambda'}\,\,(\ell'^+\ell'^-)+X\nnb\,,\\[0.1cm]
\,(\PZ\PZ)\,\,\,:\qquad \Pp\Pp & \,\, \rightarrow \,\,\PZ_{\lambda} \,\, (\ell^+\ell^-)&\,\,\PZ_{\lambda'}\,\,(\ell'^+\ell'^-)+X\,,
\end{eqnarray}
where the polarisation states $\lambda, \lambda'$ can take the values $\{\rL,+,-,\rT,\rU\}$ (longitudinal, right-handed, left-handed, transverse, unpolarised).
We consider the fully leptonic decay channel, implying that the matching procedure only involves QCD ISR.
Although the DPA procedure would be the same as in the case of different flavours (up to suitable symmetry factors),
we do not consider the case of identical flavours, owing to subtleties in the experimental selections.

The setting and input parameters can be set in \powheg input card. 
Here we use the following default input parameters. 
The on-shell masses and widths for EW bosons are taken from \citere{ParticleDataGroup:2020ssz}:
\begin{eqnarray}\label{eq:weakpar}
  \MWOS  =  80.379  \GeV \,,\quad
  \GWOS  =   2.085  \GeV \,,\quad
  \MZOS  =  91.1876 \GeV \,,\quad
  \GZOS  =   2.4952 \GeV \,,
\end{eqnarray}
and converted into pole values according to the well-known relations \cite{Bardin:1988xt},
\begin{eqnarray}\label{eq:weakpar}
  M_V &=&{ \MVOS }/{ \sqrt{1+{(\GVOS/\MVOS)}^2 }}\,,\quad V=\PW,\,\PZ\,.
\end{eqnarray}
The electroweak-boson treatment is carried out in the complex-mass scheme
\cite{Denner:2005fg,Denner:2006ic,Denner:2019vbn}
and the electroweak coupling is computed in the $G_\mu$ scheme \cite{Denner:2000bj},
\beq
\alpha\,=\,\frac{G_{\rm F}\sqrt{2}}{\pi}\MW^2\left(1-\frac{\MW^2}{\MZ^2}\right)\,,\qquad G_{\rm F}=1.16638\cdot 10^{-5}\GeV^{-2}\,.
\eeq
All leptons are considered massless.
The five-flavour scheme is employed for all diboson processes. However, $\Pb$-quark-induced processes are excluded in the $\PW^+\PW^-$ case. %
A unit quark-mixing matrix is considered. 
The mass and width of the Higgs boson and the top quark do not enter any of the computations.
The PDFs of the proton and the value of the strong coupling constant $\as$ are evaluated
with the \textsc{LHAPDF} interface \cite{Buckley:2014ana}.
The \sloppy \textsc{NNPDF31\_nlo\_as\_0118} PDF set \cite{Ball:2017nwa} is used for both LO and NLO QCD results, which uses $\alpha_s(M_Z) = 0.118$.
The renormalisation and factorisation scales are both set to the arithmetic average of the boson pole masses
involved in the calculation,
\beq
\mu_{\rR}=\mu_{\rF}=\frac{M_{V_1} + M_{V_2}}2\,.
\eeq

Throughout the whole paper, leptons are understood as dressed with possible photon radiation using a cone dressing
with resolution radius $R=0.1$. Since we only consider NLO QCD corrections, QED radiation can only come from
the parton shower.

\subsection{Fixed-order validation}\label{subsec:valid}
To validate the calculation at fixed order, we have compared the \powheg results
with those obtained with the private Monte Carlo program \mocanlo at LO and at
NLO QCD in an inclusive phase space, characterised only by an invariant-mass cut
applied to same-flavour, opposite-sign lepton pairs coming from Z-boson decays,
\beq\label{eq:mll_cut}
81\GeV < M_{\ell^+\ell^-}<101\GeV\,.
\eeq
The newly developed \powheg code and the \mocanlo one both rely on the \mbox{\recola1} SM-amplitude provider.
The \recola1 amplitudes at tree-level and at one-loop (both QCD and EW) have been
widely validated in the context of fixed-order polarisation studies performed with \mocanlo
\cite{Denner:2020bcz,Denner:2020eck,Denner:2021csi,Denner:2022riz}, by means of internal checks
(EW Ward identities, UV finiteness) and through the comparison against {\sc MadLoop} \cite{Hirschi:2011pa}.

In Table~\ref{tab:polbeams1} we show the integrated cross sections for the
unpolarised and doubly polarised signals for $\PW^+\PW^-,\,\PW^+\PZ$ and $\PZ\PZ$
calculated with the two Monte Carlo programs.
Perfect agreement was found within integration uncertainties for all diboson production channels.
Also the theory uncertainties from 7-point QCD-scale variations, i.e. by varying $\mu_{\rR}$ and $\mu_{\rF}$ independently around the central value with the constraint $\frac12< \mu_{\rF}/\mu_{\rR}<2$, were found in good agreement with
\mocanlo for all processes and polarisation modes.
\begin{table}
  \begin{center}
    \begin{tabular}{ccc}
      \hline\rule{0ex}{2.7ex}
      \cellcolor{blue!9} state & \cellcolor{blue!9}\powheg & \cellcolor{blue!9}\mocanlo \\
      \hline\rule{0ex}{2.7ex}
      \!\!$\PW^+_{\rU}\PW^-_{\rU}$  & $1248.8(9)^{+3.8 \%} _{-3.1 \%}$  & $1249.2(6) ^{+ 3.8 \%}_{- 3.0 \%}$  \\[0.1cm]
      $\PW^+_{\rL}\PW^-_{\rL}$      & $65.88(9)^{+3.2 \%} _{-2.7 \%} $  & $65.90(8)  ^{+ 3.2 \%}_{- 2.8 \%}$  \\[0.1cm]
      $\PW^+_{\rL}\PW^-_{\rT}$      & $158.65(6) ^{+5.2 \%} _{-4.1 \%} $  & $158.60(7) ^{+ 5.1 \%}_{- 4.2 \%}$  \\[0.1cm]
      $\PW^+_{\rT}\PW^-_{\rL}$      & $163.04(7)^{+5.3 \%} _{-4.3 \%}$  & $162.91(7) ^{+ 5.3 \%}_{- 4.3 \%}$  \\[0.1cm]
      $\PW^+_{\rT}\PW^-_{\rT}$      & $861.8(3)^{+3.3 \%} _{-2.6 \%} $  & $860.1(5)  ^{+ 3.3 \%}_{- 2.6 \%}$  \\[0.1cm]
      \hline
    \end{tabular}\qquad
    \begin{tabular}{ccc}
      \hline\rule{0ex}{2.7ex}
       \cellcolor{blue!9} state & \cellcolor{blue!9}\powheg & \cellcolor{blue!9}\mocanlo \\
      \hline\rule{0ex}{2.7ex}
      $\PW^+_{\rU}\,\,\PZ^{\,}_{\rU}$  &$  97.20(3) ^{+4.8 \%} _{-3.9 \%} $   &     $  97.19(3) ^{+4.8 \%} _{-3.9 \%}  $     \\[0.1cm]
      $\PW^+_{\rL}\,\,\PZ^{\,}_{\rL}$  &$   4.499(1) ^{+2.8 \%} _{-2.3 \%}$   &     $  4.496(2) ^{+2.8 \%} _{-2.3 \%}  $    \\[0.1cm]
      $\PW^+_{\rL}\,\,\PZ^{\,}_{\rT}$    & $ 13.151(6)^{+7.0 \%} _{-5.7 \%}$  &     $  13.132(4) ^{+7.0 \%} _{-5.6 \%} $     \\[0.1cm]
      $\PW^+_{\rT}\,\,\PZ^{\,}_{\rL}$    & $ 12.724(6)^{+7.3 \%} _{-5.9 \%}$  &     $  12.716(4) ^{+7.3 \%} _{-5.9 \%} $    \\[0.1cm]
      $\PW^+_{\rT}\,\,\PZ^{\,}_{\rT}$    & $ 66.88(3) ^{+4.0 \%} _{-3.3 \%} $ &     $  66.84(3) ^{+4.0 \%} _{-3.3 \%} $    \\[0.1cm]      
      \hline
    \end{tabular}\\[0.4cm]
    \begin{tabular}{ccc}
      \hline\rule{0ex}{2.7ex}
      \cellcolor{blue!9} state & \cellcolor{blue!9}\powheg & \cellcolor{blue!9}\mocanlo \\
      \hline\rule{0ex}{2.7ex}
      $\PZ^{\,}_{\rU}\,\,\PZ^{\,}_{\rU}$ &  $28.22(1)   ^{+2.9 \%} _{-2.3 \%}$ & $ 28.21(2) ^{+ 2.9 \%}_{- 2.4 \%}$ \\[0.1cm]
      $\PZ^{\,}_{\rL}\,\,\PZ^{\,}_{\rL}$ &  $1.664(1)   ^{+3.0 \%} _{-2.4 \%}$ & $ 1.664(2) ^{+ 3.0 \%}_{- 2.5 \%}$ \\[0.1cm]
      $\PZ^{\,}_{\rL}\,\,\PZ^{\,}_{\rT}$ &  $3.551(2)   ^{+3.7 \%} _{-2.9 \%}$ & $ 3.548(1) ^{+ 3.6 \%}_{- 3.0 \%}$ \\[0.1cm]
      $\PZ^{\,}_{\rT}\,\,\PZ^{\,}_{\rL}$ &  $3.554(2)   ^{+3.7 \%} _{-3.0 \%}$ & $ 3.548(2) ^{+ 3.6 \%}_{- 3.0 \%}$ \\[0.1cm]
      $\PZ^{\,}_{\rT}\,\,\PZ^{\,}_{\rT}$ &  $19.46(1)   ^{+2.6 \%} _{-2.1 \%}$ & $ 19.45(1) ^{+ 2.6 \%}_{- 2.1 \%}$ \\[0.1cm]
      \hline
    \end{tabular}
  \end{center}
  \caption{
    Comparison between \powheg and \mocanlo 
    inclusive cross sections (fb) for unpolarised (U), longitudinal (L), and transverse (T) bosons
    in $\PW^+\PW^-$ \cite{Denner:2020bcz}, $\PW^+\PZ$ \cite{Denner:2020eck}, and $\PZ\PZ$ \cite{Denner:2021csi} at NLO QCD in the DPA.
    The only cut of Eq.~\ref{eq:mll_cut} is applied to lepton pairs from Z-boson decays.
    The errors in parentheses denote that Monte Carlo integration uncertainties, while the percentages in subscripts and superscripts
    are the uncertainties from 7-point QCD-scale variations. 
    \label{tab:polbeams1}
  }
\end{table}
Giving a global look at Table~\ref{tab:polbeams1}, it can be noticed that in a very inclusive setup
the LL cross section accounts for the 5\% of the unpolarised one, the TT one for the 65-70\%,
and the mixed states (LT, TL) for 13\% each. This is true in all production channels, in spite of
differences in the contributing diagrams in the SM. The interference terms are compatible with zero,
as expected in the absence of selection cuts on individual decay products of the bosons.

\begin{figure}[h]
  \centering
  \subfigure[Transverse momentum of $\mu^+$ in $\PW^+\PZ$\label{fig:valid_wz_1}]{\includegraphics[width=0.48\textwidth]{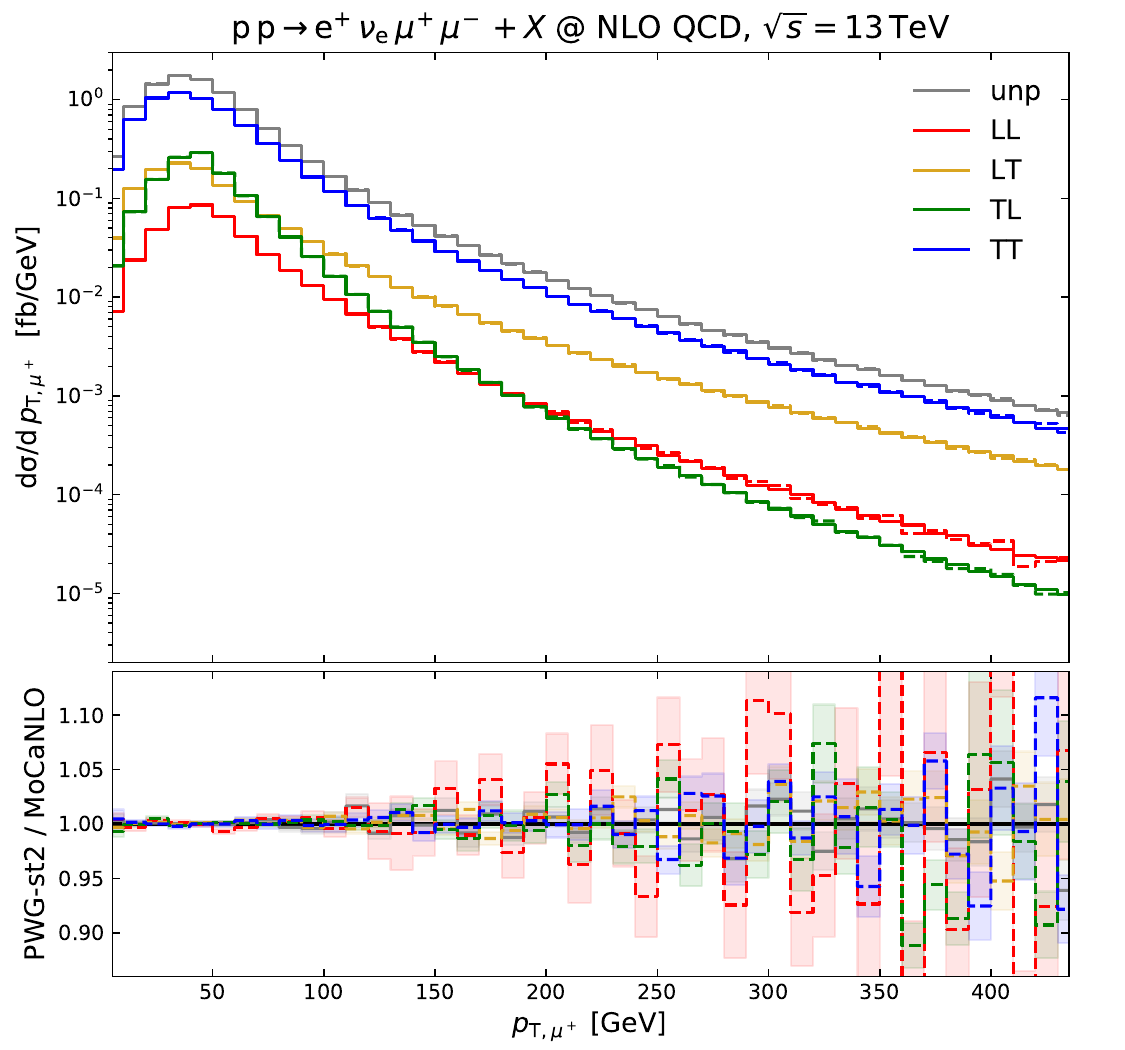}}
  \subfigure[Rapidity separation between $\Pe^+$ and $\mu^-$ in $\PW^+\PZ$\label{fig:valid_wz_3}]{\includegraphics[width=0.48\textwidth]{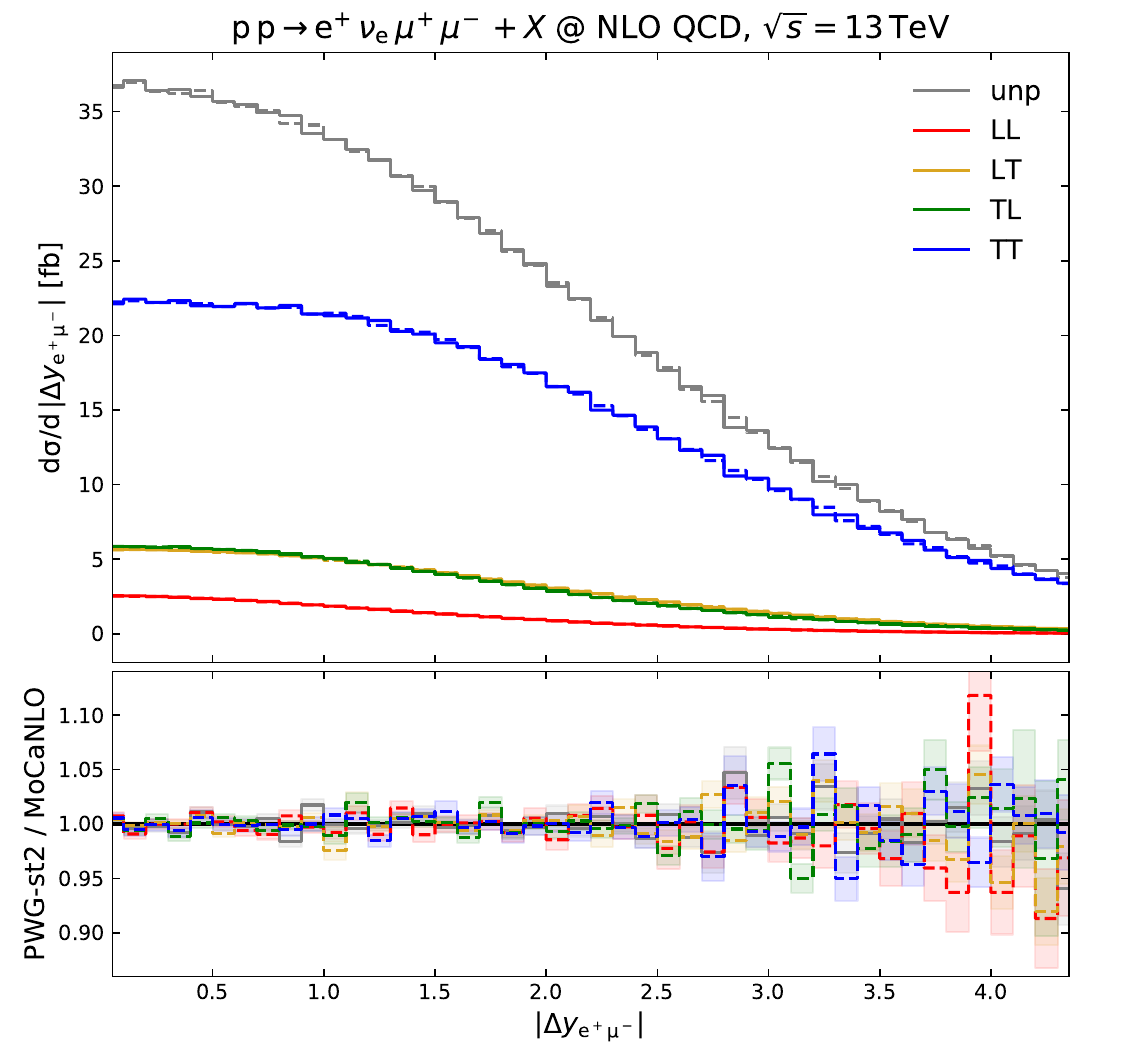}}
  \subfigure[Transverse momentum of the $\Pe^+\Pe^-$ pair in $\PZ\PZ$ \label{fig:valid_zz_1}]{\includegraphics[width=0.48\textwidth]{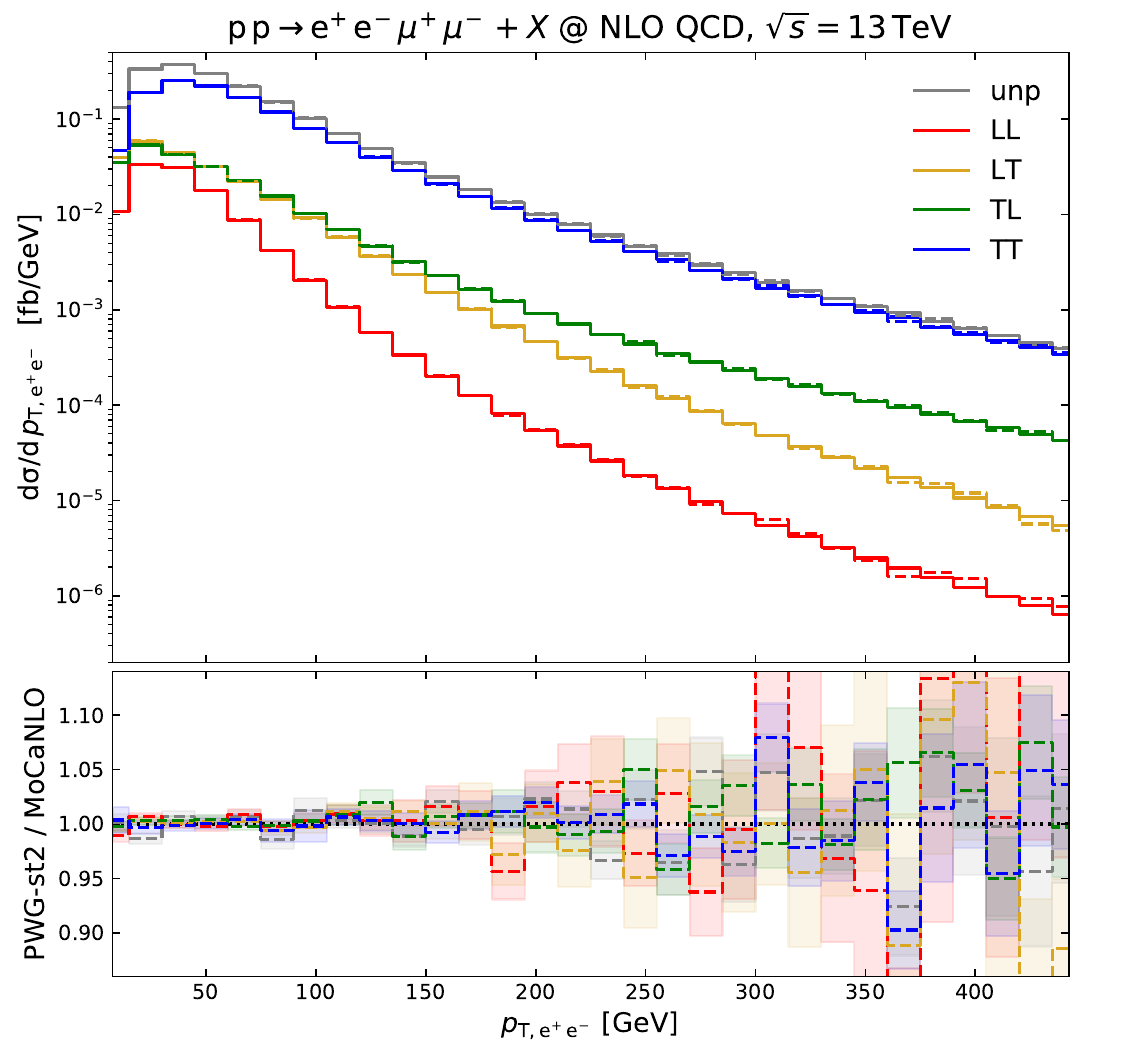}}
  \subfigure[Invariant mass of the $\Pe^+\nu_{\Pe}$ pair in $\PW^+\PW^-$\label{fig:valid_ww_2}]{\includegraphics[width=0.48\textwidth]{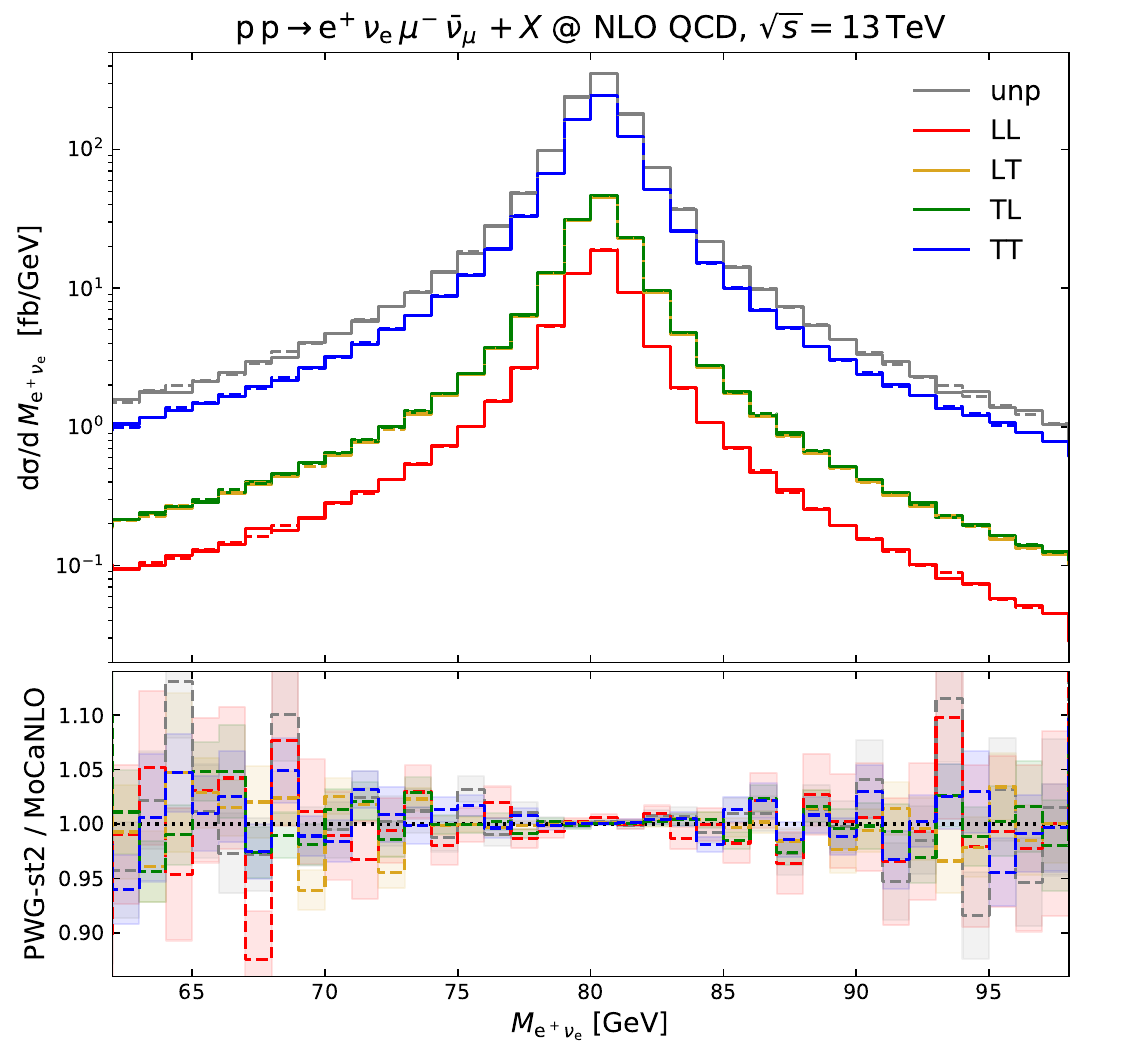}}
  \caption{
    Differential results at NLO QCD for $\PW^+\PZ$, $\PZ\PZ$ and $\PW^+\PW^-$ production at the LHC in the DPA.
    The only cut of Eq.~\ref{eq:mll_cut} is applied to lepton pairs from Z-boson decays.
    The curves obtained with \powheg (dashed) are compared with those obtained
    with \mocanlo (solid) for the unpolarised (gray), $\rL\rL$ (red), $\rL\rT$ (yellow),
    $\rT\rL$ (green) and $\rT\rT$ (blue) states. Absolute distributions and the ratios of
    \powheg results over the \mocanlo ones are shown in the upper and lower panel, respectively.
    Shaded bands in the lower panels represent 1$\sigma$ Monte Carlo uncertainties.
  }\label{fig:pol_2}
\end{figure}

The agreement between the two codes has been checked also at differential level for a large number of LHC observables and
MC-truth variables, be means of a bin-by-bin comparison.
In \reffi{fig:pol_2} we show the results of this validation stage for four variables.
The differences between the results obtained with the two codes is within MC-integration errors in each bin,
not only in the bulk of the calculation (low transverse momentum, angular variables), but also where the cross section is
suppressed, \ie in the tails of transverse-momentum and invariant-mass distribution. This holds for both unpolarised and doubly polarised predictions.

The detailed and successful validation of the new \powheg implementation at fixed order enables to proceed with
phenomenological studies matching NLO QCD predictions to parton shower.

\section{Results for the WZ process}\label{sec:WZ}
Amongst diboson production channels, $\PW\PZ$ is the best-suited one for polarisation studies,
as witnessed by the recent measurement of individual \cite{Aaboud:2019gxl,CMS:2021icx}
and joint \cite{ATLAS:2022oge} polarisation fractions by
the ATLAS and CMS collaborations with Run-2 data.
On top of a fairly large production rate (larger than $\PZ\PZ$, lower than $\PW\PW$),
the three-lepton decay channel features a good signal purity and enables to access the
complete final state by means of single-neutrino reconstruction.
The experimental community, after the pioneering measurements in rather inclusive setups \cite{Aaboud:2019gxl,CMS:2021icx,ATLAS:2022oge},
will have the possibility to refine the existing polarisation analyses and to perform new ones in more exclusive phase-space volumes
with Run-3 and High-Luminosity data.
For these motivations, the polarisation structure of the $\PW\PZ$ processes has also received a lot of attention by the theory community \cite{Denner:2020eck,Le:2022lrp,Le:2022ppa,Dao:2023pkl,SherpaPOL}.

In this section, we perform a phenomenological analysis of $\PW^+\PZ$ production in the five-flavour scheme (even though no bottom-induced contributions are present up to NLO QCD), in the positron--muon--antimuon decay channel,
\beq
\Pp\Pp \rightarrow \PW^+\,(\Pe^+\nu_{\Pe})\, \PZ\,(\mu^+\mu^-)+X\,.
\eeq
When considering the doubly polarised signals, we make use of the L or T labels for longitudinal or transverse boson, respectively,
with a first index referring to the polarisation state of the $\PW^+$ boson, and a second index referring the the polarisation state of the $\PZ$ boson.
For example, the TL state means a transversely polarised $\PW^+$ and a longitudinally polarised $\PZ$ boson.

\subsection{Selection cuts and integrated results}\label{subsec:wzsetups}
Four different setups are considered in our phenomenological analysis of $\PW^+\PZ$ events:
\begin{enumerate}
\item The \emph{inclusive setup} corresponds to the one used for validation purposes in \refse{subsec:valid}, only featuring a 10-GeV invariant-mass cut around the $\PZ$-boson pole mass
for the muon-antimuon pair.
\item As a \emph{fiducial setup}, we employ the one used in a recent ATLAS analysis \cite{ATLAS:2022oge}, with the following selection cuts:
\beqn\label{eq:fiddef}
&& \pt{\Pe^+}>20\GeV\,,\quad |y_{\Pe^+}|<2.5\,,\nnb\\
&& \pt{\mu^\pm}>15\GeV\,,\quad \quad|y_{\mu^\pm}|<2.5\,,\nnb\\
&& {\rm \Delta} R_{\mu^+\mu^-}>0.2\,,\quad {\rm \Delta} R_{\Pe^+\mu^\pm}>0.3\nnb\\
&& 81\GeV < M_{\mu^+\mu^-} <101\GeV\,,\nnb\\
&& M_{\rm T,W}>30\GeV\,,
\eeqn
where the transverse mass of the $\PW$ is defined as,
\beq
M_{\rm T,W} = \sqrt{2\pt{\Pe^+}\pt{\rm mis}(1-\cos{\rm \Delta}\phi_{\Pe^+,\rm mis})}\,.
\eeq
\item 
The \emph{radiation-zero} setup is characterised by an additional transverse-momentum cut on the four-lepton system
that effectively acts as a hadronic-activity veto:
\beq\label{eq:razdef}
\pt{\PW\PZ}<70\GeV\,.
\eeq
\item The \emph{boosted} setup is characterised by an additional transverse-momentum cut on the reconstructed $\PZ$ system,
on top of the fiducial cuts of Eq.~\eqref{eq:fiddef} and the one in Eq.~\eqref{eq:razdef}
\beq\label{eq:boodef}
\pt{\PZ}>200\GeV\,.
\eeq
\end{enumerate}
The last two setups, already studied at NLO both in the SM \cite{Dao:2023pkl} and in the
presence of BSM effects \cite{Franceschini:2017xkh}, are motivated by the approximate amplitude-zero effect \cite{Ohnemus:1991gb,Baur:1994ia}
and by the enhancement of the purely longitudinal contribution \cite{Franceschini:2017xkh} in the high-energy regime, for
central Born-like configurations.

In Table~\ref{tab:polWZ_allresults} we show our \powheg results for the full off-shell, unpolarised and doubly polarised integrated cross sections.
\begin{table}[h]
\begin{center}
\begin{tabular}{ccccc}
\hline\\[-0.3cm]
\cellcolor{blue!9} state & \cellcolor{blue!9}$\sigma[\fb]$ LHE & \cellcolor{blue!9} ratio [/unp., \%] LHE & \cellcolor{blue!9} $\sigma[\fb]$ PS+hadr & \cellcolor{blue!9} ratio [/unp., \%] PS+hadr\\[0.1cm]
\hline\\[-0.3cm]
\multicolumn{5}{c}{\cellcolor{green!9} Inclusive setup}\\
\hline\\[-0.1cm]
full off-shell 	&	$98.36(3) ^{+4.8 \%} _{-3.9 \%}$ 	&	$101.20 $	&	$95.27(3) ^{+4.9 \%} _{-3.9 \%}$		&	$101.28 $	\\[0.1cm]
unpolarised 	&	$97.20(3)  ^{+4.8 \%} _{-3.9 \%}$ 	&	$100$ 	&	$94.07(3) ^{+4.9 \%} _{-3.9 \%}$		&	$100$	\\[0.1cm]
LL 	&	$4.499(2)  ^{+2.8 \%} _{-2.3 \%}$ 	&	$4.63  ^{+0.13 } _{-0.13 }$ 	&	$4.359(2) ^{+2.8 \%} _{-2.2 \%}$		&	$4.63  ^{+0.13 } _{-0.13 }$	\\[0.1cm]
LT 	&	$13.151(4) ^{+7.0 \%} _{-5.7 \%}$ 	&	$13.53  ^{+0.28 } _{-0.27 }$ 	&	$12.730(5)   ^{+7.0 \%} _{-5.7 \%}$		&	$13.53  ^{+0.28 } _{-0.28 }$	\\[0.1cm]
TL 	&	$12.724(4) ^{+7.3 \%} _{-5.9 \%}$ 	&	$13.09  ^{+0.32 } _{-0.31 }$ 	&	$12.314(5)   ^{+7.4 \%} _{-5.9 \%}$		&	$13.09  ^{+0.31 } _{-0.32 }$	\\[0.1cm]
TT 	&	$66.88(2) ^{+4.0 \%} _{-3.3 \%}$ 	&	$68.81  ^{+0.47 } _{-0.51 }$ 	&	$64.74(2) ^{+4.1 \%} _{-3.2 \%}$		&	$68.82  ^{+0.46 } _{-0.51 }$	\\[0.1cm]
interference 	&	$-0.058 $	&	$-0.06 $	&	$-0.069 $	&	$-0.06 $	\\[0.1cm]
\hline\\[-0.3cm]
\multicolumn{5}{c}{ \cellcolor{green!9} Fiducial setup}\\
\hline\\[-0.1cm]
full off-shell 	&	$35.40(5) ^{+5.2 \%} _{-4.2 \%}$ 	&	$102.15 $	&	$34.04(5) ^{+5.3 \%} _{-4.2 \%}$		&	$102.20 $	\\[0.1cm]
unpolarised 	&	$34.65(5) ^{+5.2 \%} _{-4.2 \%}$ 	&	$100$ 	&	$33.30(5) ^{+5.2 \%} _{-4.2 \%}$		&	$100$	\\[0.1cm]
LL 	&	$1.965(3)  ^{+2.7 \%} _{-2.2 \%}$ 	&	$5.67  ^{+0.17 } _{-0.18 }$ 	&	$1.892(3)  ^{+2.7 \%} _{-2.2 \%}$		&	$5.68  ^{+0.18 } _{-0.18 }$	\\[0.1cm]
LT 	&	$5.344(7) ^{+7.3 \%} _{-5.9 \%}$ 	&	$15.42  ^{+0.31 } _{-0.30 }$ 	&	$5.140(7)  ^{+7.3 \%} _{-5.9 \%}$		&	$15.43  ^{+0.31 } _{-0.30 }$	\\[0.1cm]
TL 	&	$5.083(7) ^{+7.4 \%} _{-5.9 \%}$ 	&	$14.67  ^{+0.30 } _{-0.30 }$ 	&	$4.888(6)  ^{+7.4 \%} _{-6.0 \%}$		&	$14.68  ^{+0.30 } _{-0.31 }$	\\[0.1cm]
TT 	&	$22.04(3) ^{+4.5 \%} _{-3.6 \%}$ 	&	$63.60  ^{+0.40 } _{-0.45 }$ 	&	$21.16(3)  ^{+4.6 \%} _{-3.5 \%}$		&	$63.55  ^{+0.51 } _{-0.40 }$	\\[0.1cm]
interference 	&	$0.223 $	&	$0.64 $	&	$0.217 $	&	$0.64 $	\\[0.1cm]
\hline\\[-0.3cm]
\multicolumn{5}{c}{\cellcolor{green!9} Radiation-zero setup}\\
\hline\\[-0.1cm]
full off-shell 	&	$25.19(4) ^{+4.4 \%} _{-3.6 \%}$ 	&	$102.24 $	&	$24.18(4) ^{+4.4 \%} _{-3.6 \%}$		&	$102.24 $	\\[0.1cm]
unpolarised 	&	$24.64(4)  ^{+4.4 \%} _{-3.6 \%}$ 	&	$100$ 	&	$23.65(4) ^{+4.4 \%} _{-3.6 \%}$		&	$100$	\\[0.1cm]
LL 	&	$1.779(3) ^{+2.6 \%} _{-2.2 \%}$ 	&	$7.22  ^{+0.15 } _{-0.15 }$ 	&	$1.711(3) ^{+2.6 \%} _{-2.1 \%}$		&	$7.23  ^{+0.16 } _{-0.16 }$	\\[0.1cm]
LT 	&	$3.243(6)^{+5.9 \%} _{-4.7 \%}$ 	&	$13.16  ^{+0.18 } _{-0.16 }$ 	&	$3.113(6) ^{+5.9 \%} _{-4.8 \%}$		&	$13.16  ^{+0.19 } _{-0.16 }$	\\[0.1cm]
TL 	&	$3.114(6) ^{+6.0 \%} _{-4.8 \%}$ 	&	$12.64  ^{+0.19 } _{-0.17 }$ 	&	$2.993(5) ^{+6.0 \%} _{-4.9 \%}$		&	$12.66  ^{+0.19 } _{-0.16 }$	\\[0.1cm]
TT 	&	$16.80(2) ^{+3.6 \%} _{-2.9 \%}$ 	&	$68.17  ^{+0.44 } _{-0.51 }$ 	&	$16.10(3) ^{+3.7 \%} _{-2.8 \%}$		&	$68.07  ^{+0.56 } _{-0.46 }$	\\[0.1cm]
interference 	&	$-0.292 $	&	$-1.19 $	&	$-0.266 $	&	$-1.19 $	\\[0.1cm]
\hline\\[-0.3cm]
\multicolumn{5}{c}{\cellcolor{green!9} Boosted setup}\\
\hline\\[-0.1cm]
full off-shell 	&	$0.452(5)^{+7.3 \%} _{-5.6 \%}$ 	&	$103.56 $	&	$0.436(5) ^{+7.7 \%} _{-5.6 \%}$		&	$104.14 $	\\[0.1cm]
unpolarised 	&	$0.437(5) ^{+7.2 \%} _{-5.5 \%}$ 	&	$100$ 	&	$0.418(5) ^{+7.3 \%} _{-4.7 \%}$		&	$100$	\\[0.1cm]
LL 	&	$0.1031(7) ^{+2.6 \%} _{-1.7 \%}$ 	&	$23.61  ^{+0.96 } _{-1.02 }$ 	&	$0.0993(7) ^{+2.4 \%} _{-1.8 \%}$		&	$23.73  ^{+0.73 } _{-1.08 }$	\\[0.1cm]
LT 	&	$0.0223(6) ^{+7.4 \%} _{-5.7 \%}$ 	&	$5.11  ^{+0.03 } _{-0.03 }$ 	&	$0.0214(5) ^{+8.3 \%} _{-6.0 \%}$		&	$5.12  ^{+0.10 } _{-0.07 }$	\\[0.1cm]
TL 	&	$0.0207(5) ^{+6.7 \%} _{-5.1 \%}$ 	&	$4.75  ^{+0.02 } _{-0.02 }$ 	&	$0.0200(5) ^{+6.3 \%} _{-5.5 \%}$		&	$4.77  ^{+0.11 } _{-0.04 }$	\\[0.1cm]
TT 	&	$0.293(3) ^{+8.4 \%} _{-6.5 \%}$ 	&	$66.98  ^{+0.73 } _{-0.69 }$ 	&	$0.281(3)  ^{+8.9 \%} _{-6.4 \%}$		&	$67.14  ^{+1.00 } _{-1.22 }$	\\[0.1cm]
interference 	&	$-0.002 $	&	$-0.45 $	&	$-0.003 $	&	$-0.45 $	\\[0.1cm]
\hline
\end{tabular}
\end{center}
\caption{
  Off-shell, unpolarised and doubly polarised predictions for $\PW^+\PZ$ in the fully leptonic decay channel at the LHC@13TeV,
  with NLO QCD accuracy. Results are provided at Les Houches Event (LHE) level, as well as matched to {\sc Pythia8} QCD+QED shower and hadronisation effects (PS+hadr).
  Integrated cross sections (in fb) and ratios normalised to the unpolarised result are presented.
  Monte--Carlo-integration (in parentheses) and QCD-scale (percentages in subscripts and superscripts) uncertainties are shown for the integrated cross sections.
  Correlated QCD-scale uncertainties are also shown for the polarised ratios.\label{tab:polWZ_allresults}}
\end{table}
The results labeled by LHE come from the Les-Houches events which already embed the Sudakov form factor resumming QCD emissions at low transverse momentum of
the colour singlet (the diboson system). The results labeled by PS+hadr include the matching of the NLO QCD calculation with the {\sc Pythia8} QCD+QED parton
shower \cite{Sjostrand:2014zea}, including hadronisation effects. The considered {\sc Pythia8} settings are the ones used in \citere{Chiesa:2020ttl}, corresponding
to the predictions labeled ``NLO$_{\rm QCD}+$PS$_{\rm QCD,\,QED}$'' therein.
The integrated cross sections are given in absolute numbers, as well as normalised to the unpolarised results. The uncertainties in subscripts and superscripts
come from 7-point variations of the renormalisation and factorisation scale, both for absolute cross sections and for ratios.
In the case of ratios, the scale variations are performed in a correlated way between the numerator (polarised) and the denominator (unpolarised).

A common aspect to all numbers reported in Table~\ref{tab:polWZ_allresults} is the decrease of the cross sections
between LHE and PS+hadr level, mostly due to the negative $4\%$ shift given by the leading-logarithmic QED corrections that are included
in the showered predictions. On the contrary, QCD-shower and hadronisation have milder effects on the integrated results.
This effect is rather independent of the polarisation state and of the specific setup, giving as a result polarisation fractions that
do not change sizeably between LHE and PS+hadr level.

For each setup, the interference contribution is computed taking the difference between the unpolarised cross section and the sum of doubly polarised ones.
The interferences, which vanish in the inclusive setup, range between $0.6\%$ and $-1.2\%$ in the other setups. In particular, interferences
become negative in the case of exclusive setups (radiation-zero and boosted).
The off-shell effects, often dubbed non-resonant background \cite{Denner:2020bcz}, can be estimated taking the difference between the full off-shell cross section
and the unpolarised one. Similarly to the interference terms, the off-shell effects are small but not negligible, ranging between $+1\%$ and $+4\%$.
Interference and off-shell effects, although at the percent level in integrated cross sections may be larger in the case of differential
distributions.

It can be appreciated that going from inclusive to more exclusive setup, in spite of a dramatic decrease of the absolute cross sections, the
LL contribution increases from 4.6\% in the inclusive setup up to 23.7\% in the boosted setup.
The enhancement of the LL polarisation fraction at large transverse momenta of the two bosons could be even more marked if an additional constraint is
imposed on the rapidity separation between the positron and the $\PZ$ system \cite{Dao:2023pkl}, resulting in topologies with a scattering
angle (in the diboson CM frame) close to $\pi/2$ where the LL signal is expected to be larger than all other helicity combinations \cite{Franceschini:2017xkh}.
In the boosted setup, it is also unavoidable to include NLO EW corrections which amount at $-12\%$ \cite{Dao:2023pkl}.

The TT fraction is the largest one in all setups, as expected from general spin-balance considerations~\cite{Bern:2011ie,Stirling:2012zt}, and
increases from 63\% in the fiducial setup to 68\% in the radiation-zero setup.
Notice that the hadronic-activity veto ($\pt{\PW\PZ}< 70\GeV$), enhances not only the TT fraction, but also the LL one (from $5.7\%$ to $7.2\%$).
  A tighter constraint on the QCD activity, \eg a lower $\pt{\PW\PZ}$-cut value, would further suppress QCD real radiation which is known to
  break the approximate amplitude-zero effect in $\PW\PZ$ \cite{Ohnemus:1991gb,Baur:1994ia}, but would also make the 
  the NLOPS description inadequate, due to the appearance of large logarithms in the $p_{\rm T}$-veto cut that need to be resummed \cite{Monni:2014zra,Becher:2014aya,Campbell:2023cha}.
For large diboson energy, the equivalence theorem \cite{Cornwall:1974km,Vayonakis:1976vz,Chanowitz:1985hj,Gounaris:1986cr} implies
that the contribution of mixed states (LT, TL) is suppressed. In fact, the
contribution of both states decreases from 15\% in the fiducial setup, down to 5\% in the boosted setup.

It is interesting to compare the NLO QCD results at fixed order and matched to PS in the radiation-zero and boosted setups. 
Although not shown in the table, the fixed-order polarisation fractions obtained with \powheg reproduce almost perfectly the ones
of \citere{Dao:2023pkl} (see Table~1 therein), where the same selections but a different set of PDFs and SM parameters are employed.
The fixed-order LL fraction amounts at $6.9\%$ and $22.6\%$ in the two setups, which are $0.3\%$ and $1\%$ lower than the corresponding
values after PS matching. This effect comes from the Sudakov form factor that regularises the low-transverse-momentum spectrum
of the diboson system, with different behaviours for the various polarisation states. Owing to the $\pt{\PW\PZ}<70\GeV$ cut,
the matched cross section is typically smaller than the fixed-order one. More in details, the LL and TT cross sections
at fixed-order decrease by roughly 1\% at LHE level, while the TL and LT ones decrease by 5-6\%. This results in a different
polarisation balance after PS matching, which tends to favour the LL and TT states.

A comment is worth regarding the QCD-scale uncertainties found for the integrated cross sections. Owing to the dominance of the TT state in all setups, the scale variations
for the TT signal reflect those for the unpolarised and off-shell process, ranging from 3-4\% in the fiducial setup to 7-8\% in the
boosted setup. The mixed states show similar QCD-scale uncertainties in all setups, ranging between 5\% and 7\%.
Owing to much smaller real-radiation contributions, the LL polarisation state is characterised by smaller uncertainties than the other
states, amounting at 2-3\% in all considered setups. It is a known effect \cite{Denner:2020eck,Ballestrero:2020qgv} in diboson inclusive
production and scattering that systems of two longitudinal bosons $VV$ are preferably produced
with small $\pt{VV}$ and typically receive small real corrections.
Up to small differences, the overall picture of QCD-scale uncertainties given in Table~\ref{tab:polWZ_allresults}
is rather close to the one found at fixed order
\cite{Denner:2020eck,Le:2022lrp,Le:2022ppa,Dao:2023pkl}.

In the final part of this subsection, we compare our NLOPS integrated results obtained with \powheg against the recently presented approximate ones (nLO+PS)
of \citere{SherpaPOL}, in the same fiducial setup. The cross sections are shown in Table~\ref{tab:compSherpa}.
Note that both results include QED+QCD PS and hadronisation effects.
\begin{table}[h]
  \begin{center}
    \begin{tabular}{cccc}
      \hline\\[-0.3cm]
      \cellcolor{blue!9} state & \cellcolor{blue!9}NLOPS &  \cellcolor{blue!9}nLO+PS \cite{SherpaPOL} &  \cellcolor{blue!9} $\frac{\textrm{nLO+PS}}{\textrm{NLOPS}}-1$ \\[0.1cm]
      \hline\\[-0.1cm]
full off-shell 	       	&	$34.04(5)$       &  $ 33.80(4) $        &       $-0.7\%$       \\[0.1cm]
unpolarised 	        &	$33.30(5)$       &  $ 33.46(3) $        &       $+0.5\%$       \\[0.1cm]
LL 	          	&	$1.892(3)$   	 &  $ 1.902(2) $        &       $+0.5\%$       \\[0.1cm]
LT 	          	&	$5.140(7)$   	 &  $ 5.241(4) $   	&       $+1.9\%$       \\[0.1cm]
TL 	          	&	$4.888(6)$   	 &  $ 5.002(4) $   	&       $+2.3\%$       \\[0.1cm]
TT 	          	&	$21.16(3)$   	 &  $ 21.10(2) $   	&       $-0.3\%$       \\[0.1cm]
interference 	        & 	$0.217 $         &  $ 0.215    $        &       $-0.9\%$       \\[0.1cm]
      \hline
    \end{tabular}
  \end{center}
  \caption{
    Comparison between \powheg (NLOPS) and \Sherpa (nLO+PS) results for integrated cross sections (in fb) in the off-shell, unpolarised and doubly polarised calculations of $\PW^+\PZ$ at the LHC.
    Both predictions are matched to QED+QCD PS and hadronisation effects.
    The setup described in Eq.~\eqref{eq:fiddef} is understood.
  }\label{tab:compSherpa} 
\end{table} 
There are two notable differences between the two approaches:
\begin{itemize}
\item[-] our \powheg predictions for the unpolarised and polarised process are calculated in the DPA (see \refse{subsec:polepol}), while the \Sherpa computation
  makes use of a spin-correlated narrow-width approximation (NWA) which involves a smearing procedure to recover partial off-shell effects \cite{Richardson:2001df};
\item[-] for polarised signals, the \Sherpa calculation involves approximate virtual corrections (see Section 3.3 in \citere{SherpaPOL}), while
  our \powheg implementation makes use of exact one-loop matrix elements provided by the \recola1 library.
\end{itemize}
A fair agreement between the two calculations is found, with a sub-percent discrepancy for the LL and TT states. A slightly larger discrepancy (2\%) characterises the LT and TL
polarisation states. This comparison highlights that using approximate virtual corrections works reasonably well compared to the exact calculation of polarised signals.
In addition, the off-shell effects, i.e. the difference between full off-shell and unpolarised in Tab.~\ref{tab:compSherpa}, are mildly smaller when using the NWA (1\%) compared to the DPA (2\%), although in both cases they are of the expected order of magnitude, $\mc O({\rm \Gamma}_{V}/M_{V})$.
Similar differences have been noticed also for $\PW\PW$ inclusive production \cite{Poncelet:2021jmj}.

\subsection{Differential distributions}
\begin{figure}[h]
  \centering
  \includegraphics[width=1.0\textwidth]{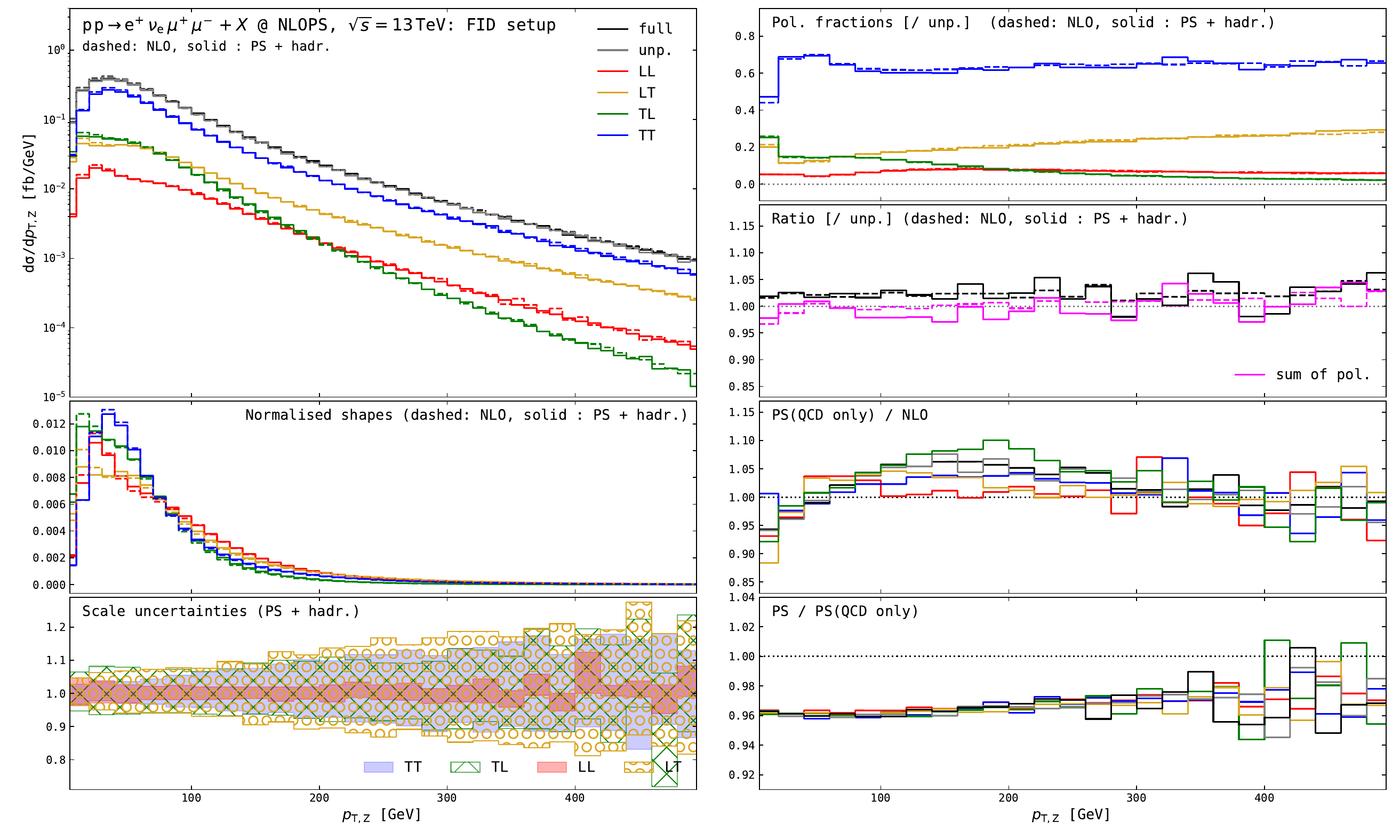}
  \caption{
    Distributions in the transverse momentum of the $\PZ$ boson for $\PW^+\PZ$ production at the LHC.
    The fiducial setup defined in Eq.~\eqref{eq:fiddef} is understood.
    The colour code reads: full off-shell (black), unpolarised (gray), LL (red), LT (goldenrod), TL (green), TT (blue), sum of polarised (magenta).
    The figure is structured in 7 panels.
    Top left: absolute distributions at fixed order (NLO, dashed) and matched to QCD+QED PS and hadronisation (PS+hadr, solid).
    Middle left: normalised distributions (to have unit integral) for polarised states at NLO (dashed) and PS+hadr (solid).
    Bottom left: theory uncertainties for the polarised states from 7-point renormalisation and factorisation-scale variations of PS+hadr predictions, relative to the corresponding central values.
    Top right: polarisation fractions (ratio of polarised results over unpolarised one) at NLO (dashed) and PS+hadr (solid).
    Middle-top right: ratio of the full off-shell and sum of polarised cross sections over the unpolarised one at NLO (dashed) and PS+hadr (solid).
    Middle-bottom right: ratio of NLO QCD cross sections matched to QCD PS over the fixed-order ones.
    Bottom right: ratio of NLO QCD cross sections matched to QCD+QED PS over the ones matched to QCD PS.
  }\label{fig:WZfid_1}
\end{figure}

\begin{figure}
  \centering
  \includegraphics[width=1.0\textwidth]{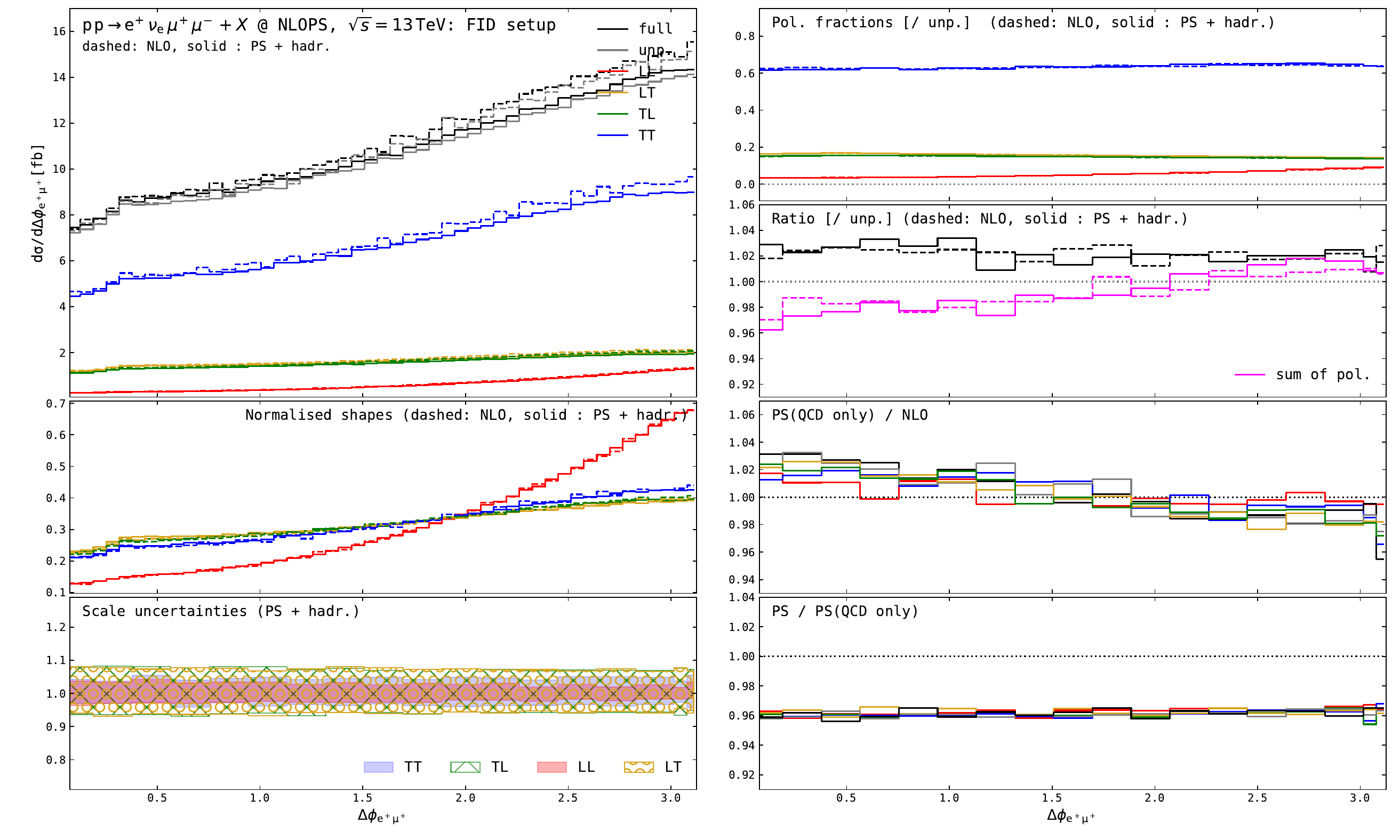}
  \caption{
    Distributions in the azimuthal-angle separation between the positron and the antimuon for $\PW^+\PZ$ production at the LHC.
    The fiducial setup described in \refse{subsec:setup} is understood.
    Same structure as \reffi{fig:WZfid_1}.    
  }\label{fig:WZfid_12}
\end{figure}

The polarisation structure of $\PW\PZ$ becomes even more interesting when considering differential results.
In \reffis{fig:WZfid_1}--\ref{fig:WZfid_12} we show the behaviour of unpolarised and polarised distributions for two LHC observables
in the fiducial setup detailed in Eq.~\eqref{eq:fiddef}. Additional results for other kinematic variables can be found in Appendix~\ref{app:WZ}.
The figure structure enables to appreciate the differences between the fixed-order description (NLO) and the matched predictions (PS+hadr).
For completeness, we have also studied intermediate descriptions of the $\PW\PZ$ process, by matching to QCD PS only, and in the absence of hadronisation effects.
Since however hadronisation effects are negligible, results without hadronisation effects are not shown separately in the plots.

The transverse momentum of the $\PZ$ boson, \ie of the muon--antimuon system, is considered in \reffi{fig:WZfid_1}.
The unpolarised distribution is dominated by the TT contribution, with a fraction which remains rather constant for values above $50\GeV$.
In the lowest-$p_{\rm T}$ region allowed by the implicit cuts (induced by lepton cuts), the sum of mixed-state (LT, TL) cross sections is comparable to the TT one.
At large $\PZ$-boson transverse momenta ($\pt{\PZ}\gtrsim 400\GeV$), the LT contribution amounts at half of the TT one, while the LL and TL are suppressed by
one order of magnitude. The crossing of the red and green curve around 200\GeV highlights the fact that if the longitudinal $\PZ$ boson has a large transverse momentum
the $\PW$ is more likely to be longitudinal, favoured by the small real QCD corrections for the LL state, while in the TL case the $\PZ$-boson
transverse momentum is shared between the typically hard real radiation and the transverse boson \cite{Denner:2020eck,Le:2022lrp,Denner:2022riz}. 
A similar reasoning motivates the large difference between the LT and TL states. The $\pt{\PZ}$ is shared between the $\PW$ boson and QCD real radiation, therefore
configurations with large $\pt{\PZ}$ tend to favour a longitudinal $\PW$ which is typically softer than a transverse one.
The much narrower QCD-scale band for the LL state compared to the others confirms that the LL state receives small real QCD corrections,
resulting in a truly NLO scale dependence. On the contrary, the mixed and TT states are dominated by hard real radiation which induces a LO-like scale dependence.
Both off-shell and interference effects are at the percent level and rather constant for this observable.
As depicted in the bottom right panels of \reffi{fig:WZfid_1},
the inclusion of QCD PS changes all NLO shapes just in the low-$p_{\rm T}$ region, with larger effect ($\approx 5-10\%$ in the lowest bin allowed by the fiducial cuts).
At moderate transverse momentum ($\gtrsim 100\GeV$), the QCD-shower corrections are almost vanishing for the LL state, while they are not negligible  for
other states (above 5\% for the TL state). The inclusion of multiple QED (photon) radiation in the PS shifts the cross section by roughly 4\%, with
basically no differences amongst the various polarisation states.

In \reffi{fig:WZfid_12} we present distributions in the azimuthal separation between the two positively charged leptons.
While the cross section is smaller, the LL distribution shape substantially differs from the other polarised shapes,
providing a marked discrimination power to be exploited in template fits of LHC data. 
For this angular observable, while off-shell effects (estimated as the difference between the black curve and 1 in the middle-top right panel of \reffi{fig:WZfid_12}) are rather constant over the whole spectrum and compatible with
the integrated result, the interference effects (estimated as the difference between 1 and the magenta curve) decrease with almost constant slope from 0 to $\pi$, ranging from $+4\%$ in the least populated region to $-2\%$ in the most populated one.
While QED-shower effects are independent of the polarisation state and constant over the whole spectrum, the QCD PS introduces a shape distortion to the NLO curve that is slightly more marked for the
TT and mixed states than for the LL state. Although this effect is at the percent level, it stresses the importance of performing separate PS-matched simulations for each polarised signal, not only for
transverse-momentum-dependent observables (as shown in \reffi{fig:WZfid_1}), but also for angular ones.

\subsection{Comparison with the ATLAS measurement}
The ATLAS Collaboration has recently measured \cite{ATLAS:2022oge} the joint polarisation fractions in $\PW\PZ$ production, considering the fully leptonic decay channel,
finding fair agreement with SM predictions.
In Table~\ref{tab:WZfid_ATLAS} we compare the results of our new \powheg implementation against the measurement result and theoretical predictions used therein.
\begin{table}[h]
  \begin{center}
    \begin{tabular}{ccccc}
      \hline\\[-0.3cm]
      \cellcolor{blue!9}fraction         &  \cellcolor{blue!9}PS+hadr (this work) & \cellcolor{blue!9}TH1 \cite{ATLAS:2022oge}  & \cellcolor{blue!9}TH2 \cite{ATLAS:2022oge}  & \cellcolor{blue!9}measured \cite{ATLAS:2022oge} \\[0.1cm] 
      \hline\\[-0.2cm]      
      ${\rL\rL}$              &  $ 5.68  ^{+0.18 } _{-0.18 }$  &  $ 5.7 \pm 0. 2$   &  $ 5.83 \pm 0.12 $   & $ 7.2 \pm 1.6$   \\[0.1cm] 
      ${\rL\rT}$              &  $15.43  ^{+0.31 } _{-0.30 }$  &  $ 15.5 \pm 0. 3$   &  $14.84 \pm 0 .22$     & $11.9 \pm 3.4$   \\[0.1cm] 
      ${\rT\rL}$              &  $14.68  ^{+0.30 } _{-0.31 }$  &  $ 14.7 \pm 0. 3$   &  $14.61 \pm 0 .22$      & $15.2 \pm 3.3$   \\[0.1cm] 
      ${\rT\rT}$              &  $63.55  ^{+0.51 } _{-0.40 }$  &  $ 63.5 \pm 0. 4$   &  $64.72 \pm 0 .26 $     & $66.0 \pm 4.0$   \\[0.1cm] 
\hline
    \end{tabular}
  \end{center}
  \caption{
    Theoretical predictions for joint polarisation fractions (LL, LT, TL, TT) in $\PW^+\PZ$ production, normalised to the unpolarised cross section,
    compared to the ATLAS results reported in Table~2 of \citere{ATLAS:2022oge}.
    The fractions obtained with the new \powheg implementation (PS+hadr) are computed at NLO QCD and matched to {\sc Pythia8} QCD+QED shower with hadronisation effects.
    The uncertainties associated to the PS+hadr predictions come from correlated 7-point QCD-scale variations. 
    The SM predictions quoted by ATLAS and labeled by TH1 were generated with \mocanlo \cite{Denner:2020eck},
    while those labeled by TH2 were extracted by means of an event-by-event reweighting of an unpolarised
    sample generated with {\scshape{Powheg-Box-V2}} \cite{Melia:2011tj,Nason:2013ydw} matched to {\scshape{Pythia8}}.
  }\label{tab:WZfid_ATLAS} 
\end{table} 

The measurement methodology of \citere{ATLAS:2022oge}, relying on a deep-neural-network-based template fit with individual templates for each doubly polarised state, also accounts for interference and off-shell effects.
The results were compared against two different SM predictions.
A first prediction reflects the fixed-order \mocanlo calculation of \citere{Denner:2020eck} in the DPA. A second prediction comes from an a-posteriori reweighting of
unpolarised events generated with  {\scshape{Powheg-Box-V2}}~\cite{Melia:2011tj,Nason:2013ydw} matched to {\scshape{Pythia8}} \cite{Sjostrand:2014zea}.
The extracted fractions (rightmost column of Table~\ref{tab:WZfid_ATLAS}), as well as those predicted with the reweighting procedure (labeled TH2), are effective fractions whose sum is 1 by construction (interferences and off-shell effects already subtracted).
The fixed-order (NLO QCD) fractions (labeled TH1) are instead defined as fractions of the unpolarised-signal cross section, therefore their sum differs from 1 due to interference effects (which amount to a positive 0.6\% in the considered fiducial setup). The same normalisation is used to present our results.
We provide NLO-accurate predictions for the polarisation fractions obtained with our new \powheg implementation,
matched to the QED+QCD {\scshape{Pythia8}} parton-shower \cite{Sjostrand:2014zea,Sjostrand:2019zhc} and including hadronisation effects (PS+hadr).

Our predictions confirm those calculated at fixed order with \mocanlo (TH1), while they differ by a few percent from those
obtained with the reweighting technique. The measured fractions, which still suffer from rather large experimental uncertainties of both systematic (modelling)
and statistical type, agree with our predictions by $1\sigma$.

\section{Results for the ZZ process}\label{sec:ZZ}
In this section, we consider the LHC production of two $\PZ$ bosons, which has a radically different spin structure compared to $\PW\PZ$.
Owing to the absence of triple-gauge-boson couplings in quark-induced LO contributions, the purely longitudinal state is suppressed in the
high-energy regime by the equivalence theorem \cite{Willenbrock:1987xz}. This peculiar aspect of the LL state makes the QCD and EW corrections
to the transverse-momentum distribution tails very large, as observed in \citere{Denner:2021csi}. In $\PZ\PZ$ production, the PDF-enhanced
loop-induced contribution with gluons in the initial state also changes the polarisation picture in a sizeable manner \cite{Denner:2021csi}.
The $\PZ\PZ$ process in the four-charged-lepton decay channel has been widely investigated as it represents a very clean signature, in spite
of a rather small branching ratio of the decay $\PZ\rightarrow \ell^+\ell^-$ (rightly 3.4\% for a single lepton flavour).
This decay channel enables the complete reconstruction of the diboson system, offering an optimal environment to investigate the polarisation structure
from the decay angles. Polarisation studies in the four-charged-lepton decay channel were not only carried out in the in the so-called \emph{on-shell region}
\cite{Bierweiler:2013dja,Denner:2021csi}, but also in Higgs decays \cite{Maina:2020rgd,Maina:2021xpe}.

The specific process considered in this section is $\PZ\PZ$ production in the five-flavour scheme, in the decay channel with an electron--positron and a muon--antimuon pair,
\beq
\Pp\Pp   \rightarrow          \PZ\,(\Pe^+\Pe^-)\, \PZ\,(\mu^+\mu^-)+X\,.
\eeq
In the following, when considering doubly polarised signals, the first polarisation label (L, T) refers to the boson decaying to $\Pe^+\Pe^-$.

\subsection{Selection cuts and integrated results}
Two different setups are considered in our phenomenological analysis of $\PZ\PZ$ events in the four-charged-lepton decay channel.
Like for $\PW\PZ$, the \emph{inclusive setup} corresponds to the one used for validation purposes in \refse{subsec:valid}, featuring a 10-GeV invariant-mass cut around the $\PZ$-boson pole mass
for both pairs of opposite-sign, same-flavour charged leptons.
The \emph{fiducial setup} mimics the one of the most recent ATLAS analysis of polarised ZZ production~\cite{ATLAS:2023zrv}, with the following selections:
\beqn\label{eq:fiddefZZ}
&&  \pt{\Pe^\pm}>7\GeV\,,\quad |y_{\Pe^\pm}|<2.47\,,\quad \pt{\mu^\pm}>5\GeV\,,\quad |y_{\mu^\pm}|<2.7\,,\nnb\\
&&  \pt{\ell_{1(2)}}>20\GeV\,,{\textrm{ with }} \ell_{1(2)}= {\textrm{(sub)leading lepton}}\,,\nnb\\
&&  {\rm \Delta} R_{\ell\ell'}>0.05\,, {\textrm{ with }} \ell,\ell'=\Pe^\pm,\mu^\pm\nnb\\
&&  81\GeV < M_{\ell^+\ell^-} <101\GeV\,, {\textrm{ with }} \ell=\Pe,\mu\,,\nnb\\
&&  M_{\rm 4\ell}>180\GeV\,.
\eeqn
where the leading and sub-leading leptons are sorted in transverse momentum.

In Table~\ref{tab:polZZ_allresults} we show the integrated cross section for the full off-shell, unpolarised and doubly polarised $\PZ\PZ$ process in
the inclusive and fiducial setups.
\begin{table}
\begin{center}
\begin{tabular}{ccccc}
\hline\\[-0.3cm]
\cellcolor{blue!9} state & \cellcolor{blue!9}$\sigma[\fb]$ LHE & \cellcolor{blue!9} ratio [/unp., \%] LHE & \cellcolor{blue!9} $\sigma[\fb]$ PS+hadr & \cellcolor{blue!9} ratio [/unp., \%] PS+hadr\\[0.1cm]
\hline\\[-0.3cm]
\multicolumn{5}{c}{\cellcolor{green!9} Inclusive setup}\\
\hline\\[-0.1cm]
full off-shell	&	$28.63(1)   ^{+2.9 \%} _{-2.3 \%}$ 	&	$101.43 $	                &	$26.84(2)  ^{+2.9 \%} _{-2.3 \%}$		&	$101.59 $                   	\\[0.1cm]
unpolarised 	&	$28.22(1)   ^{+2.9 \%} _{-2.3 \%}$ 	&	$100$ 	                        &	$26.42(1)  ^{+2.8 \%} _{-2.3 \%}$		&	$100$	                        \\[0.1cm]
LL 	&	$1.664(1)   ^{+3.0 \%} _{-2.4 \%}$ 	&	$5.90  ^{+0.03 } _{-0.04 }$ 	&	$1.555(1)  ^{+2.9 \%} _{-2.5 \%}$		&	$5.89  ^{+0.03 } _{-0.04 }$	\\[0.1cm]
LT 	&	$3.551(2)   ^{+3.7 \%} _{-2.9 \%}$ 	&	$12.58  ^{+0.09 } _{-0.08 }$ 	&	$3.320(2)  ^{+3.7 \%} _{-2.9 \%}$		&	$12.57  ^{+0.10 } _{-0.08 }$	\\[0.1cm]
TL 	&	$3.554(2)   ^{+3.7 \%} _{-3.0 \%}$ 	&	$12.59  ^{+0.09 } _{-0.08 }$ 	&	$3.326(2)  ^{+3.6 \%} _{-3.1 \%}$		&	$12.59  ^{+0.09 } _{-0.10 }$	\\[0.1cm]
TT 	&	$19.46(1)   ^{+2.6 \%} _{-2.1 \%}$ 	&	$68.95  ^{+0.16 } _{-0.19 }$ 	&	$18.23(1)  ^{+2.7 \%} _{-2.2 \%}$		&	$69.01  ^{+0.11 } _{-0.13 }$	\\[0.1cm]
interference 	&	$-0.0060 $	                &	$-0.02 $	                &	$-0.0141 $                          	&	$-0.02 $                   	\\[0.1cm]
\hline\\[-0.3cm]
\multicolumn{5}{c}{\cellcolor{green!9} Fiducial setup}\\
\hline\\[-0.1cm]
full off-shell 	&	$15.15(2)   ^{+3.0 \%} _{-2.4 \%}$ 	&	$101.14 $	                &	$14.21(2)   ^{+3.0 \%} _{-2.4 \%}$		&	$101.32 $               	\\[0.1cm]
unpolarised 	&	$14.98(1)   ^{+3.0 \%} _{-2.4 \%}$ 	&	$100$                    	&	$14.02(1)   ^{+3.0 \%} _{-2.4 \%}$		&	$100$	                        \\[0.1cm]
LL 	&	$0.877(1)  ^{+3.0 \%} _{-2.6 \%}$ 	&	$5.86  ^{+0.03 } _{-0.05 }$ 	&	$0.819(1)   ^{+2.9 \%} _{-2.7 \%}$		&	$5.84  ^{+0.03 } _{-0.05 }$	\\[0.1cm]
LT 	&	$1.908(2)   ^{+3.6 \%} _{-2.9 \%}$ 	&	$12.74  ^{+0.08 } _{-0.07 }$ 	&	$1.783(2)   ^{+3.6 \%} _{-2.8 \%}$		&	$12.71  ^{+0.08 } _{-0.06 }$	\\[0.1cm]
TL 	&	$1.904(2)   ^{+3.5 \%} _{-2.8 \%}$ 	&	$12.71  ^{+0.07 } _{-0.06 }$ 	&	$1.782(2)   ^{+3.5 \%} _{-3.0 \%}$		&	$12.70  ^{+0.06 } _{-0.08 }$	\\[0.1cm]
TT 	&	$10.10(1)   ^{+2.7 \%} _{-2.2 \%}$ 	&	$67.42  ^{+0.13 } _{-0.15 }$ 	&	$9.47(1)    ^{+2.8 \%} _{-2.2 \%}$		&	$67.52  ^{+0.09 } _{-0.13 }$	\\[0.1cm]
interference 	&	$0.191 $	                &	$1.28 $	                        &	$0.171 $                                &	$1.28 $	                        \\[0.1cm]
\hline
\end{tabular}
\end{center}
\caption{
  Off-shell, unpolarised and doubly polarised predictions for $\PZ\PZ$ in the four-charged-lepton decay channel at the LHC@13TeV,
  with NLO QCD accuracy. Same structure as in Table~\ref{tab:polWZ_allresults}.
  The inclusive and fiducial setups, defined in Eq.~\eqref{eq:mll_cut} and Eq.~\eqref{eq:fiddefZZ} respectively, are considered.\label{tab:polZZ_allresults}}
\end{table}
The rate drop between the inclusive and fiducial setup is not dramatic, owing to rather loose transverse-momentum fiducial cuts on the charged leptons.
This motivates an overall picture of polarisation fractions which is very similar in the two setups.
The LL state gives a 6\% contribution, the mixed ones sum up to a 25\%, while the TT one dominates the fiducial region accounting for almost 68\% of the
total.
Comparing the LHE results with those including PS matching and hadronisation effects, the change in the polarisation fraction is negligible, as
already seen in Table~\ref{tab:polWZ_allresults} for $\PW\PZ$. The QED PS gives a negative correction to all cross sections which roughly amounts
at 5-6\%.
The off-shell effects are at the $1.5\%$ level. Interference effects are compatible with zero in the absence of cuts on individual-lepton kinematics,
and they amount at $1.3\%$ of the unpolarised cross section in the presence of fiducial cuts.
The QCD-scale uncertainties are at the $3\%$ level for all polarised cross sections, though with mild differences (slightly larger for mixed states).
It is interesting to notice that the QCD-scale uncertainties of the polarisation fractions (correlated) are at the sub-percent level for $\PZ\PZ$, while for $\PW\PZ$
they amount at 1-$2\%$

Notice that in our calculation we do not include any loop-induced contributions in the $\Pg\Pg$ partonic channel. It is known \cite{Denner:2021csi} that
this contribution (calculated at LO) is sizeable for the LL and TT states, giving  10-15\% increase to the NLO QCD cross sections, and also introduces larger QCD-scale uncertainties.
The matching of the $\Pg\Pg\rightarrow \PZ\PZ$ contribution to PS with NLO QCD corrections has been carried out for the off-shell case \cite{Alioli:2021wpn,Buonocore:2021fnj}, and it is desirable
to extend it to the case of fixed-polarisation $\PZ$ bosons. This is left for future work.

\subsection{Differential results}
The four-charged-lepton decay channel offers a large number of angular variables which potentially discriminate between transverse and longitudinal states.
In particular it allows to reconstruct exactly the polarisation angles, \ie the lepton decay angles in the corresponding $\PZ$-boson rest frame, starting from
the Lorentz frame where polarisation are defined (the diboson CM frame). In addition to the results described in this section, more differential distributions
for $\PZ\PZ$ can be found in Appendix~\ref{app:ZZ}.

In \reffi{fig:ZZfid_1} we show differential results for the cosine of the positron polar angle in the rest frame of the positron-electron system.
\begin{figure}[h]
  \centering
  \includegraphics[width=1.0\textwidth]{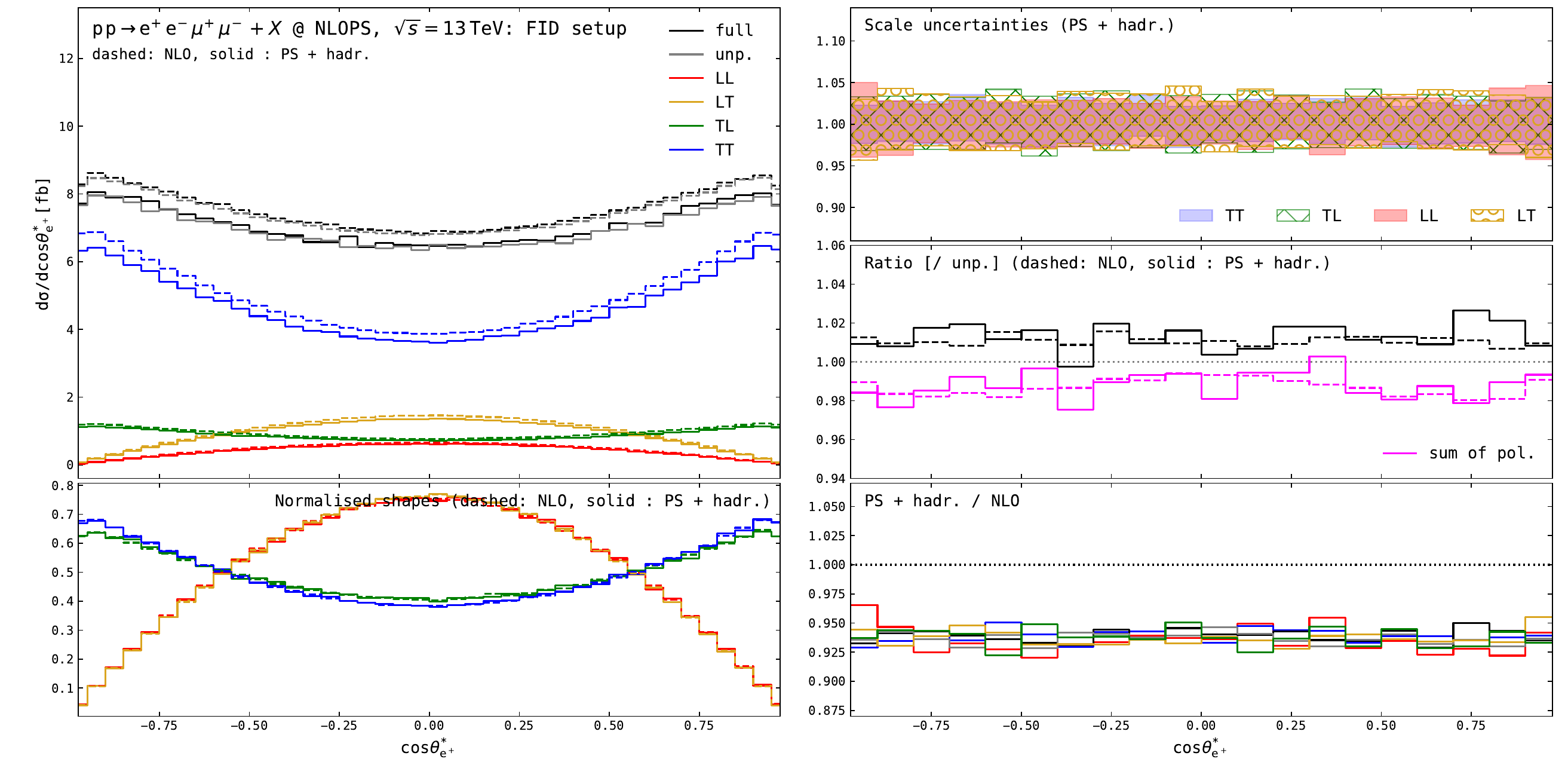}
  \caption{
    Distributions in the polar decay angle of the positron in the corresponding $\PZ$-boson rest frame for $\PZ\PZ$ production at the LHC.    
    The fiducial setup defined in Eq.~\eqref{eq:fiddefZZ} is understood.
    The colour code reads: full off-shell (black), unpolarised (gray), LL (red), LT (goldenrod), TL (green), TT (blue), sum of polarised (magenta).
    The figure is structured in 5 panels.
    Top left: absolute distributions at fixed order (NLO, dashed) and matched to QCD+QED PS and hadronisation (PS+hadr, solid).
    Bottom left: normalised distributions (to have unit integral) for polarised states at NLO (dashed) and PS+hadr (solid).
    Top right: theory uncertainties for the polarised states from 7-point renormalisation and factorisation-scale variations of PS+hadr predictions, relative to the corresponding central values.
    Middle right: ratio of the full off-shell and sum of polarised cross sections over the unpolarised one at NLO (dashed) and PS+hadr (solid).
    Bottom right: ratio of NLO QCD cross sections matched to QCD+QED PS and hadronisation over the fixed-order ones.
  }\label{fig:ZZfid_1}
\end{figure}
As expected \cite{Bern:2011ie,Stirling:2012zt}, this angular observable is sensitive
to the polarisation state of the $\PZ$ boson decaying in the $\Pe^+\Pe^-$ pair, while is
rather agnostic of the polarisation state of the other boson.
The expected differential rate in $\cos\tl$ for a longitudinal (L) and transverse (T) boson, in a fully inclusive phase space, has a closed analytic form at tree level,
 \begin{align}\label{eq:costhetaINCL}  
   \frac1{\sigma_{\rm unp}}\frac{\rd\sigma_{\rL}}{\rd\cos\tl}={}&
   \frac{3}{4} f_{\rL}\,\,\left(1-\cos^2\tl\right) \nnb\\
   \frac1{\sigma_{\rm unp}}\frac{\rd\sigma_{\rT}}{\rd\cos\tl}={}&
    \frac{3}{8} \left(1+\cos^2\tl\right)\, (f_{+}+f_{-})+\frac{3}{4} \left(\frac{c^2_{\mathrm{L},\ell}-c^2_{\mathrm{R},\ell}}{c^2_{\mathrm{L},\ell}+c^2_{\mathrm{R},\ell}}\right)\,\cos\tl\,(f_{+}-f_{-})\,,
 \end{align}
where  $\sigma_{\rm unp}$ is the unpolarised cross section, $f_{\rL},f_{+},f_{-}$ are the longitudinal, left-handed and right-handed fractions of the $\PZ$ boson ($\rightarrow \Pe^+\Pe^-$), and $c_{\mathrm{L},\ell},c_{\mathrm{R},\ell}$  are the left- and right-handed couplings of the $\PZ$ boson to leptons.
From the normalised shapes in the bottom-left panel of \reffi{fig:ZZfid_1},
it is clear that the shape is only mildly distorted by transverse-momentum and rapidity cuts at the end points, especially in the case of a transverse boson.
The longitudinal shape is very close to the analytic one in Eq.~\eqref{eq:costhetaINCL}.
The off-shell effects are rather constant over the angular spectrum.
Interferences are characterised by a mild shape, with a size variation from $1\%$ in the central region to $2\%$ in more collinear/anti-collinear topologies. 
The QCD-scale bands do not feature sizeable changes, apart from
a mild increase of their size at the end-points in the case of a longitudinal boson (LL, LT), where the real corrections are a bit enhanced.
This observable is not too sensitive to PS effects, as can be observed in the bottom-right panel.

\begin{figure}
  \centering
  \includegraphics[width=1.0\textwidth]{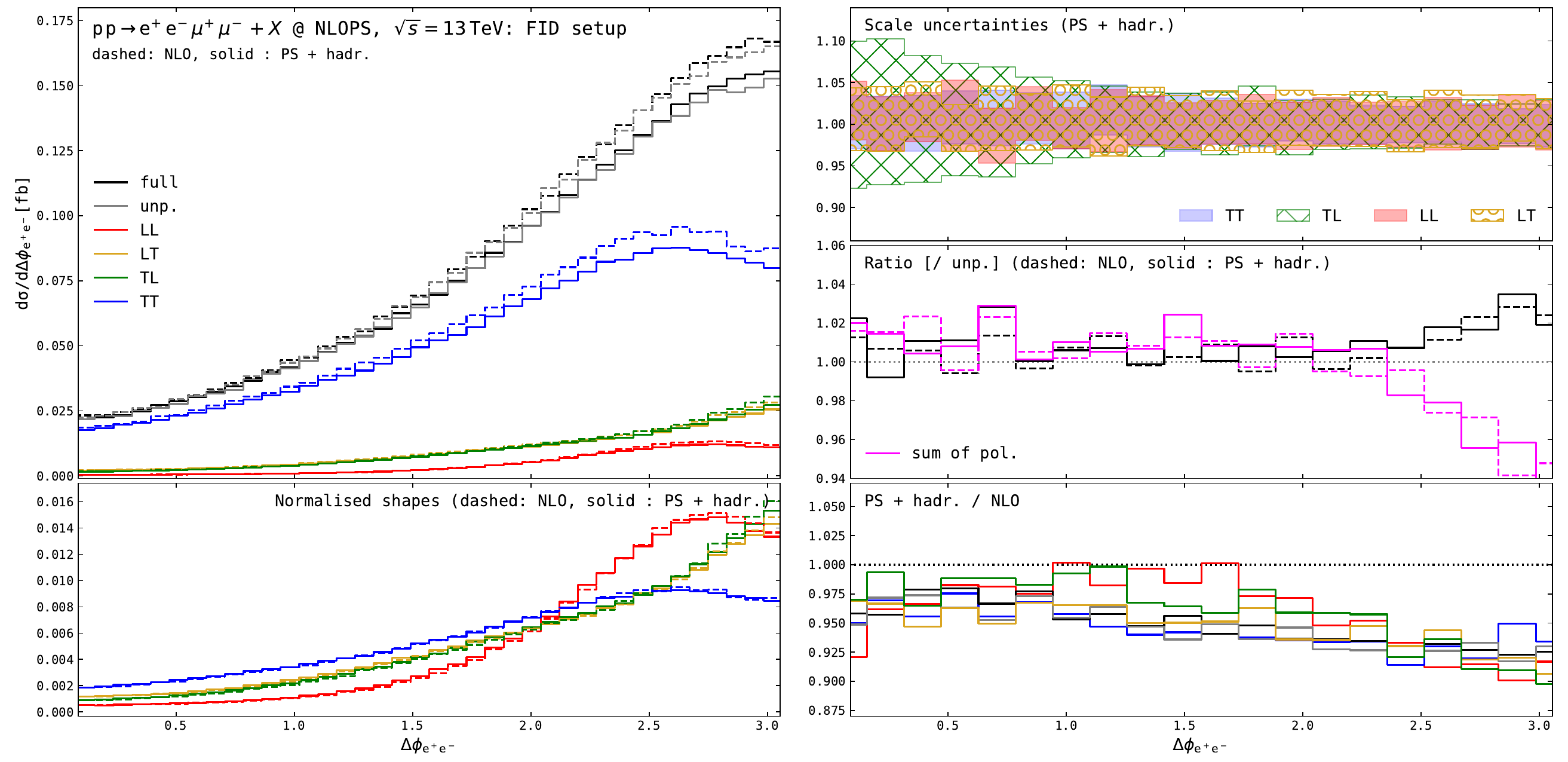}
  \caption{
    Distributions in the azimuthal-angle separation between the positron and the electron for $\PZ\PZ$ production at the LHC.
    The fiducial setup is understood. Same structure as \reffi{fig:ZZfid_1}.    
  }\label{fig:ZZfid_10}
\end{figure}

In \reffi{fig:ZZfid_10} another angular variable is considered, namely the azimuthal separation between the positron and the electron (computed in the laboratory frame).
Although one could naively expect this variable to be only sensitive to the decay of the first boson ($\PZ\rightarrow \Pe^+\Pe^-$),
it features a marked discrimination power for doubly polarised states.
In fact, the LT and TL states are almost indistinguishable and feature a maximum at $\Delta\phi_{\Pe^+\Pe^-}=\pi$. The LL and TT
normalised shapes have a maximum located around $8\pi/9$ and $5\pi/6$, respectively.
The scale uncertainties are similar for all polarisation state, up to a notable increase for the TL state in the least populated region.
Off-shell effects are compatible with the integrated result, while the interference effects are negative and small in the left side of the spectrum,
positive and increasing in size towards the most populated region, reaching up to $10\%$ at $\Delta\phi_{\Pe^+\Pe^-}=\pi$.
This azimuthal separation is also sensitive to PS effects, with sizeable differences amongst polarisation states in the
least populated region. The QED+QCD PS not only shifts the cross sections down by a $6\%$, but also gives a shape distortion
with the largest effect for the LL state.

\subsection{Comparison with the ATLAS measurement}\label{subsec:ATLASzz}
Very recently, the ATLAS collaboration has measured the purely longitudinal $\PZ\PZ$ signal \cite{ATLAS:2023zrv},
confirming the SM predictions at fixed order (NLO QCD+EW for the $q\bar{q}$ channels, LO for the $\Pg\Pg$ one) of \citere{Denner:2021csi}.
In Table~\ref{tab:ZZfid_ATLAS} we perform a comparison of the results obtained with our new \powheg implementation against the measurement result \cite{ATLAS:2023zrv}
and the theoretical predictions used therein, in the fiducial setup described in Eq.~\eqref{eq:fiddefZZ}.
\begin{table}[h]
  \begin{center}
    \begin{tabular}{ccccccc}
      \hline\\[-0.3cm]
      \cellcolor{blue!9}fraction  &\cellcolor{blue!9}  PS+hadr (this work) &\cellcolor{blue!9} TH-QCD \cite{Denner:2021csi} &\cellcolor{blue!9} TH \cite{Denner:2021csi} &\cellcolor{blue!9} pre-fit \cite{ATLAS:2023zrv}  & \cellcolor{blue!9}post-fit \cite{ATLAS:2023zrv} \\[0.1cm] 
      \hline\\[-0.2cm]      
      ${\rL\rL}$ 	         &	$5.84  ^{+0.03 } _{-0.05 } $ &  5.9    & 5.8 & $6.1 \pm 0.4$    &  $7.1 \pm 1.7$ 	\\[0.1cm]
      ${\rL\rT}+{\rT\rL}$    	 &	$25.41  ^{+0.08 } _{-0.07 } $ & 25.3     & 23.2 &$22.9 \pm 0.9$    & $22.8 \pm 1.1$  	\\[0.1cm]
      ${\rT\rT}$  	     	 &	$67.52  ^{+0.09 } _{-0.13 }$ &   67.4    & 69.8 &$69.9\pm 3.9$    &  $69.0\pm 2.7$ 	\\[0.1cm]
      ${\rm interference}$       &	$1.28 $	                  &  1.3    & 1.2  &$1.1\pm 0.1$   &  $1.1\pm 0.1$      \\[0.1cm]
\hline
    \end{tabular}
  \end{center}
  \caption{
    Theoretical predictions for joint polarisation fractions (in percentages) in $\PZ\PZ$ production, normalised to the unpolarised cross section, 
    compared to the ATLAS results reported in Table~1 of \citere{ATLAS:2023zrv}.
    Our \powheg results fractions (PS+hadr) are computed at NLO QCD and matched to {\sc Pythia8} QCD+QED shower and hadronisation effects.
    The uncertainties associated to the PS+hadr predictions come from correlated 7-point QCD-scale variations. 
    The theory predictions labeled by ``pre-fit'' are those used in \citere{ATLAS:2023zrv}. The ``post-fit'' values are the results of the measurement.
    Notice that for both pre-fit and post-fit, we have normalised the results of the first four rows of Table~1 of \citere{ATLAS:2023zrv} to the sum
    of polarised and interference contributions. The results are also compared to the NLO QCD polarisation fractions (labeled TH-QCD) of \citere{Denner:2021csi}
    and those combining NLO QCD, NLO EW and loop-induced, gluon-initiated contributions (also from \citere{Denner:2021csi}) in a very similar setup as the one considered here.
  }\label{tab:ZZfid_ATLAS} 
\end{table}
Notice that the fiducial setup of \citere{Denner:2021csi} slightly differs from the one of Eq.~\eqref{eq:fiddefZZ} by a looser transverse-momentum cut on
the subleading lepton ($10\GeV$ instead of $20\GeV$) and a tighter $R$ separation between two leptons (0.1 instead of 0.05).
In spite of these differences, the polarisation fractions at NLO QCD reported in \citere{Denner:2021csi} are very well confirmed (at the sub-percent level) by our \powheg predictions,
which are matched with QCD+QED PS and hadronisation effects.
The mild but not negligible effects induced by NLO EW corrections and by the gluon-induced channel makes it essential to embed them
in a matched calculation, which represents a natural continuation of our effort and is left for future work.
Nevertheless, the NLOPS predictions presented in this paper provide a fairly good description of the polarisation fractions
measured by ATLAS.

\section{Results for the WW process}\label{sec:WW}
The last production channel to be considered is $\PW^+\PW^-$. This diboson process features a higher rate compared to $\PW\PZ$ and $\PZ\PZ$, but is
characterised by overwhelming backgrounds from the single and pair production of top quarks.
This is why we have excluded bottom-induced contributions (although performing the calculation in the five-flavour scheme) and assumed a perfect b-jet veto in the phenomenological analysis.
Polarisation measurements in this channel are also hampered by the presence of two neutrinos, which impedes the reconstruction of the two bosons separately.
Nonetheless, a number of angular observables are known to be sensitive
to polarisation states \cite{Denner:2020bcz,Poncelet:2021jmj}.

Since in \citeres{Denner:2020bcz,Poncelet:2021jmj}, the laboratory frame was chosen for the polarisation-vector definition, the one
presented here is the first calculation of doubly polarised $\PW^+\PW^-$ production with polarisations defined in the diboson CM frame.
We consider the positron--muon decay channel,
\beq
\Pp\Pp   \rightarrow          \PW^+\,(\Pe^+\nu_{\Pe}) \PW^-\,(\mu^-\bar{\nu}_\mu)+X\,.
\eeq
We have studied a fully \emph{inclusive setup}, which avoids any selection on the final state, as well as a \emph{fiducial setup} that
mimics the selections of a recent CMS analysis \cite{CMS:2020mxy}:
\beqn\label{eq:fiddefWW}
&&  \pt{\ell_{1\,(2)}}>20\,(10)\GeV\,,{\textrm{ with }} \ell_{1(2)}= {\textrm{(sub)leading lepton}}\,,\nnb\\
&&  |\eta_{\Pe^+}|<2.5\,,\quad |\eta_{\mu^-}|<2.4\,,\nnb\\
&&  \pt{\Pe^+\mu^-}>30\GeV\,,\quad \pt{\rm mis}>20\GeV\,,\nnb\\
&&  M_{\Pe^+\mu^-} > 20\GeV\,.
\eeqn
where the leading and sub-leading leptons are sorted in transverse momentum.

It is instructive to compare the polarisation fractions computed in the CM frame (with our \powheg implementation) with those computed
in the laboratory frame (as in \citeres{Denner:2020bcz,Poncelet:2021jmj}). For this comparison, \mocanlo has been used to compute
polarised signals defined in the laboratory. The results are shown in Table~\ref{tab:polWW_allresults_inc} for the inclusive setup (no cuts).
\begin{table}[h]
\begin{center}
\begin{tabular}{ccccc}
\hline\\[-0.3cm]
\cellcolor{blue!9} state & \cellcolor{blue!9}$\sigma[\fb]$ & \cellcolor{blue!9} ratio [/unp., \%] & \cellcolor{blue!9} $\sigma[\fb]$ & \cellcolor{blue!9} ratio [/unp., \%] \\[0.1cm]
\hline\\[-0.3cm]
 & \multicolumn{2}{c}{\cellcolor{green!9} CM definition} & \multicolumn{2}{c}{\cellcolor{orange!9} laboratory definition}\\
\hline\\[-0.1cm]
full off-shell 	&	$1267.0(3)^{+3.8 \%} _{-3.0 \%}$ 	&	$101.46$                	   &   $ 1266.8(11)^{+ 3.8 \%}_{- 3.0 \%}$   &   $101.41$   \\[0.1cm]
unpolarised 	&	$1248.8(4) ^{+3.8 \%} _{-3.1 \%}$ 	&	$100$                    	   &   $ 1249.2(5)^{+ 3.8 \%}_{- 3.1 \%}$   &  $100$    \\[0.1cm]
LL 	        &	$65.88(3) ^{+3.2 \%} _{-2.7 \%}$ 	&	$5.28 $ 	                   &   $ 41.50(2)^{+ 7.2 \%}_{- 3.8 \%}$   &  $3.32$    \\[0.1cm]
LT 	        &	$158.65(6) ^{+5.2 \%} _{-4.1 \%}$ 	&	$12.70$ 	                   &   $ 285.8(2)^{+ 3.3 \%}_{- 2.6 \%}$   &   $22.88$   \\[0.1cm]
TL 	        &	$163.04(7) ^{+5.3 \%} _{-4.3 \%}$ 	&	$13.06$ 	                   &   $ 336.0(2)^{+ 3.0 \%}_{- 2.4 \%}$   &   $26.90$   \\[0.1cm]
TT 	        &	$861.8(3) ^{+3.3 \%} _{-2.6 \%}$ 	&	$69.01$ 	                   &   $ 585.4(2)^{+ 4.2 \%}_{- 3.4 \%}$  &   $46.86$   \\[0.1cm]
interference 	&	$-0.54 $	                &	$-0.04$                      	   &   $0.50$                        &   $ 0.04 $   \\[0.1cm]
\hline
\end{tabular}
\end{center}
\caption{
  Off-shell, unpolarised and doubly polarised predictions for $\PW^+\PW^-$ in the two-charged-lepton decay channel at the LHC@13TeV,
  with NLO QCD accuracy. Results computed with  \powheg (polarisations defined in the CM frame) and with
  \mocanlo (polarisation defined in the laboratory frame) are compared at fixed order.
  Integrated cross sections (in fb) and ratios normalised to the unpolarised result are presented.
  Monte--Carlo-integration (in parentheses) and QCD-scale (percentages in subscripts and superscripts) uncertainties are shown for the integrated cross sections.
  The inclusive setups (no cuts) is understood.  \label{tab:polWW_allresults_inc}}
\end{table}
As expected, the full off-shell and unpolarised results are independent of the frame where polarisation vectors are defined.
On the contrary, marked differences are there for the polarised cross sections. In particular, defining polarisations
in the laboratory frame sizeably increases the LT and TL fractions compared to the CM definition, while the LL and TT fractions diminish.
A similar pattern was found also in $\PW\PZ$ \cite{Denner:2020eck} and $\PZ\PZ$ inclusive production \cite{Denner:2021csi},
as well as in vector boson scattering (VBS) processes \cite{Ballestrero:2020qgv}.
This effect \cite{Denner:2021csi} comes from the fact that in the CM definition, the longitudinal polarisation vectors only depend on the
boson momenta and are back to back. The transverse polarisation vectors are defined to be orthogonal to the corresponding longitudinal ones.
When boosting from the CM to the laboratory frame, part of the longitudinal component becomes transverse (with a more involved
kinematic dependence) and vice-versa, resulting in increased mixed-state fractions (LT, TL) and reduced spin-diagonal ones (LL, TT).
From a theoretical point of view, defining polarisations in the diboson CM frame is more appropriate as there is a natural
interpretation in terms of the underlying $q\bar{q}\rightarrow VV$ process at LO.

In Table~\ref{tab:polWW_allresults_fid} we show integrated results obtained with \powheg in the fiducial setup of Eq.~\eqref{eq:fiddefWW}.
We show results both at LHE level and matched to QCD+QED PS and hadronisation effects.
\begin{table}[h]
\begin{center}
\begin{tabular}{ccccc}
\hline\\[-0.3cm]
\cellcolor{blue!9} state & \cellcolor{blue!9}$\sigma[\fb]$ LHE & \cellcolor{blue!9} ratio [/unp., \%] LHE & \cellcolor{blue!9} $\sigma[\fb]$ PS+hadr & \cellcolor{blue!9} ratio [/unp., \%] PS+hadr\\[0.1cm]
\hline\\[-0.3cm]
\multicolumn{5}{c}{ \cellcolor{green!9} Fiducial setup }\\
\hline\\[-0.1cm]
full off-shell 	&	$383.1(6) ^{+4.4 \%} _{-3.5 \%}$ 	&	$102.65 $	&	$377.7(6) ^{+4.4 \%} _{-3.5 \%}$		&	$102.58 $	\\[0.1cm]
unpolarised 	&	$373.2(6) ^{+4.4 \%} _{-3.5 \%}$ 	&	$100$ 	&	$368.2(6) ^{+4.3 \%} _{-3.6 \%}$		&	$100$	\\[0.1cm]
LL 	&	$23.88(4) ^{+3.2 \%} _{-2.7 \%}$ 	&	$6.40  ^{+0.13 } _{-0.14 }$ 	&	$23.51(3)  ^{+3.2 \%} _{-2.7 \%}$		&	$6.39  ^{+0.13 } _{-0.13 }$	\\[0.1cm]
LT 	&	$56.25(8)  ^{+5.5 \%} _{-4.4 \%}$ 	&	$15.07  ^{+0.16 } _{-0.14 }$ 	&	$55.47(8) ^{+5.4 \%} _{-4.5 \%}$		&	$15.06  ^{+0.16 } _{-0.14 }$	\\[0.1cm]
TL 	&	$58.68(9) ^{+5.5 \%} _{-4.4 \%}$ 	&	$15.73  ^{+0.17 } _{-0.15 }$ 	&	$57.8(1) ^{+5.5 \%} _{-4.5 \%}$		&	$15.71  ^{+0.19 } _{-0.14 }$	\\[0.1cm]
TT 	&	$254.3(4) ^{+3.9 \%} _{-3.1 \%}$ 	&	$68.15  ^{+0.30 } _{-0.35 }$ 	&	$251.1(4) ^{+3.7 \%} _{-3.2 \%}$		&	$68.21  ^{+0.28 } _{-0.38 }$	\\[0.1cm]
interference 	&	$-19.94 $	&	$-5.34 $	&	$-19.75 $	&	$-5.34 $	\\[0.1cm]
\hline
\end{tabular}
\end{center}
\caption{
  Off-shell, unpolarised and doubly polarised predictions for $\PW^+\PW^-$ in the two-charged-lepton decay channel at the LHC@13TeV,
  with NLO QCD accuracy. Same structure as in Table~\ref{tab:polWZ_allresults}.
  The fiducial setups described in Eq.~\eqref{eq:fiddefWW} is understood.  \label{tab:polWW_allresults_fid}}
\end{table}
The fiducial setup does not include any jet-activity veto, which leads to quite large real-radiation corrections.
Compared to the inclusive setup, the interference effects are unexpectedly large and negative ($-5\%$), almost as large as the
purely longitudinal contribution ($+6.4\%$). 
The enhancement of interference terms is a well known effect also in the case of polarisations defined in the laboratory
frame \cite{Denner:2020bcz,Poncelet:2021jmj}, but in the CM-frame definition this effect is even larger and evident also at
the level of integrated cross sections.
This can be traced back to the purely left-handed chirality of the $\PW$-boson coupling to leptons and its interplay with
the transverse-momentum cuts on charged leptons, that introduces large
interferences between longitudinal and transverse modes of individual bosons, as well as sizeable spin correlations between
the two $\PW$'s.
The off-shell effects amount at $+2.6\%$ of the full result, owing to the fiducial transverse-momentum cuts that
slightly enhances single-resonant diagrams, which are not part of the DPA calculation.
The PS and hadronisation effects give a $1.5\%$ decrease to all cross sections, both polarised and unpolarised,
therefore not changing the polarisation balance found at LHE level.
The QCD-scale uncertainties are at the $3\%$ level for the LL and TT integrated cross sections, while they are mildly larger ($5\%$) for
mixed states. The corresponding uncertainties on polarisation fractions are at the $1\%$ level.

\begin{figure}[h]
  \centering
  \includegraphics[width=1.0\textwidth]{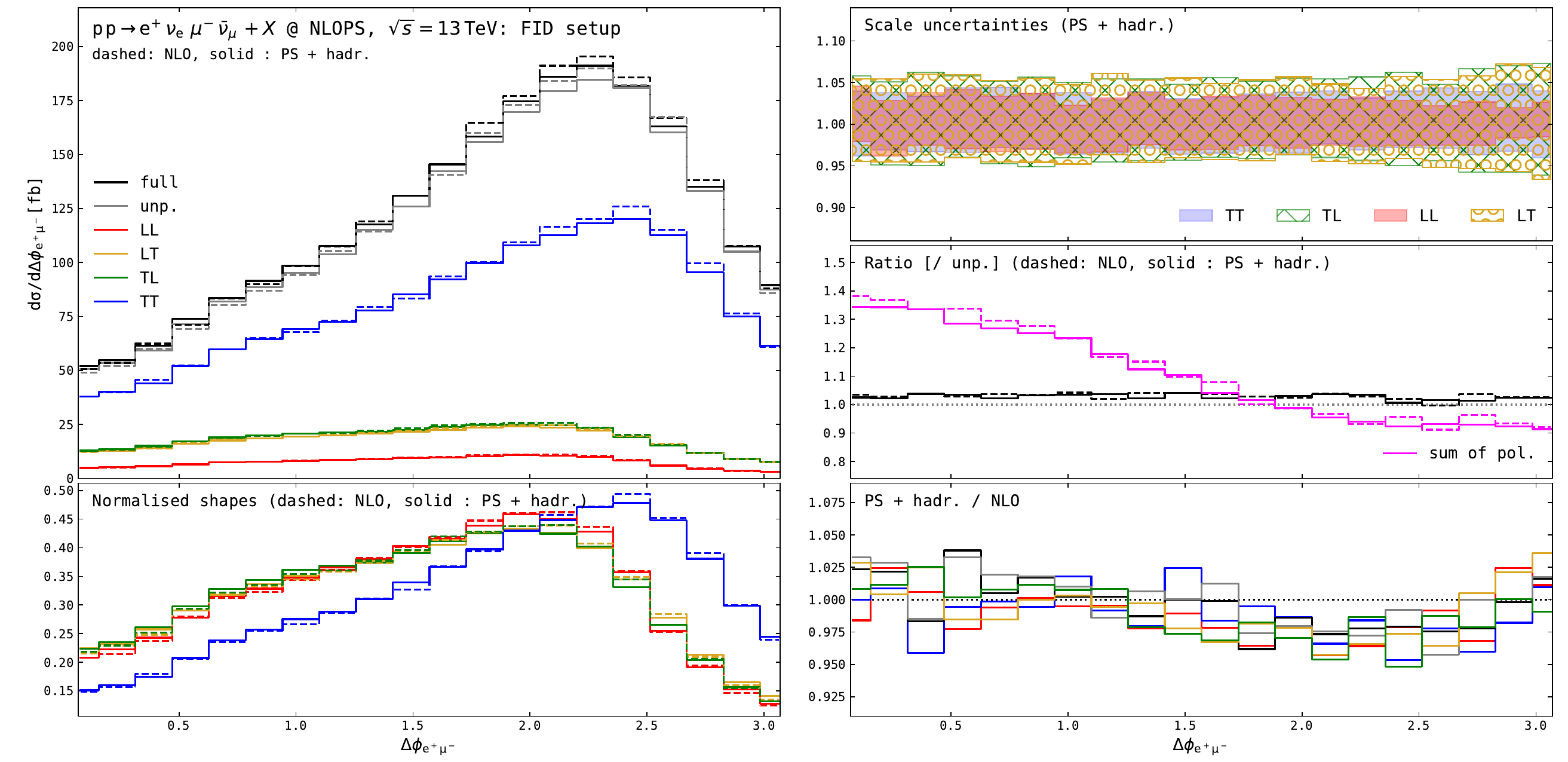}
  \caption{
    Distributions in the azimuthal-angle separation between the positron and the muon for $\PW^+\PW^-$ production at the LHC.    
    The fiducial setup defined in Eq.~\eqref{eq:fiddefWW} is understood.
    The colour code and the structure is the same as the one of \ref{fig:ZZfid_1}.
  }\label{fig:WWfid_1}
\end{figure}

The interference effects observed at integrated level are way larger when considering differential results in
angular observables. In \reffi{fig:WWfid_1} we show the fiducial results for the azimuthal-angle separation between
the two charged leptons.
In spite of a marked shape difference between the TT polarisation modes and the others,
the most interesting feature of the plot is the deviation of the sum of polarised distributions
from the unpolarised result, highlighting very large interference and spin-correlation effects
that vary from $-25\%$ at $\Delta\phi_{\Pe^+\mu^-}=0$ to $+10\%$ at $\Delta\phi_{\Pe^+\mu^-}=\pi$.
The QCD-scale bands, as well as the off-shell effects, reflect the integrated results without
notable changes in any part of the spectrum.
The PS effects, although varying between $+4\%$ and $-5\%$ over the angular range, are not sizeably different
for the various polarisation states.

\begin{figure}[h]
  \centering
  \includegraphics[width=1.0\textwidth]{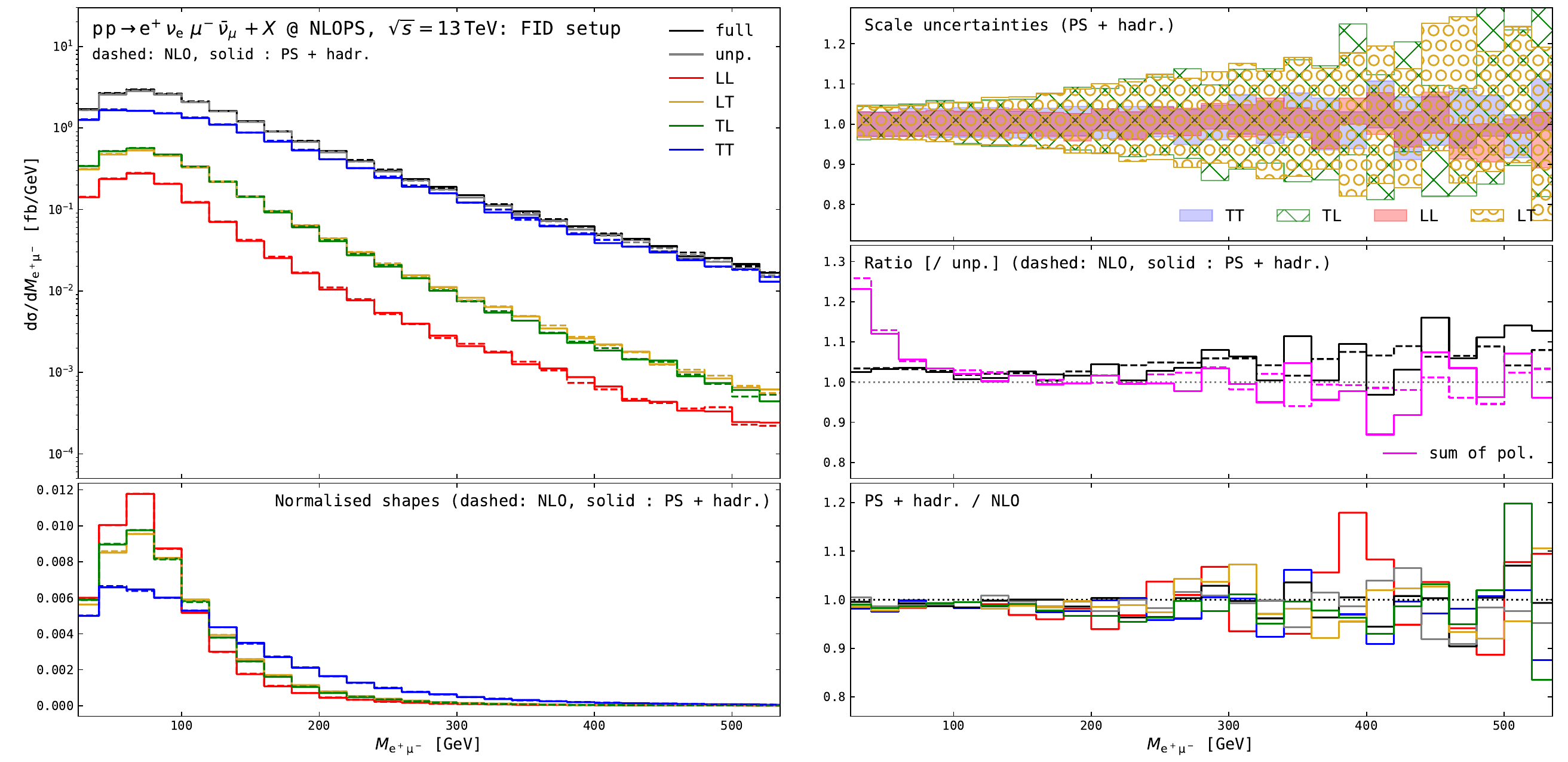}
  \caption{
    Distributions in invariant mass of the positron--muon system for $\PW^+\PW^-$  production at the LHC.
    The fiducial setup defined in Eq.~\eqref{eq:fiddefWW} is understood.
    Same structure as \reffi{fig:WWfid_1}.    
  }\label{fig:WWfid_2}
\end{figure}

In \reffi{fig:WWfid_2} we consider the differential distributions in the invariant mass of the positron--muon system.
The TT contribution dominates over the whole spectrum, with the LL and mixed states suppressed by almost two orders
of magnitude for $M_{\Pe^+\mu^-}>0.5\TeV$.
Shape-wise, the LL distribution is characterised by a narrower peak, compared to the wider spread of the TT and mixed ones.
Also this observables is affected by large and negative interference effects, especially for the low-mass region, $M_{\Pe^+\mu^-}<60\GeV$.
We observe that the same phase-space region is also affected by a sizeable Higgs background ($\Pp\Pp\rightarrow \PH\rightarrow 2\ell\, 2\nu$).
In the considered CMS fiducial volume \cite{CMS:2020mxy} the selection $M_{\Pe^+\mu^-}>20\GeV$ is applied. A tighter selection,
like done in \citere{ATLAS:2019rob}, would reduce both the Higgs background and the interference effects in $\PW\PW$ on-shell production.
The QCD-scale bands show a rather constant behaviour for the LL and TT states, while they constantly increase
towards large invariant-mass values for the LT and TL states.
No significant differences amongst polarisation states are found in the PS effects.

Additional differential results for $\PW^+\PW^-$ are shown in Appendix~\ref{app:WW}.

Although public results are not available, carrying out a polarisation analysis of Run-2 LHC data in $\PW\PW$ inclusive production
would be of extreme interest. Up to obvious challenges due to the overwhelming top-quark background, our phenomenological analysis proved that
there is room for the separation of polarised signals in $\PW\PW$ through a template fit, provided that an independent template is used for interference effects,
that are noticeably enhanced compared to other diboson channels, especially when considering differential observables.
Compared to $\PW\PZ$ and $\PZ\PZ$, in $\PW\PW$  it is not possible to properly reconstruct each boson rest frame, hindering the
access to the polarisation-density matrix. However, we demonstrated that angular observables can discriminate amongst polarisation states
and enhance interferences and spin correlations.

\section{Conclusion}\label{sec:con}
In this work we have presented the calculation  of diboson production and decay at the LHC, accounting for
definite polarisation states for intermediate EW bosons, including complete NLO QCD corrections matched to the PS
modelling in the \powheg framework. This has been achieved employing a pole-approximation approach for the separation of resonant-boson contributions in a
gauge-invariant manner, and selecting polarisation states at the amplitude level, thanks to the interface to the most recent
\recola1 library version.
This general strategy, already used in fixed-order calculations
\cite{Denner:2020bcz,Denner:2020eck,Poncelet:2021jmj,Denner:2021csi,Le:2022lrp,Le:2022ppa,Denner:2022riz,Dao:2023pkl}, has been
implemented in \powheg, building on top of the public code presented in \citere{Chiesa:2020ttl}. The resulting code
enables to simulate unpolarised, singly or doubly polarised diboson process at the LHC ($\PW^\pm\PZ, \PZ\PZ$ and $\PW^+\PW^-$) in the
fully leptonic decay channel, at NLO QCD accuracy.

The newly implemented \powheg code has undergone several successful validation stages, both at fixed order
and matched to PS, comparing against results available in the literature and other numerical tools.

A detailed phenomenological analysis of polarised signals has been performed in inclusive and fiducial setups, for $\PW^+\PZ$,
$\PZ\PZ$ and $\PW^+\PW^-$ production, focusing on shape differences amongst doubly polarised states, off-shell and
interference effects, polarisation fractions, QCD-scale uncertainties, and the effect of PS matching.

Although typically at the percent level, the QCD PS effects show some differences for the various polarisation states, especially in
some transverse momentum distributions, with the largest distortion of the NLO shapes
found at small values in transverse-momentum distributions.
Owing to the purely leptonic final state, the largest PS effect comes from the QED shower, that shifts the NLO polarised and polarised
cross sections down by a few percent ($-5\%$ for $\PW\PZ$ and $\PZ\PZ$, $-2\%$ for $\PW\PW$). The hadronisation effects are almost
negligible both at integrated and at differential level.

The overall picture of polarisation fractions does not change sizeably from the fixed-order modelling to the matching to PS,
especially when considering rather inclusive fiducial setups.
Larger effects are found in boosted setups, or in the presence of jet-activity vetoes.

In $\PW\PZ$ and $\PZ\PZ$, we have carried out comparisons against the ATLAS results of \citere{ATLAS:2022oge}
and \citere{ATLAS:2023zrv}, respectively. Our NLO QCD results matched to PS and hadronisation, up to small discrepancies
mostly coming from missing NLO EW effects (in $\PW\PZ$) and gluon-induced contributions (in $\PZ\PZ$), confirm the SM predictions
for polarisation fractions used by ATLAS, and agree by 1$\sigma$ with the measured values.

The newly implemented code, which will soon be made public on the \powheg webpage, enables the full
simulation of polarised-boson signals via unweighted-event generation at NLO QCD accuracy and enabling
multiplicative matching to PS, which is of crucial need for the experimental collaborations in order
to reduce modelling systematics in upcoming polarisation analyses.
The possibility to generate events for polarised-boson states with higher-order corrections and
matched to PS will also be helpful for further phenomenological investigations, in the direction
of polarisation taggers~\cite{Grossi:2023fqq} and of quantum observables
\cite{Barr:2022wyq,Aguilar-Saavedra:2022wam,Ashby-Pickering:2022umy,Fabbrichesi:2023cev,Morales:2023gow,Aoude:2023hxv,Bernal:2023ruk,Bi:2023uop}.

Extending the code to include NNLO QCD and NLO EW corrections in the pole approximation,
although more involved from a technical point of view, is desirable and
this work sets the foundation  
to undertake this endeavour. Such extensions and related works are
left for future investigations.

\section*{Acknowledgements}
We would like to thank Ansgar Denner for precious discussions and comments on the manuscript, and
Christoph Haitz for performing numerical checks with the BBMC Monte Carlo.
We are grateful to Dongshuo Du,  Mareen Hoppe, Oldrich Kepka, Joany Manjarres, Emanuele Re and Frank Siegert for valuable discussions.
We also acknowledge the COMETA EU COST Action (CA22130) for fruitful exchanges on polarisation measurements with LHC data.

\appendix
\newpage
\section{Additional distributions}
In this appendix we present additional distributions for the processes
considered in this work. The layout of the figures is the same as in
the plots presented in the main text. 
\subsection{Additional distributions for WZ}\label{app:WZ}
\vspace*{-0.3cm}
\begin{figure}[h]
  \centering
  \includegraphics[width=1.0\textwidth]{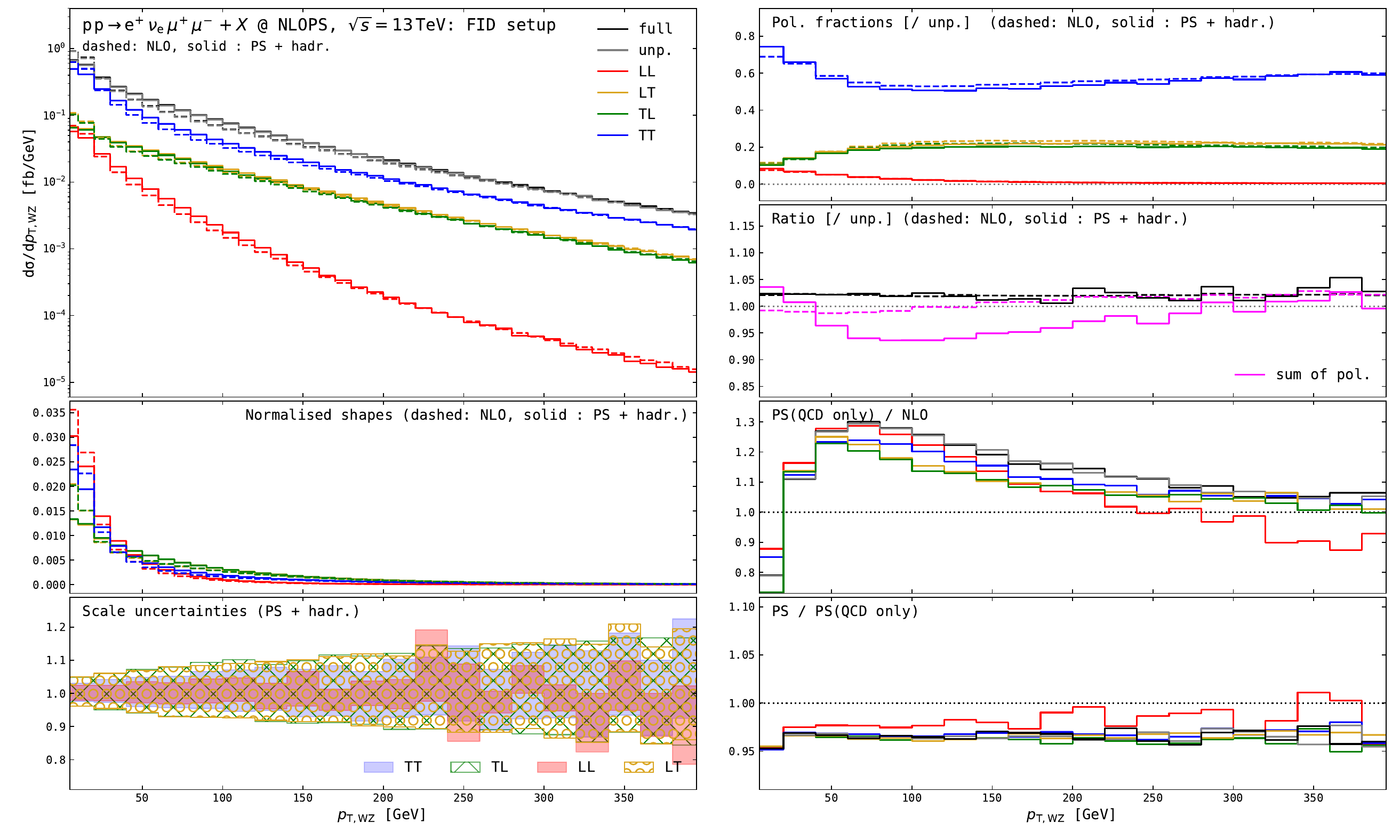}
  \caption{
    Distributions in the transverse momentum of the diboson system for $\PW^+\PZ$ production at the LHC.
    The fiducial setup described in \refse{subsec:setup} is understood. Same structure as \reffi{fig:WZfid_1}.   
  }\label{fig:WZfid_2}
\end{figure}
\begin{figure}[h]
  \centering
  \includegraphics[width=1.0\textwidth]{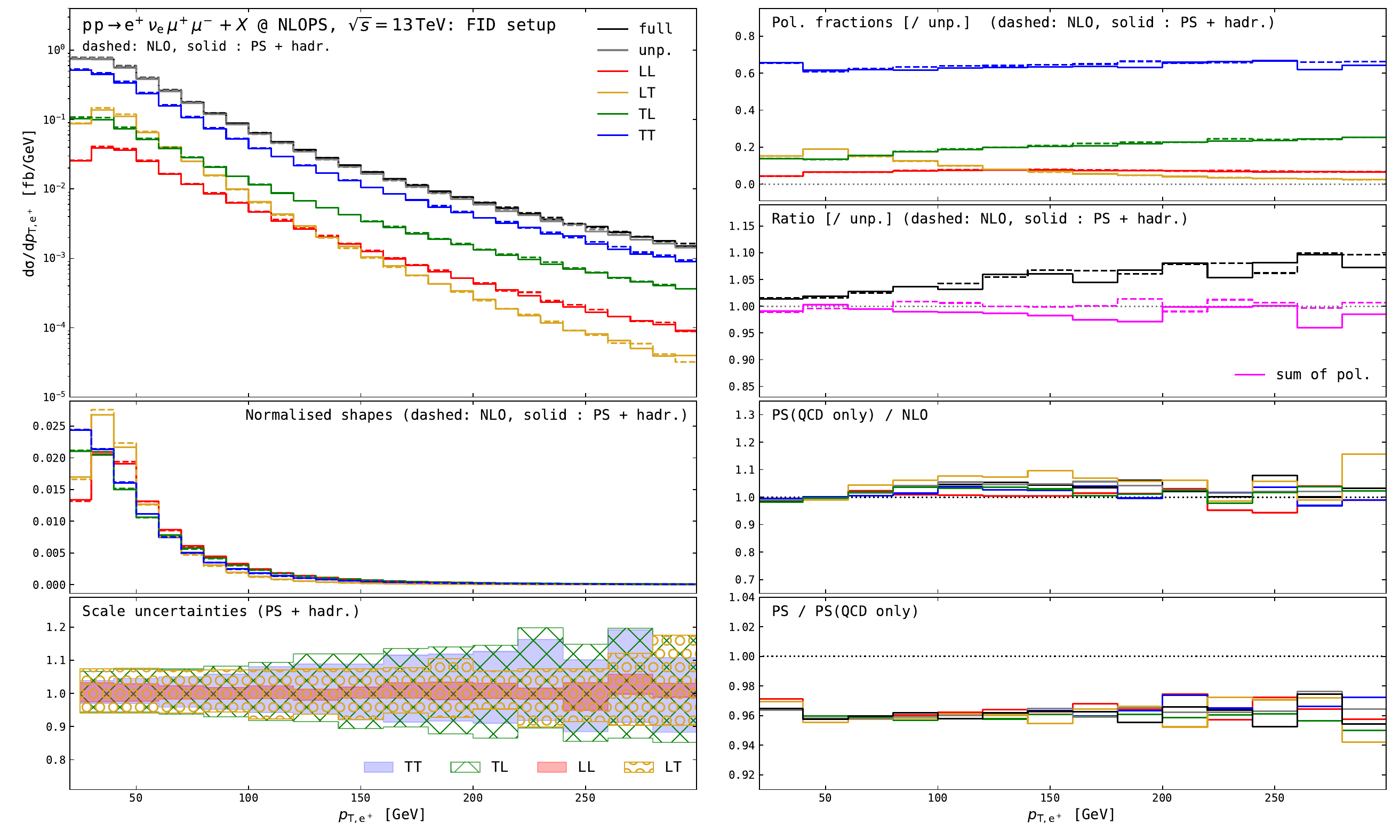}
  \caption{
    Distributions in the transverse momentum of the positron for $\PW^+\PZ$ production at the LHC.
    The fiducial setup described in \refse{subsec:setup} is understood.
    Same structure as \reffi{fig:WZfid_1}.    
  }\label{fig:WZfid_4}
\end{figure}
\newpage
\begin{figure}[h]
  \centering
  \includegraphics[width=1.0\textwidth]{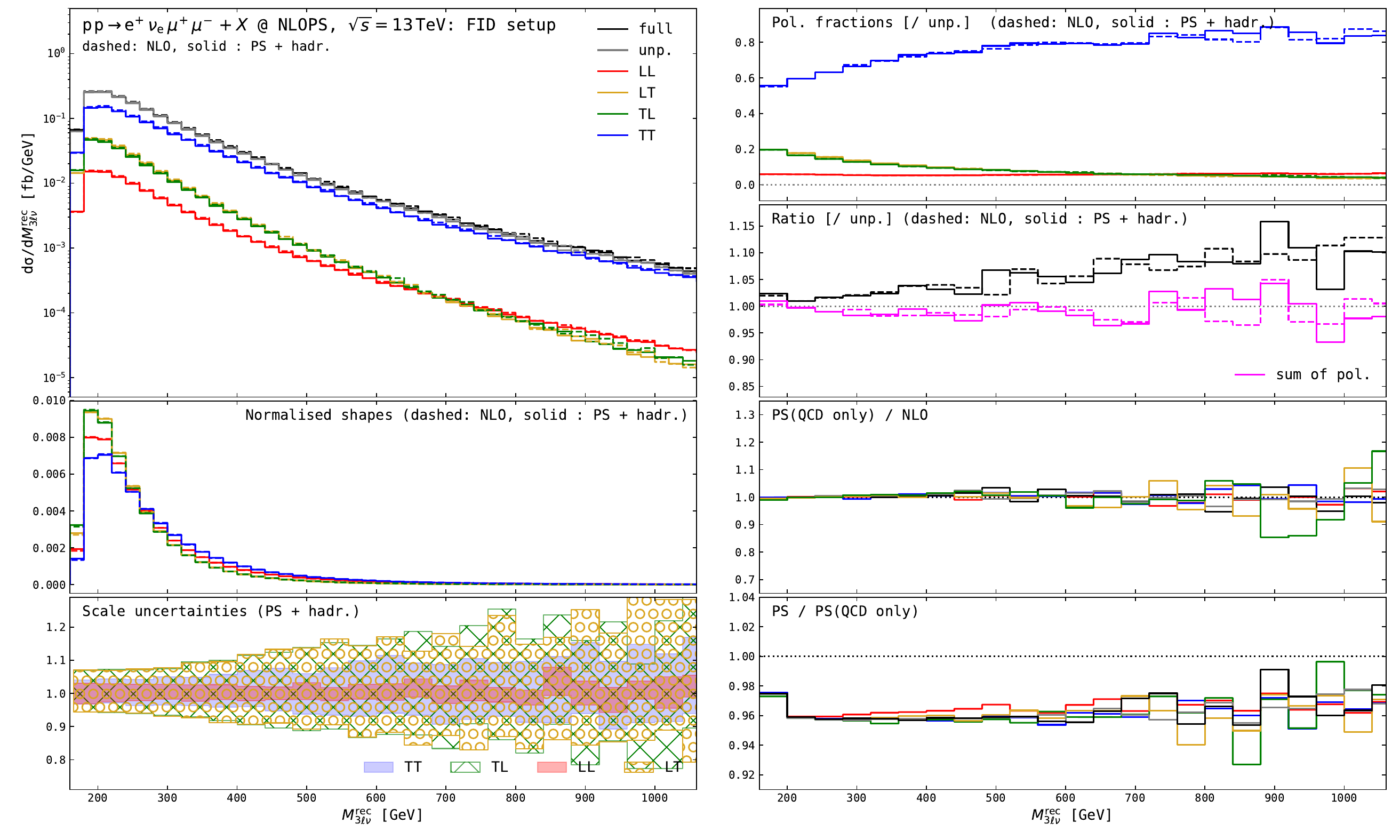}
  \caption{
    Distributions in the invariant mass of the diboson system after neutrino reconstruction for $\PW^+\PZ$ production at the LHC.
    The fiducial setup described in \refse{subsec:setup} is understood.
    Same structure as \reffi{fig:WZfid_1}.    
  }\label{fig:WZfid_6}
\end{figure}
\begin{figure}[h]
  \centering
  \includegraphics[width=1.0\textwidth]{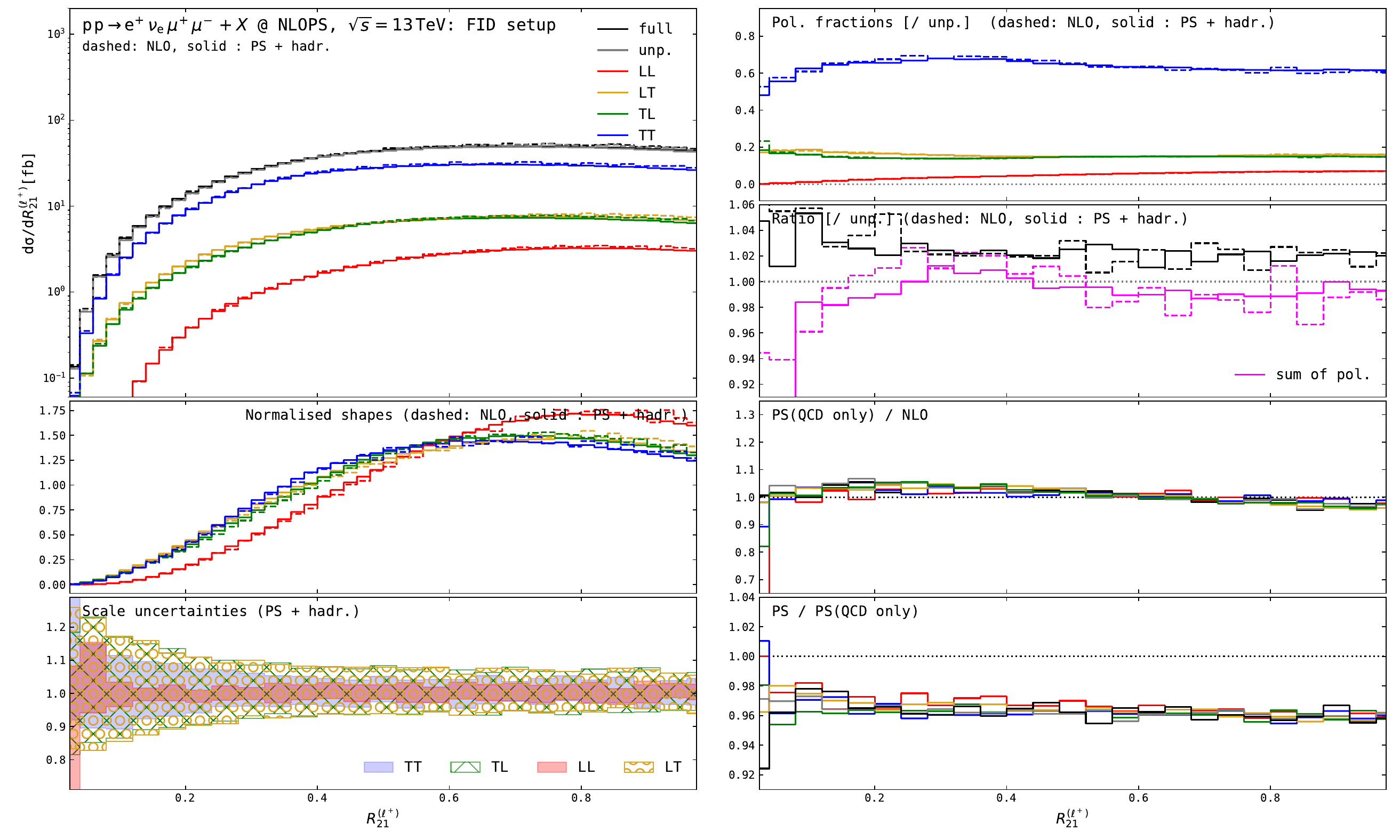}
  \caption{
    Distributions in the transverse-momentum ratio between the subleading and the leading positively charged leptons for $\PW^+\PZ$ production at the LHC.
    The fiducial setup described in \refse{subsec:setup} is understood.
    Same structure as \reffi{fig:WZfid_1}.    
  }\label{fig:WZfid_8}
\end{figure}
\newpage
\begin{figure}[h]
  \centering
  \includegraphics[width=1.0\textwidth]{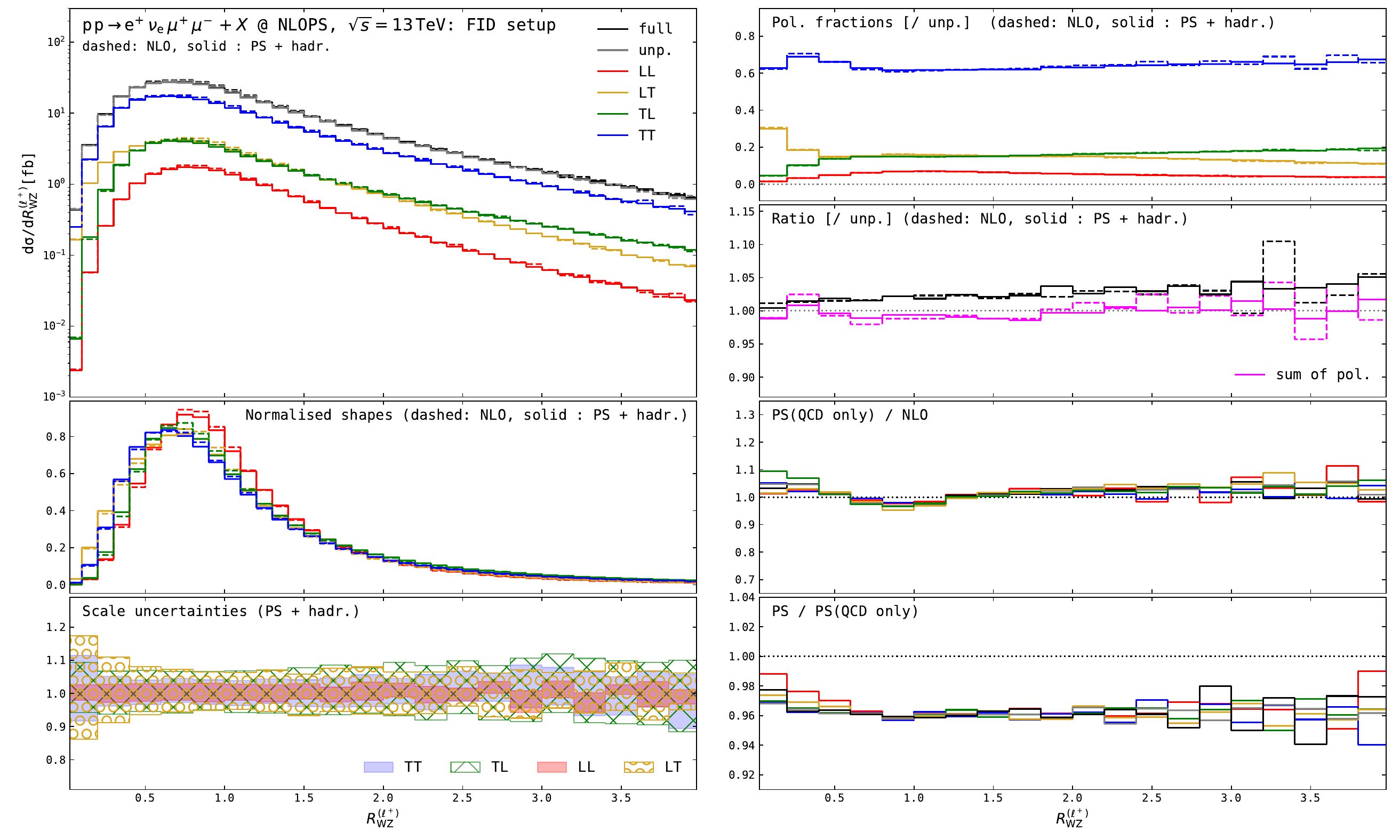}
  \caption{
    Distributions in the transverse-momentum ratio between the positron and the antimuon for $\PW^+\PZ$ production at the LHC.
    The fiducial setup described in \refse{subsec:setup} is understood.
    Same structure as \reffi{fig:WZfid_1}.    
  }\label{fig:WZfid_10}
\end{figure}
\begin{figure}[h]
  \centering
  \includegraphics[width=1.0\textwidth]{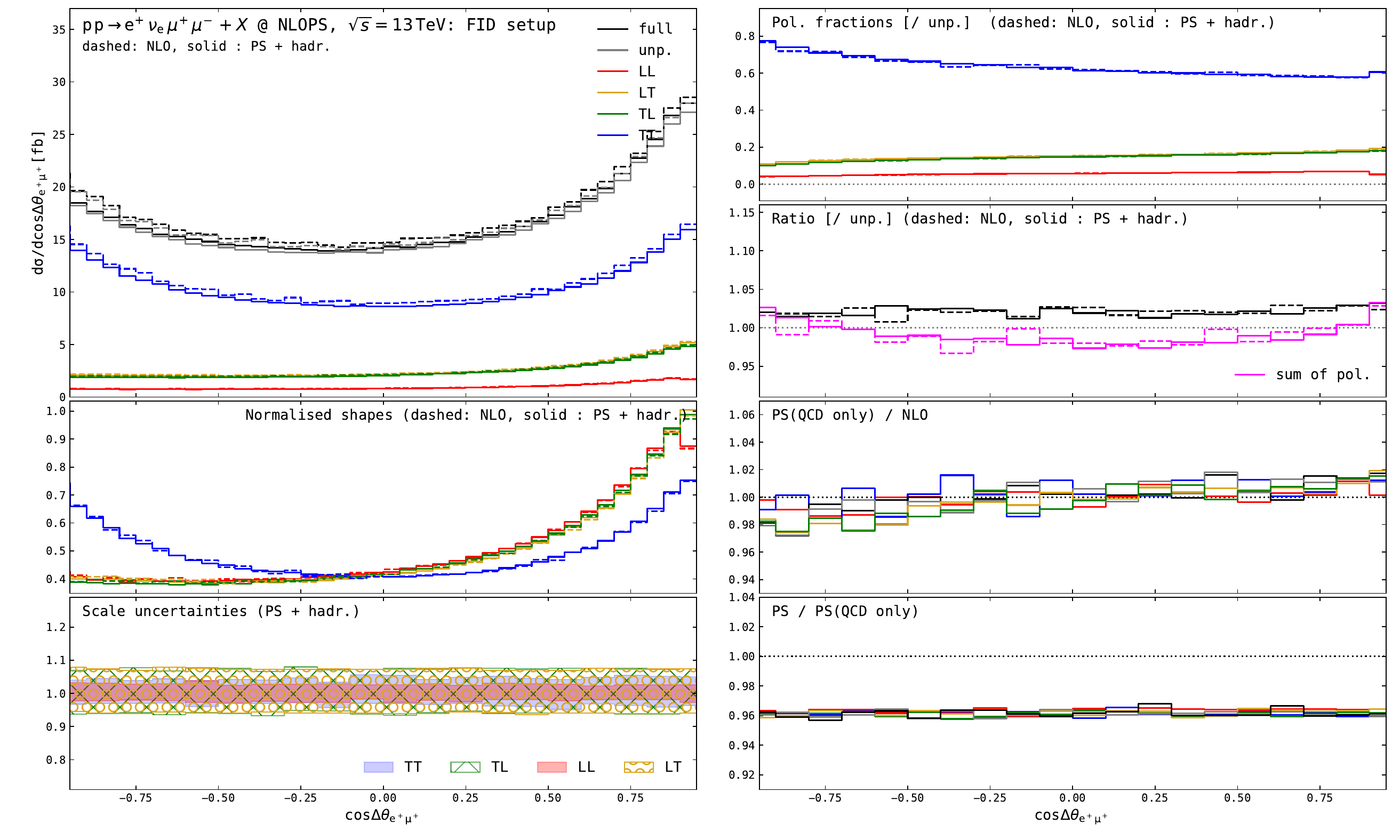}
  \caption{
    Distributions in the 3D angular separation between the positron and the antimuon for $\PW^+\PZ$ production at the LHC.
    The fiducial setup described in \refse{subsec:setup} is understood.
    Same structure as \reffi{fig:WZfid_1}.    
  }\label{fig:WZfid_11}
\end{figure}
\newpage
\begin{figure}[h]
  \centering
  \includegraphics[width=1.0\textwidth]{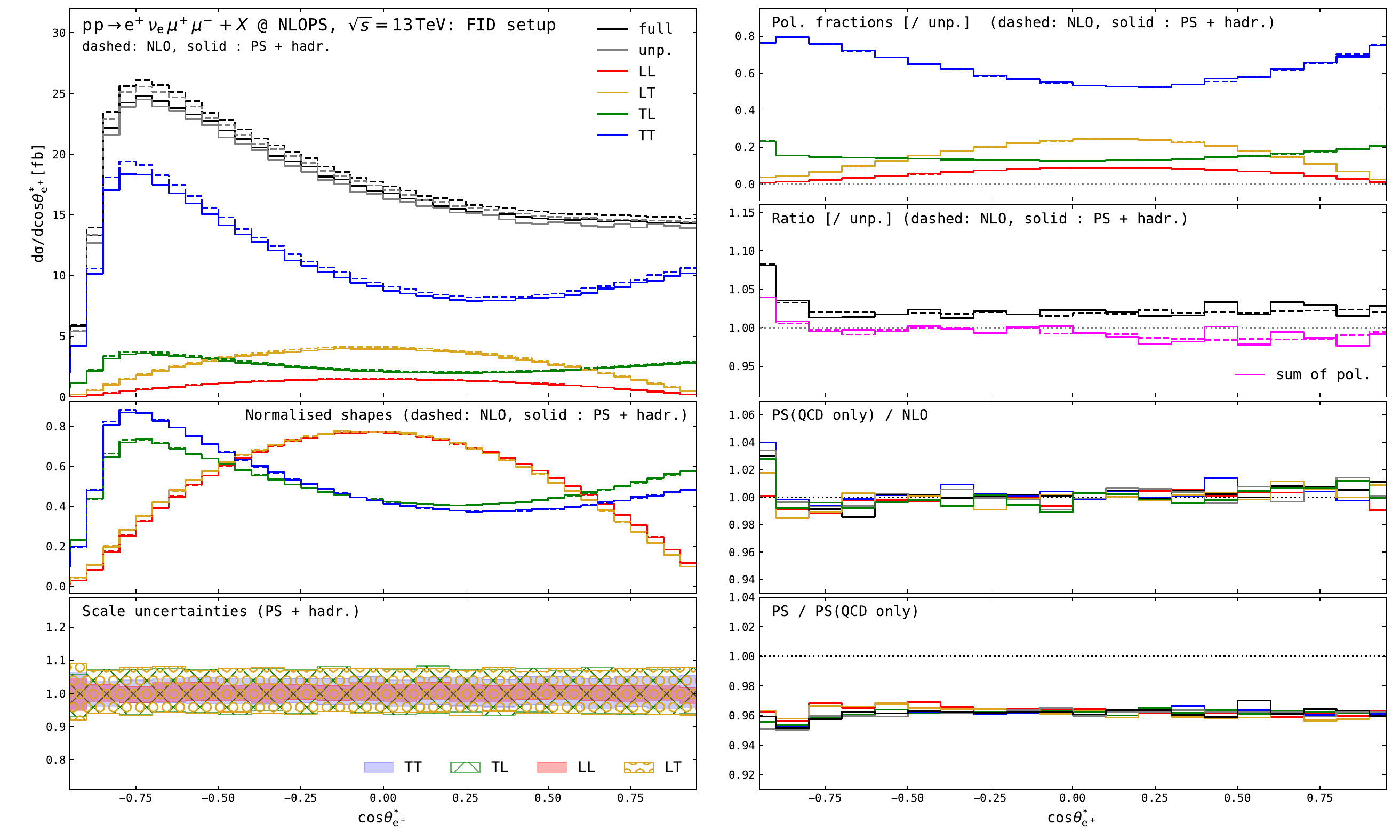}
  \caption{
    Distributions in the polar decay angle of the positron in the $\PW$ rest frame for $\PW^+\PZ$ production at the LHC.
    The fiducial setup described in \refse{subsec:setup} is understood.
    Same structure as \reffi{fig:WZfid_1}.    
  }\label{fig:WZfid_15}
\end{figure}
\begin{figure}[h]
  \centering
  \includegraphics[width=1.0\textwidth]{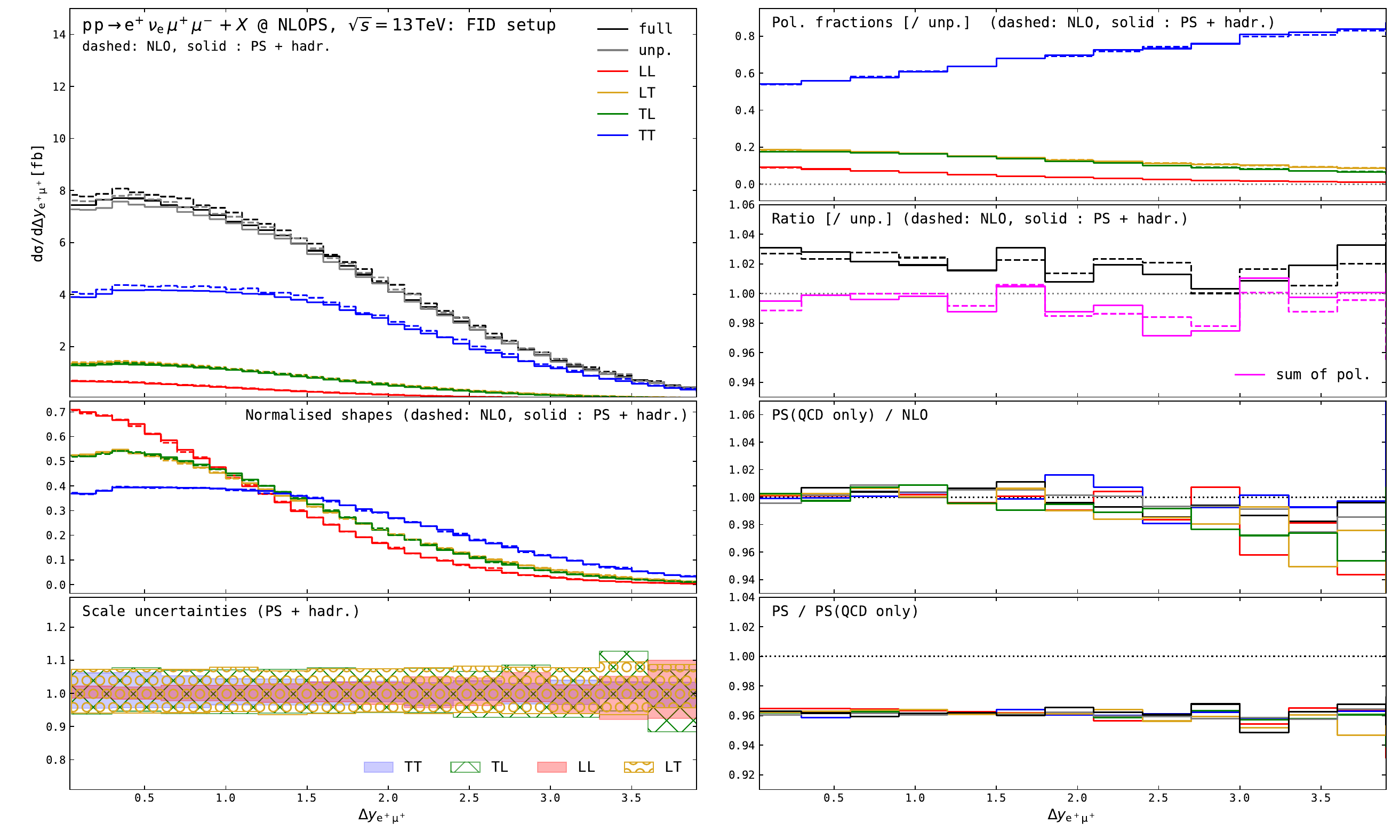}
  \caption{
    Distributions in the rapidity separation between the positron and the antimuon for $\PW^+\PZ$ production at the LHC.
    The fiducial setup described in \refse{subsec:setup} is understood.
    Same structure as \reffi{fig:WZfid_1}.    
  }\label{fig:WZfid_14}
\end{figure}

\newpage
\subsection{Additional distributions for ZZ}\label{app:ZZ}
\begin{figure}[h]
  \centering
  \includegraphics[width=1.0\textwidth]{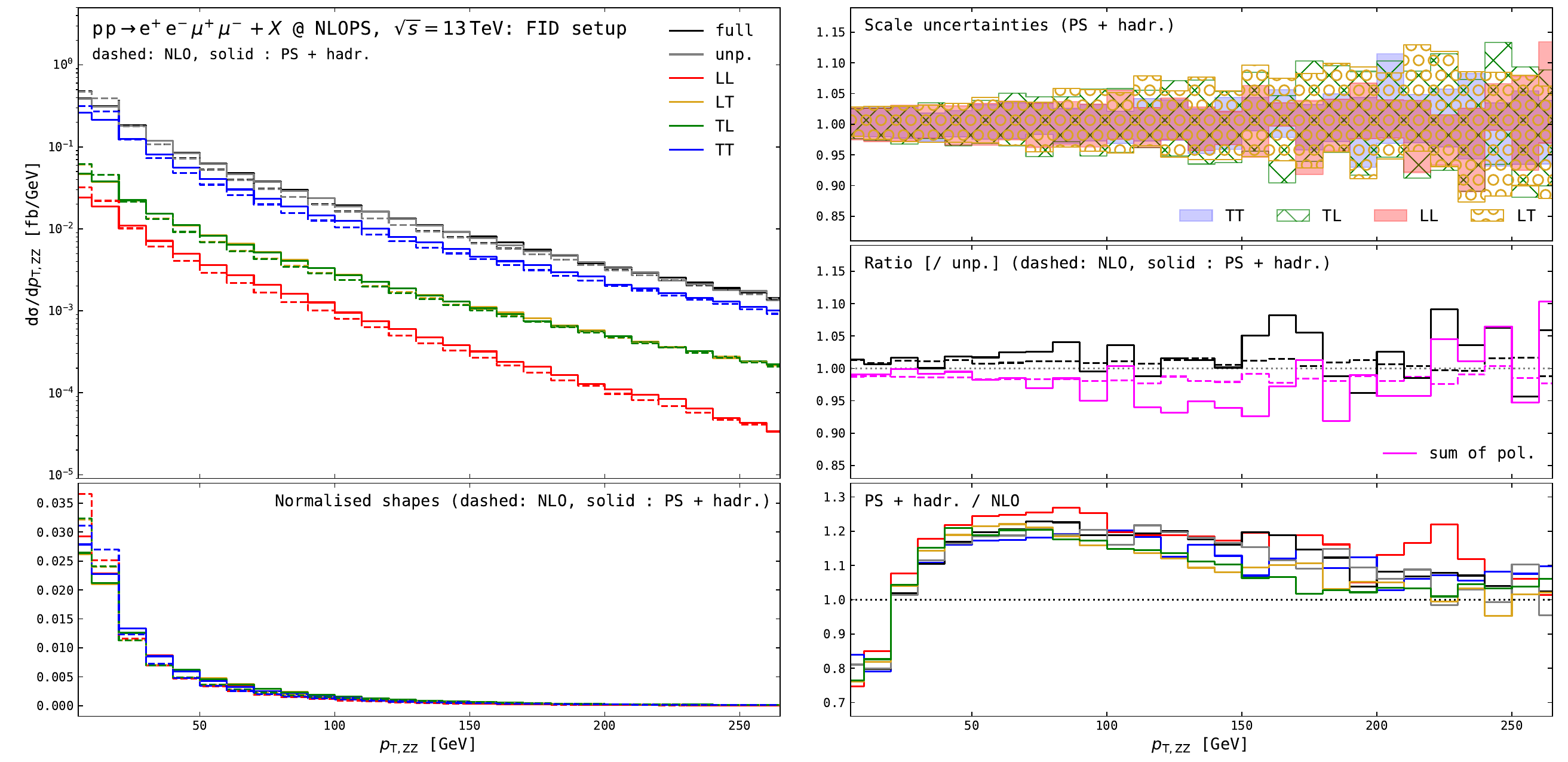}
  \caption{
    Distributions in the transverse momentum of the diboson system for $\PZ\PZ$ production at the LHC.
    The fiducial setup is understood. Same structure as \reffi{fig:ZZfid_1}. 
  }\label{fig:ZZfid_7}
\end{figure}
\begin{figure}[h]
  \centering
  \includegraphics[width=1.0\textwidth]{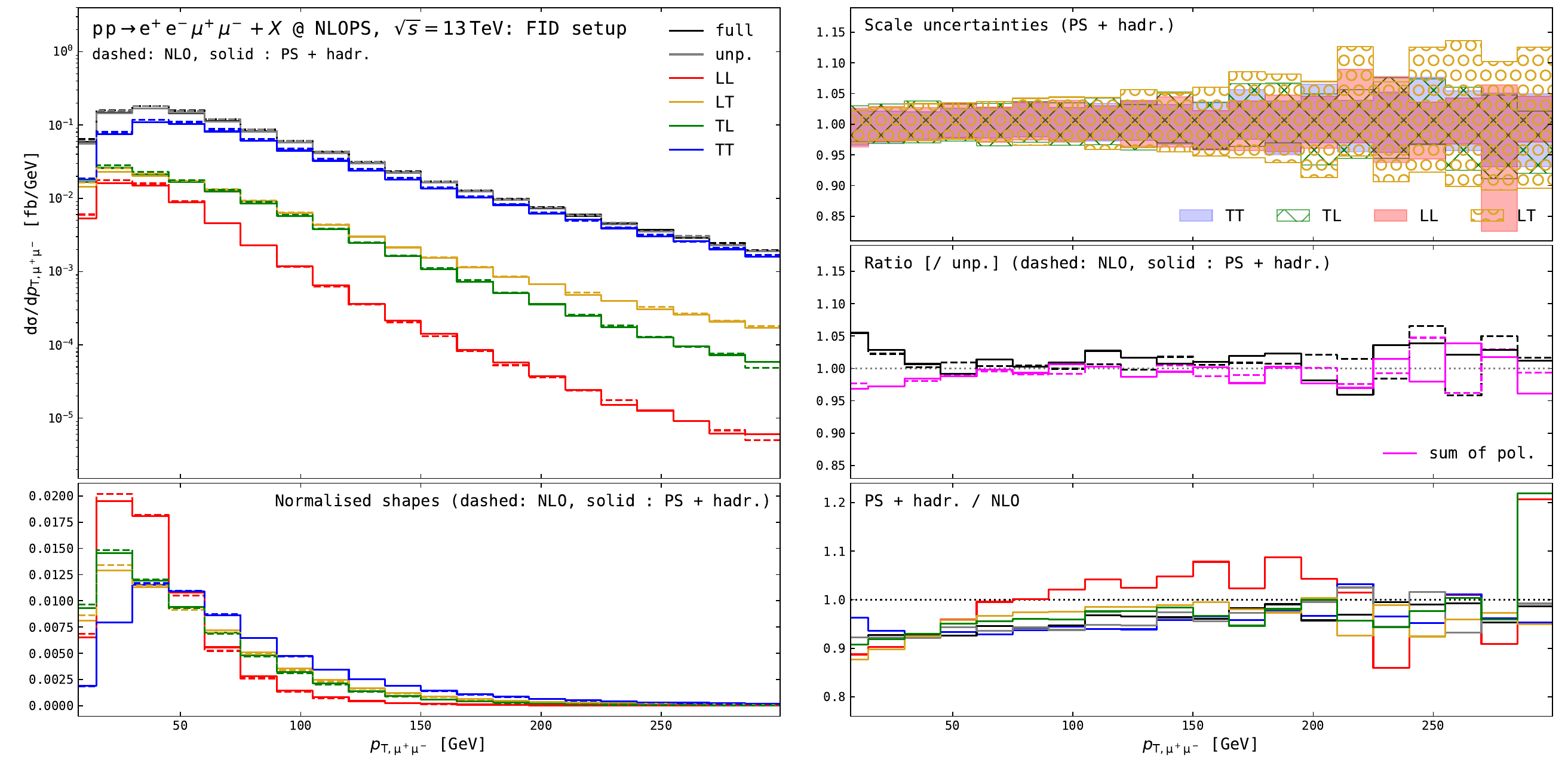}
  \caption{
    Distributions in the transverse momentum of the muon--antimuon system for $\PZ\PZ$ production at the LHC.
    The fiducial setup is understood.     
  }\label{fig:ZZfid_2}
\end{figure}
\newpage
\begin{figure}[h]
  \centering
  \includegraphics[width=1.0\textwidth]{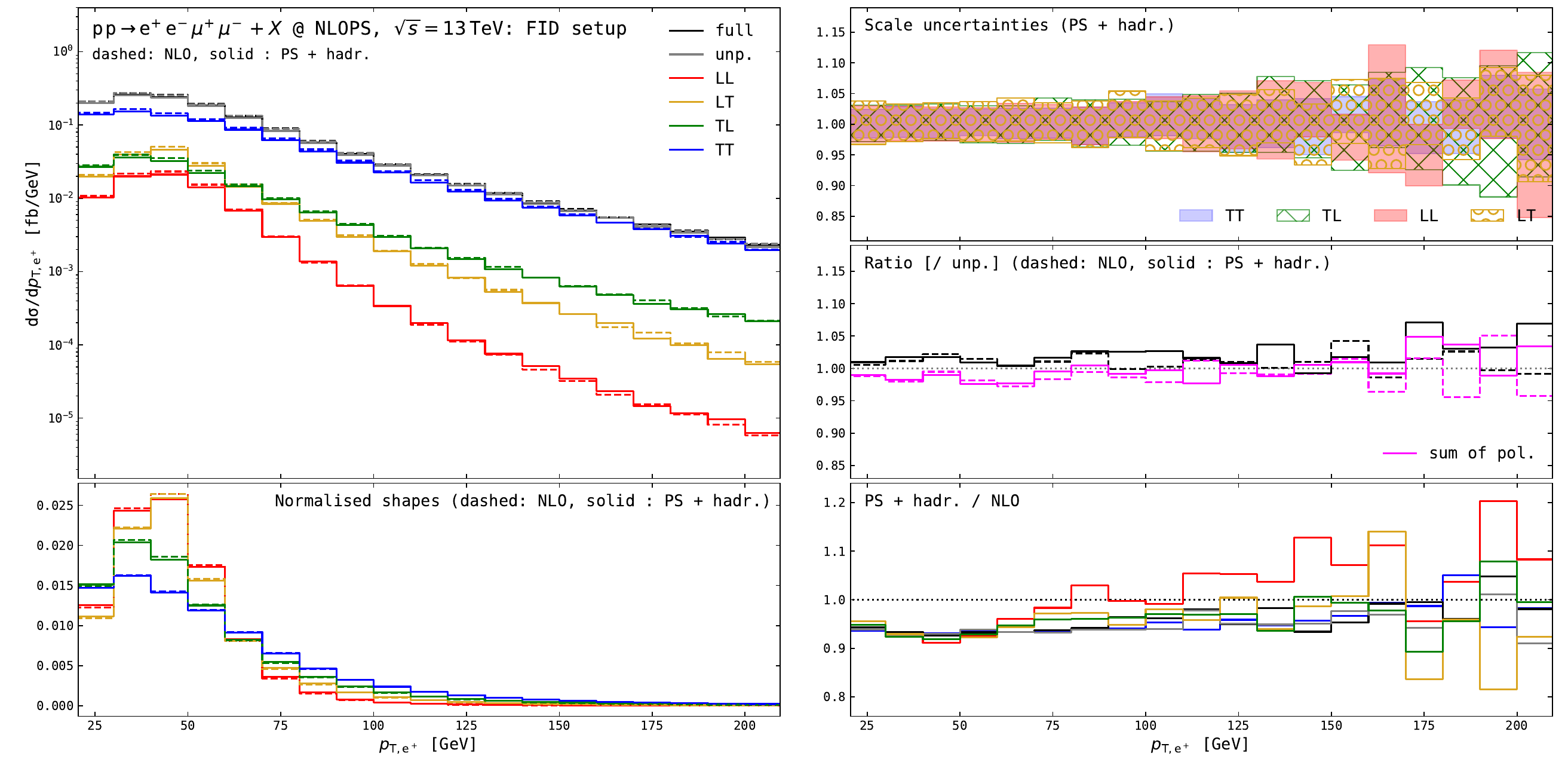}
  \caption{
    Distributions in the transverse momentum of the positron for $\PZ\PZ$ production at the LHC.
    The fiducial setup is understood. Same structure as \reffi{fig:ZZfid_1}.    
  }\label{fig:ZZfid_3}
\end{figure}
\begin{figure}[h]
  \centering
  \includegraphics[width=1.0\textwidth]{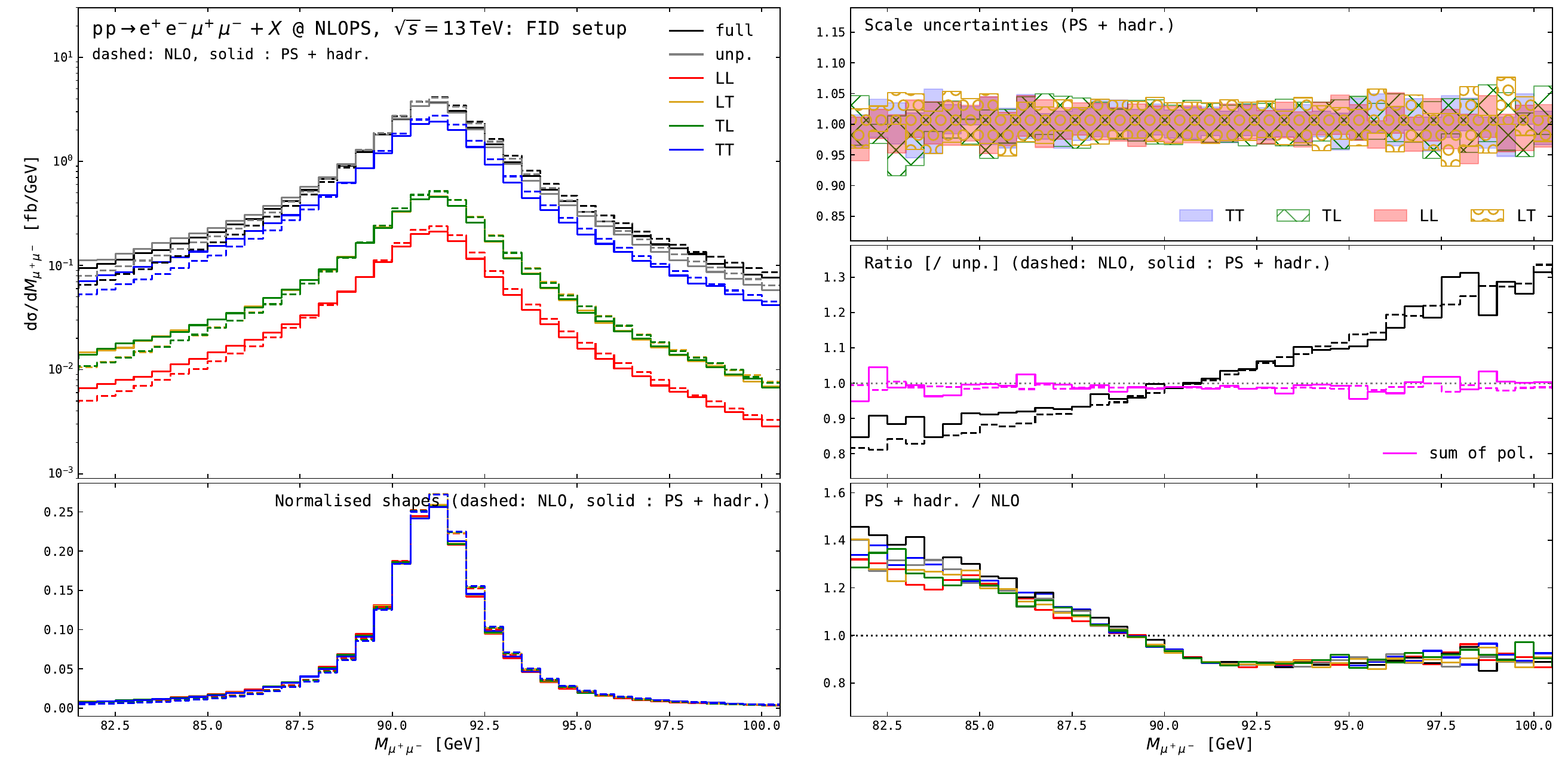}
  \caption{
    Distributions in the invariant mass of the muon--antimuon system for $\PZ\PZ$ production at the LHC.
    The fiducial setup is understood. Same structure as \reffi{fig:ZZfid_1}.    
  }\label{fig:ZZfid_4}
\end{figure}
\newpage
\begin{figure}[h]
  \centering
  \includegraphics[width=1.0\textwidth]{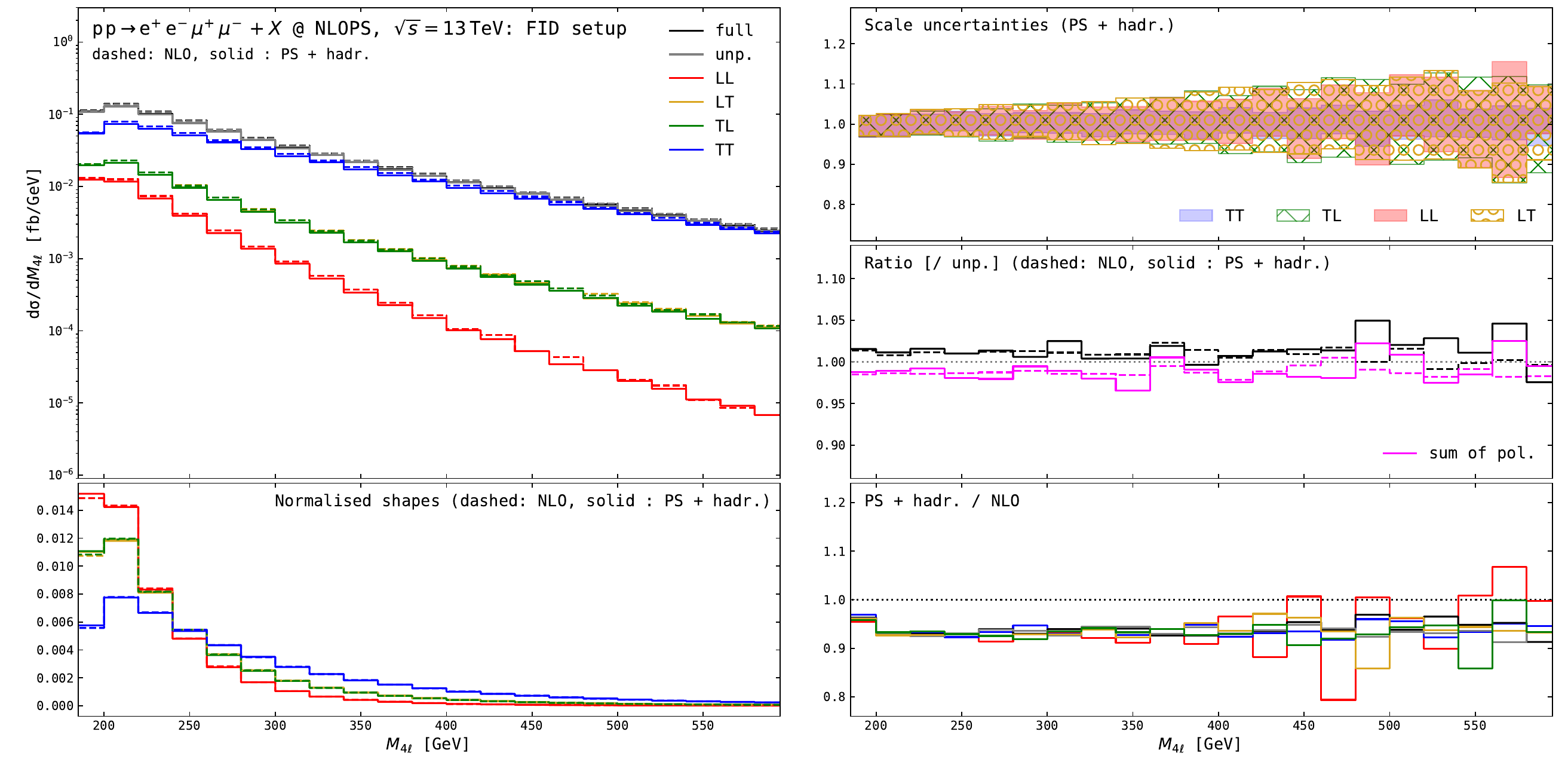}
  \caption{
    Distributions in the invariant mass of the diboson system  for $\PZ\PZ$ production at the LHC.
    The fiducial setup is understood. Same structure as \reffi{fig:ZZfid_1}.    
  }\label{fig:ZZfid_5}
\end{figure}
\begin{figure}[h]
  \centering
  \includegraphics[width=1.0\textwidth]{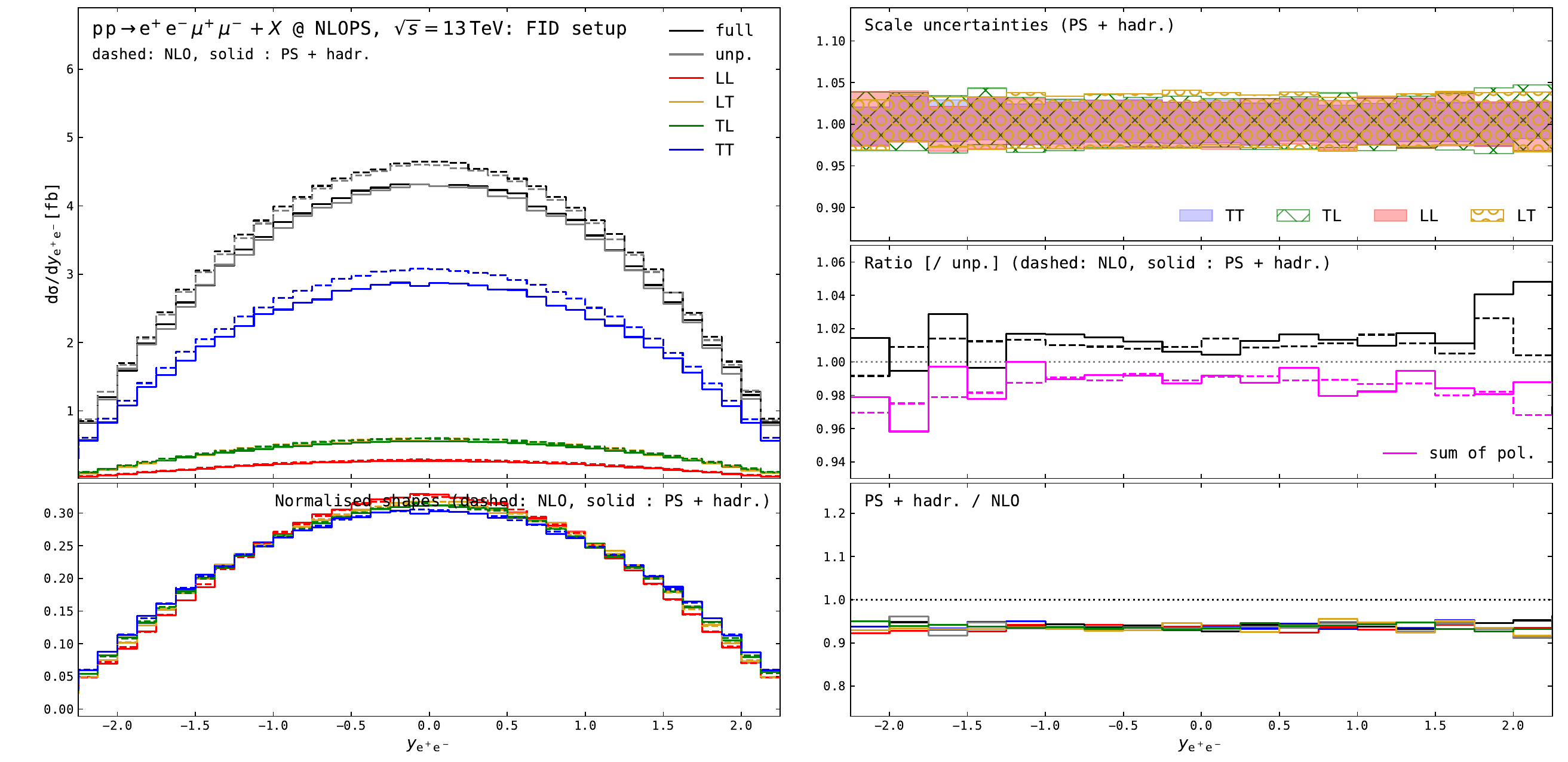}
  \caption{
    Distributions in the rapidity of the electron--positron system for $\PZ\PZ$ production at the LHC.
    The fiducial setup is understood. Same structure as \reffi{fig:ZZfid_1}.    
  }\label{fig:ZZfid_6}
\end{figure}
\newpage
\begin{figure}[h]
  \centering
  \includegraphics[width=1.0\textwidth]{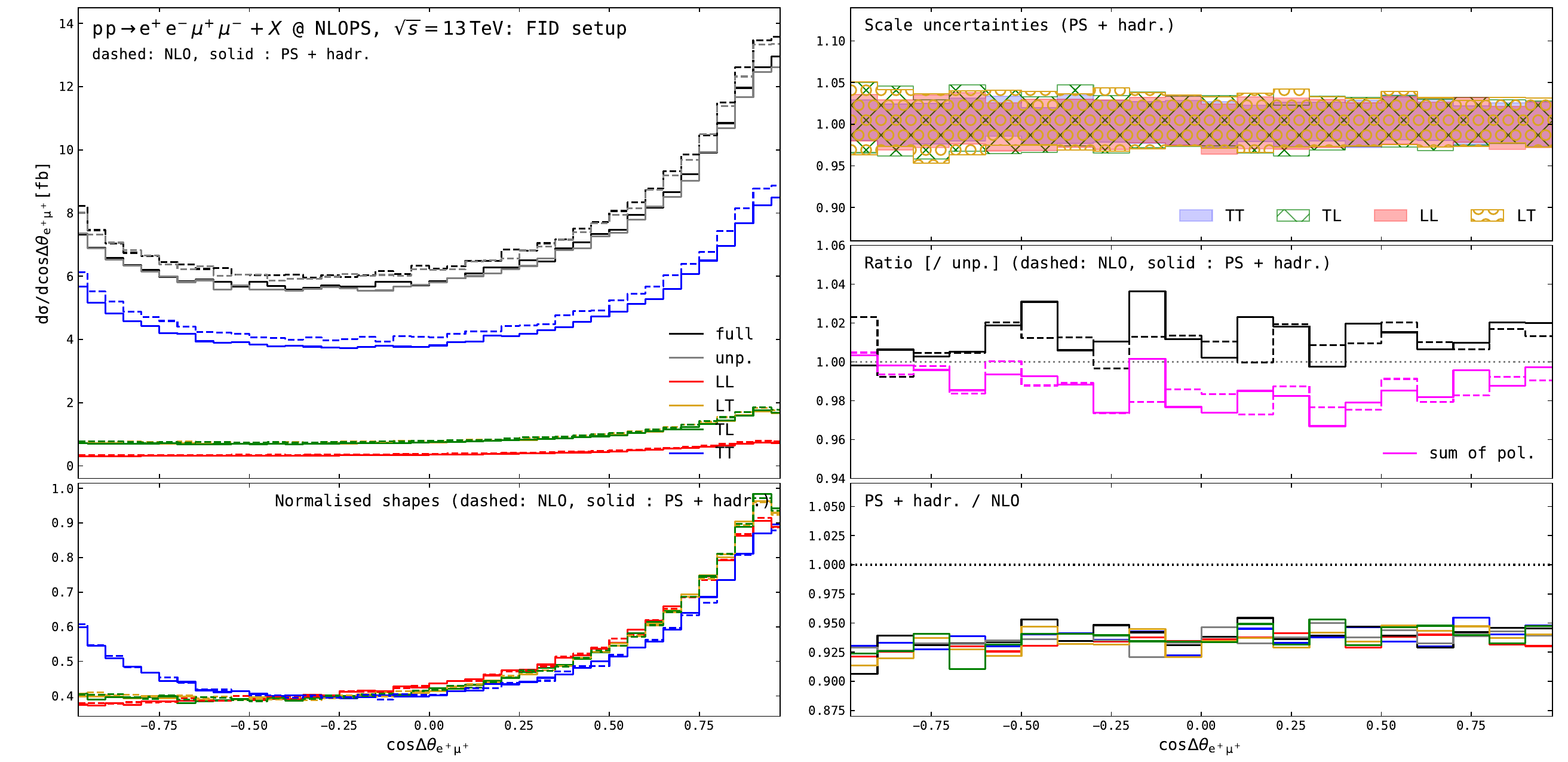}
  \caption{
    Distributions in the 3D angular separation between the positron and the antimuon for $\PZ\PZ$ production at the LHC.
    The fiducial setup is understood. Same structure as \reffi{fig:ZZfid_1}.    
  }\label{fig:ZZfid_8}
\end{figure}
\begin{figure}[h]
  \centering
  \includegraphics[width=1.0\textwidth]{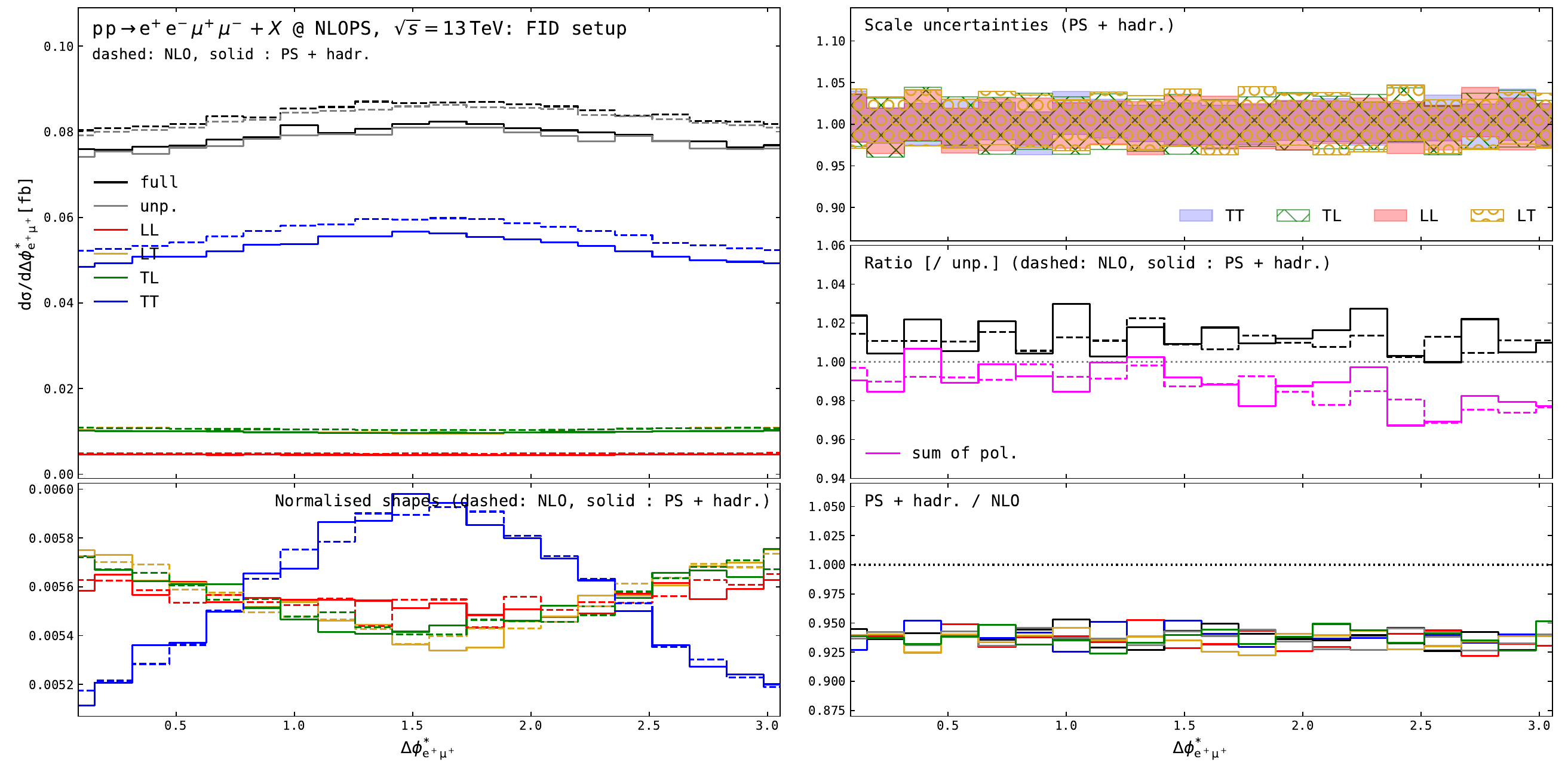}
  \caption{
    Distributions in the difference between the azimuthal decay angles of the positively charged lepton in each boson rest frame for $\PZ\PZ$ production at the LHC.
    The fiducial setup is understood. Same structure as \reffi{fig:ZZfid_1}.    
  }\label{fig:ZZfid_9}
\end{figure}

\newpage
\subsection{Additional distributions for WW}\label{app:WW}
\begin{figure}[h]
  \centering
  \includegraphics[width=1.0\textwidth]{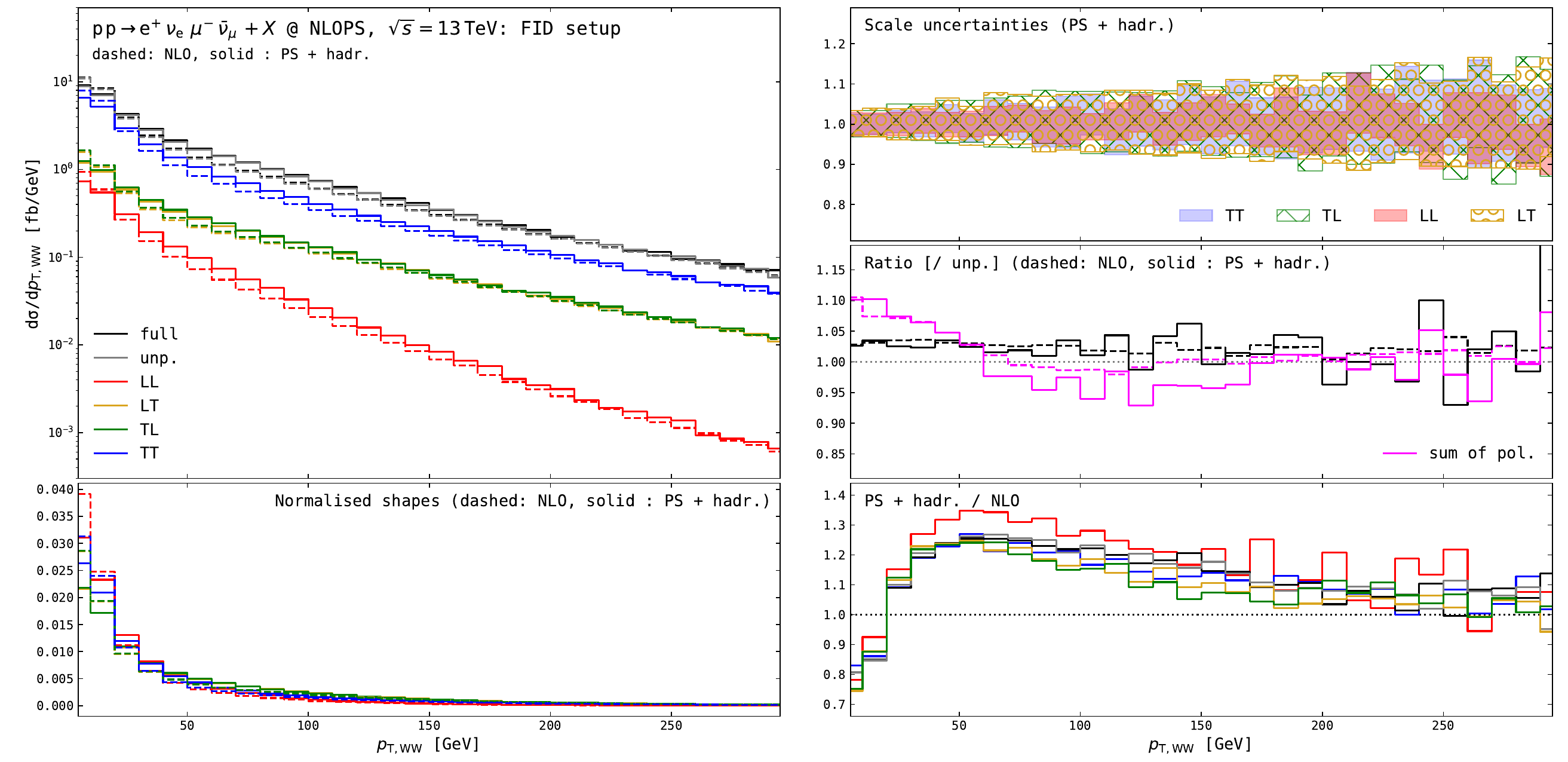}
  \caption{
    Distributions in the transverse momentum of the diboson system for $\PW^+\PW^-$ production at the LHC.
    The fiducial setup defined in Eq.~\eqref{eq:fiddefWW} is understood.
    Same structure as \reffi{fig:WWfid_1}.    
  }\label{fig:WWfid_6}
\end{figure}
\begin{figure}[h]
  \centering
  \includegraphics[width=1.0\textwidth]{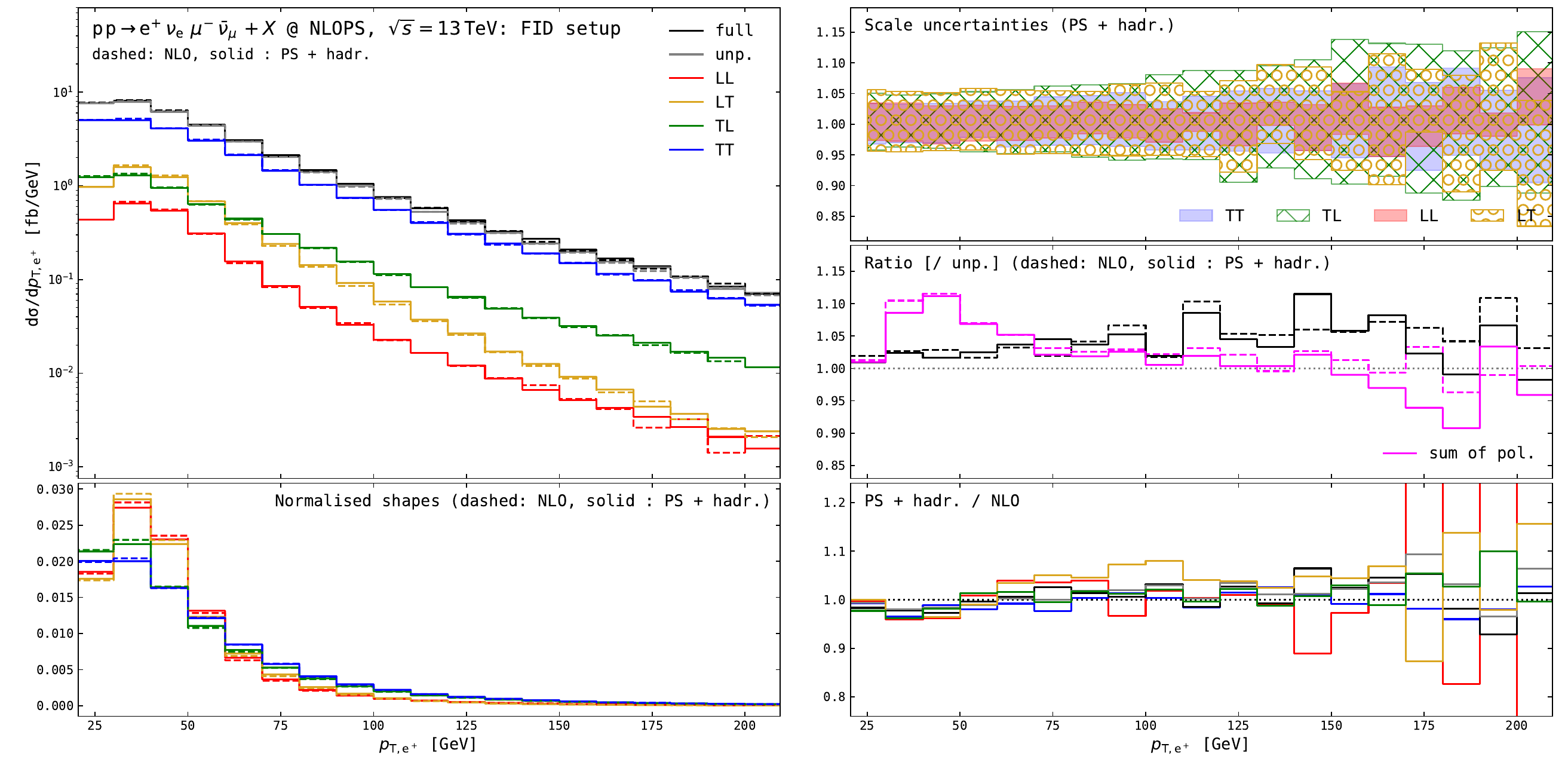}
  \caption{
    Distributions in the transverse momentum of the positron for $\PW^+\PW^-$  production at the LHC.
    The fiducial setup defined in Eq.~\eqref{eq:fiddefWW} is understood.
    Same structure as \reffi{fig:WWfid_1}.    
  }\label{fig:WWfid_4}
\end{figure}
\newpage
\begin{figure}[h]
  \centering
  \includegraphics[width=1.0\textwidth]{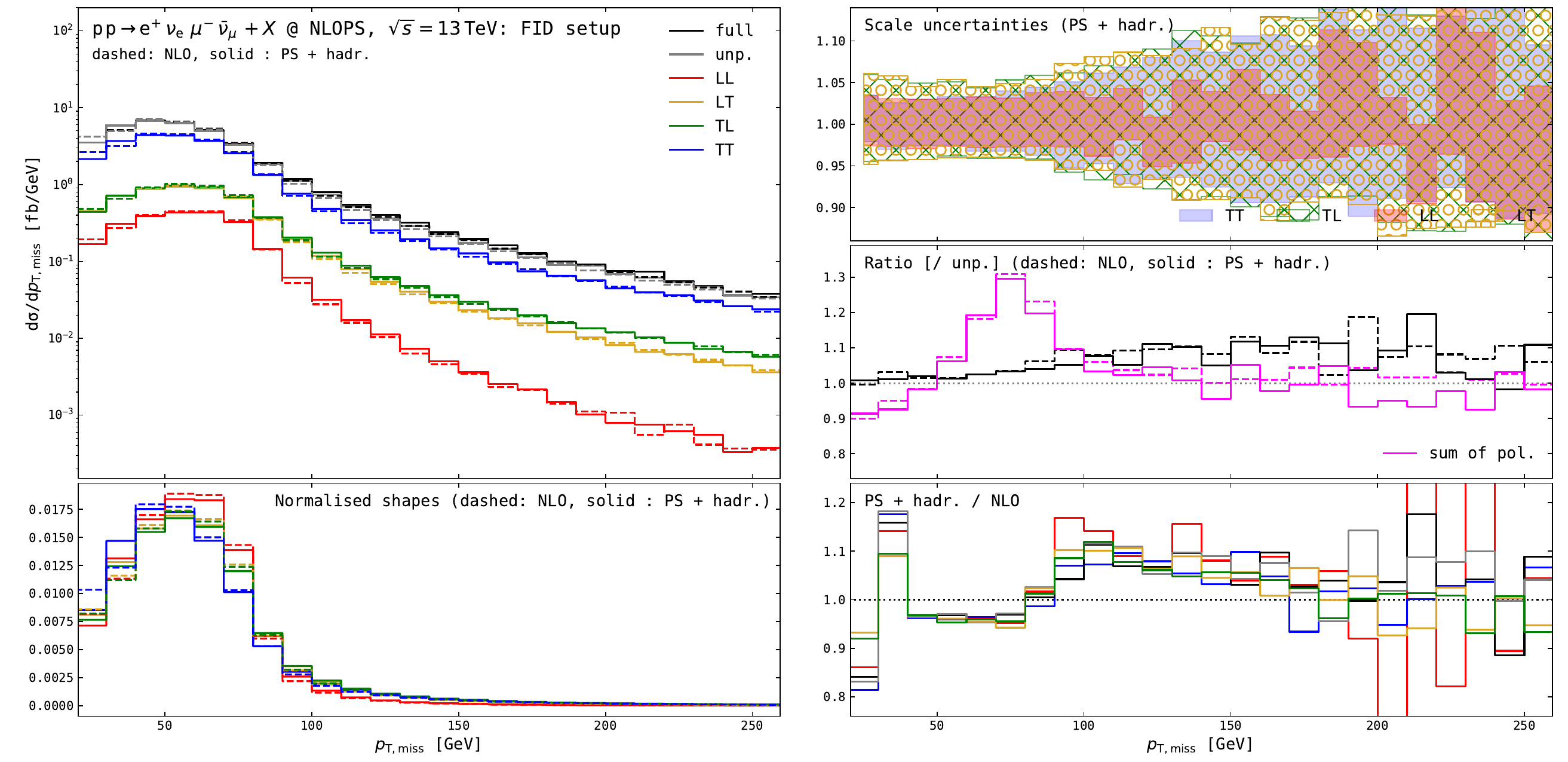}
  \caption{
    Distributions in the missing transverse momentum for $\PW^+\PW^-$  production at the LHC.
    The fiducial setup defined in Eq.~\eqref{eq:fiddefWW} is understood.
    Same structure as \reffi{fig:WWfid_1}.    
  }\label{fig:WWfid_5}
\end{figure}
\begin{figure}[h]
  \centering
  \includegraphics[width=1.0\textwidth]{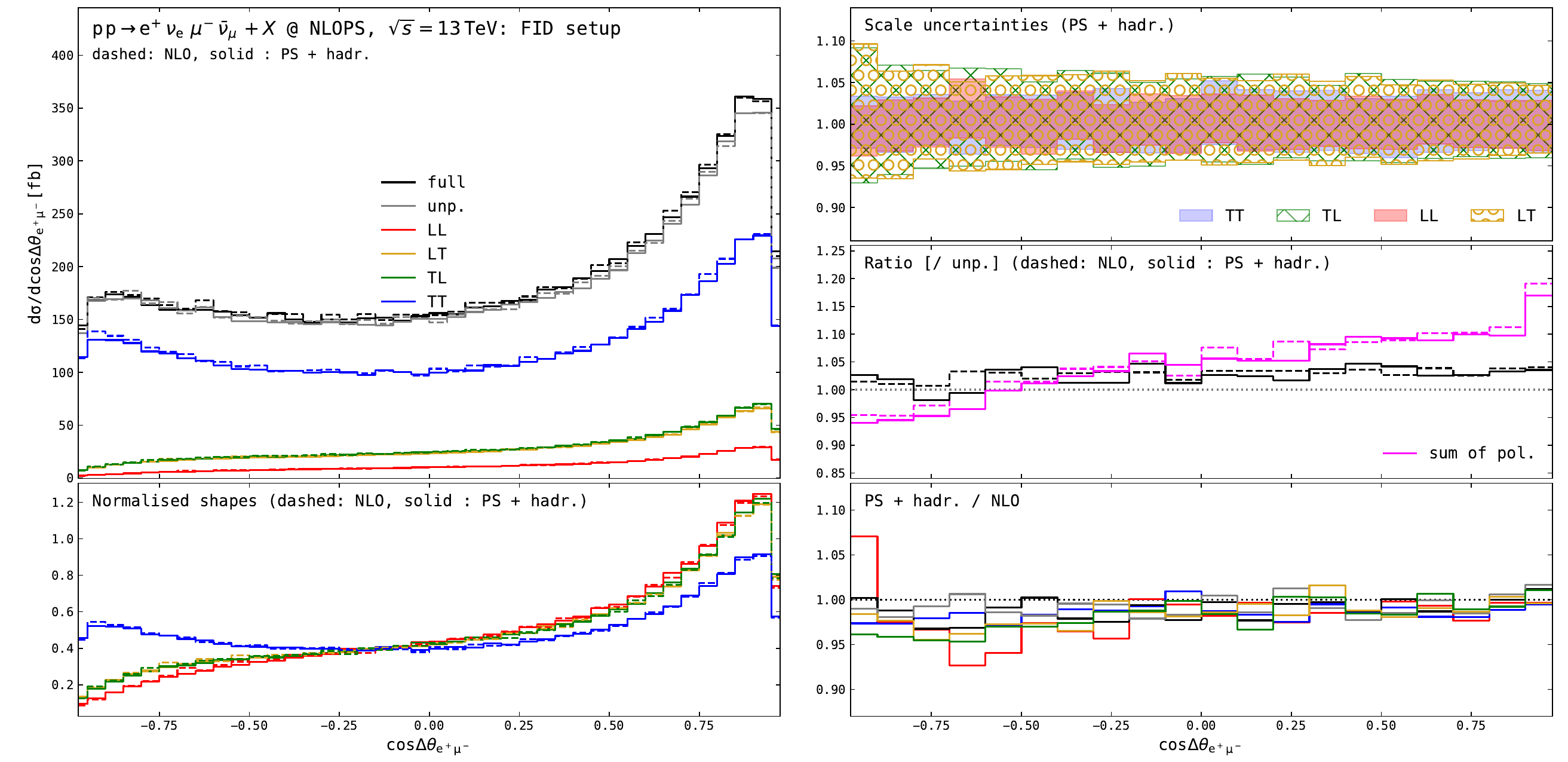}
  \caption{
    Distributions in the 3D angular separation between the positron and the muon for $\PW^+\PW^-$  production at the LHC.
    The fiducial setup defined in Eq.~\eqref{eq:fiddefWW} is understood.
    Same structure as \reffi{fig:WWfid_1}.    
  }\label{fig:WWfid_3}
\end{figure}

\newpage

\bibliographystyle{elsarticle-num_mod}
\bibliography{polvv}

\end{document}